\documentclass[10pt]{amsart}

\usepackage{amsmath,amssymb,amsthm}
\usepackage{amsmath,amssymb,amsthm,amscd}
\numberwithin{equation}{section}
\title{Existence of Supersymmetric Hermitian metrics with torsion on
non-kahler manifolds }

\author{Ji-Xiang Fu}
\author{Shing-Tung Yau}
\thanks{}

\address{Institute of Mathematics\\  Fudan University\\ Shanghai 200433, China}
\email{majxfu@fudan.edu.cn}
\address{Department of Mathematics\\ Harvard University\\ Cambridge,
MA 02138} \email{yau@math.harvard.edu}
\date{}

\newtheorem{prop}{Proposition}
\newtheorem{theo}[prop]{Theorem}
\newtheorem{lemm}[prop]{Lemma}

\newtheorem{rema}[prop]{Remark}

\DeclareMathOperator{\supp}{Supp}

\DeclareMathOperator{\Endo}{End}

\def\g{\mathfrak g}

\def\R{{\mathfrak R}}

\def\cC{{\mathcal C}}

\def\cH{{\mathcal H}}

\def\bP{{\mathbf P}}

\def\bL{{\mathbf L}}

\def\brr{{\mathbf r}}

\def\and{\quad{\rm and}\quad}

\def\bl{\bigl(}
\def\br{\bigr)}

\def\dbar{\bar\partial}

\def\pri{^{\prime}}

\def\sub{\subset}
\def\sta{^{\ast}}

\def\lab{\label}

\begin{document}

 \maketitle
\section{Introduction}
In their proposed compactification of superstring [6], Candelas,
Horowitz, Strominger and Witten took the matric product of a
maximal symmetric four dimensional spacetime $M$ with a six
dimensional Calabi-Yau vacua $X$ as the ten dimensional spacetime;
they identified the Yang-Mills connection with the SU(3)
connection of the Calabi-Yau metric and set the dilaton to be a
constant. To make this theory compatible with the standard grand
unified field theory, Witten \cite{W1} and Horava-Witten \cite{HW}
proposed to use higher rank bundles for strong coupled heterotic
string theory so that the gauge groups can be SU(4) or SU(5).
Mathematically, this approach relies on Uhlenbeck-Yau's theorem on
constructing Hermitian-Yang-Mills connections over stable bundles
\cite{UY}.

In \cite{Str}, A.~Strominger analyzed heterotic superstring
background with spacetime sypersymmetry and non-zero torsion by
allowing a scalar ``warp factor'' to multiply the spacetime
metric. He considered a ten dimensional spacetime that is the
product $M\times X$ of a maximal symmetric four dimensional
spacetime $M$ and an internal space $X$; the metric on $M\times X$
takes the form
$$e^{2D(y)}\left(%
\begin{array}{cc}
g_{ij}(y)&0\\
0&g_{\mu\nu}(x)\\
\end{array}
\right),\qquad x\in X,\quad y\in M;
$$
the connection on an auxiliary bundle is Hermitian-Yang-Mills over
$X$:
$$F\wedge\omega^2=0,\quad  F^{2,0}=F^{0,2}=0.
$$
Here $\omega$ is the hermitian form
$\omega=\frac{\sqrt{-1}}{2}g_{i\bar j}dz^i\wedge d\bar z^j$. In
this system, the physical relevant quantities are
$$h=\frac{\sqrt{-1}}{2}(\bar\partial-\partial)\omega,
$$
$$\phi=\frac{1}{8}\log\|\Omega\|+\phi_0,
$$
and
$$g_{ij}^0=e^{2\phi_0}\|\Omega\|^{\frac{1}{4}}g_{ij},
$$
for a constant $\phi_0$. The spacetime supersymmetry forces $D(y)$
to be the dilaton field.

In order for such ansatze to provide a supersymmetric
configuration, one introduces a Majorana-Weyl spinor $\epsilon$ so
that
$$
\delta \phi_j^0= \nabla^0_j\epsilon^0 +\frac{1}{48}e^{2\phi}\bl
\gamma_j^0 H^0-12 h_j^0\br \epsilon^0=0,
$$
$$
\delta\lambda^0 = \nabla^0\phi
\epsilon^0+\frac{1}{24}e^{2\phi}h^0\epsilon^0=0,
$$
$$
\delta \chi^0 = e^{\phi} F_{ij}\Gamma^{0ij}\epsilon^0=0,
$$
where $\psi^0$ is the gravitano, $\lambda^0$ is the dilatino,
$\chi^0$ is the gluino, $\phi$ is the dilaton and $h$ is the
Kalb-Ramond filed strength obeying\footnote{The curvature $F$ of
vector bundle $E$ in ref.\cite{Str} is real, i.e.,
$c_1(E)=\frac{F}{2\pi}$.  But we are used to take the curvature
$F$ satisfying $c_1(E)=\frac{\sqrt{-1}}{2\pi}F$.}
$$dh=\alpha'(\text{tr} F\wedge F-\text{tr} R\wedge R),
$$
where $\alpha'>0$. (For details of this discussion, please consult
\cite{Str,Str2}.) By suitably transforming these quantities,
Strominger showed that in order to achieve space-time
supersymmetry the internal six manifold $X$ must be a complex
manifold with a non-vanishing holomorphic three form $\Omega$; the
Hermitian form $\omega$ must obey
$$\sqrt{-1}\partial\bar\partial\omega=\alpha'(\text{tr} F\wedge
F-\text{tr} R\wedge R)
$$
and\footnote{See eq. (56) of ref.\cite{Str2}, which corrects eq.
(2.30) of ref.\cite{Str} by a minus sign.}
$$d\sta\omega=\sqrt{-1}(\dbar-\partial)\log\|\Omega\|_\omega.
$$
Accordingly, he proposed to solve the system
\begin{equation}\label{101}
F_H\wedge\omega^2=0;
\end{equation}
\begin{equation}\label{102}
F^{2,0}_H=F^{0,2}_H=0;
\end{equation}
\begin{equation}\label{103}
\sqrt{-1}\partial\dbar \omega= \alpha'(\text{tr}F_H\wedge
F_H-\text{tr} R\wedge R);
\end{equation}
\begin{equation}\label{104}
d\sta \omega=\sqrt{-1}(\dbar-\partial)\ln\|\Omega\|_\omega
\end{equation}
that are solutions of superstring with torsion that allows
non-trivial dilaton field and Yang-Mills field.  Here $\omega$ is
the Hermitian form and $R$ is the curvature tensor of the
Hermitian metric $\omega$; $H$ is the Hermitian metric and $F$ is
its curvature of a vector bundle $E$; the $\text{tr}$ is the trace
of the endomorphism bundle of either $E$ or $TX$.

In \cite{LY}, Li and Yau have proven the following useful:
\begin{lemm}
The equation (\ref{104})  is equivalent to
\begin{equation}\label{105}
d(\parallel\Omega\parallel_\omega\omega^2)=0.
\end{equation}
\end{lemm}
In their paper, Li and Yau have given the first irreducible
non-singular solution of the supersymmetric  system of Strominger
for $U(4)$ and $U(5)$ principle bundle. They obtain their
solutions by perturbing around the Calabi-Yau vacua paired with
the gauge field that  is the tangent connection.

It was speculated by M.~Reid that all Calabi-Yau manifolds can be
deformed to each other through conifold transition. To achieve
this goal, it is inevitable that we must work with non-Kahler
manifolds.

The most common examples of non-Kaehler manifolds $X$ are some
$T^2$ bundles over Calabi-Yau varieties
\cite{BBDG,BBDP,CDL,GP,GLM,KST}. Because internal six manifold $X$
is a complex manifold with a non-vanishing holomorphic three form
$\Omega$, at first we may consider the $T^2-$bundle
$(X,\omega,\Omega)$ over complex surface $(S,\omega_S,\Omega_S)$
with non-vanishing holomorphic 2-form $\Omega_S$. According to the
classification of complex surfaces by Enriques and Kodaira, such
surfaces include K3 surface and complex torus (Calabi-Yau) and
Kodaira  surface (non-Kahler). If $(X,\omega,\Omega)$ satisfies
the Strominger's equation (\ref{104}), then  by Lemma 1,
$d(\parallel\Omega\parallel_\omega\omega^2)=0$. If we let
$\omega'=\parallel\Omega\parallel_\omega^{\frac{1}{2}}\omega$,
then $d\omega'^2=0$, i.e., $\omega'$ is a balanced metric
\cite{MI}. Balanced metric is a very interesting concept. This was
studied extensively by Michelson. For example, Michelson proved
that the balanced condition  is preserved under the proper
holomorphic submersion.  Note that Alessandrini and Bassanelli
\cite{AB} proved that this condition is also preserved under the
modofications. Now $X$ is balanced and  holomorphic submersion
$\pi$ from $X$ to complex surface $S$ is proper, so $S$ is  also
balanced (actually $\pi_\ast\omega'^2$ is the balanced metric, see
proposition 1.9 in \cite{MI}). Note that when the dimension of
complex manifold is 2, the conditions of being balanced and
Kaehler concide. So $S$ is Kaehler. Then there is no solution to
Strominger's equation on $T^2$ bundles over Kodaira surface and we
should only consider the case of K3 surface and complex torus.

Up to now, only known example of the solution to Strominger's
system on non-Kahler manifold is given by Cardoso, Curio,
Dall'Agata and Lust in \cite{CDL}. By calculating the curvature,
they have given the reducible solution on the Iwasawa manifold
which is  some $T^2$  bundle over complex torus.

On the other hand, when K. Becker, M. Becker , K. Dasgupta etc.
finished their two papers on compactification of heterotic theory
on non-Kahler complex manifolds \cite{BBDG,BBDP}, M. Becker and K.
Dasgupta in their review paper \cite{BD} think that the question
of finding stable vector bundle for their manifolds (i.e., some
$T^2$ bundles over K3 surfaces) is a very important one especially
because we are no longer allowed to embed the spin connection
(i.e., the connection with torsion
$\sqrt{-1}(\bar\partial-\partial)\omega$) into the gauge
connection (hermitian connection). They think that Strominger's
equations (\ref{103}) and (\ref{104}) look very restrictive and
one might wonder if there exists any solution at all to the
equations.

 In this
 paper, we will construct this solution on some torus bundles over K3
 surface or complex torus provided by Goldstein and Prokushkin
 \cite{GP}. Let $(S,\omega_S,\Omega_S)$ be a K3 surface or complex
 torus with  kahler form $\omega_S$ and a non-vanishing holomorphic
(2,0) form $\Omega_S$. Let $\omega_1$ and $\omega_2$ are
anti-self-dual (1,1) forms such that $\frac{\omega_1}{2\pi}$ and
$\frac{\omega_2}{2\pi}$ represent integral cohomology classes.
Using these two forms, Goldstein and Prokushkin constructed the
non-Kahler manifold $X$ such that $\pi:X\rightarrow S$ is a
holomorphic $T^2$ fibration over $S$ with hermitian form
$\omega_0=\pi^*\omega_S+\frac{\sqrt{-1}}{2}\theta\wedge\bar\theta$
and holomorphic 3 form $\Omega=\Omega_S\wedge\theta$ (The
definition of $\theta$ see \cite{GP} or section 3). Now we
construct the superstring as follows.

Let $L_1$ and $L_2$ be holomorphic line bundle over $S$ such that
their curvatures are $\sqrt{-1}\omega_1$ and $\sqrt{-1}\omega_2$
respectively. Corresponding to these curvature,  there exist
hermitian metrics $h_1$ and $h_2$ on $L_1$ and $L_2$. Let
$E=L_1\oplus L_2\oplus T'S$ and let $H_0=(h_1,h_2,\omega_S)$. Then
$F_{H_0}=\text{diag}(\sqrt{-1}\omega_1,\sqrt{-1}\omega_2,R_{\omega_S})$.
Let $u$ be any smooth function on $S$ and let
\begin{equation}
\omega_u=\pi^*(e^u\omega_S)+\frac{\sqrt{-1}}{2}\theta\wedge\bar\theta.
\end{equation}
Then $(V=\pi^*E,\pi^*F_{H_0},X,\omega_u)$ satisfies the
Strominger's equation (\ref{101}),(\ref{102}) and (\ref{104}). So
we should only need to consider the equation (\ref{103}). Because
$\omega_1$ and $\omega_2$ are harmonic, locally write $\omega_1$
and $\omega_2$ as
$$\omega_1=\bar\partial\xi=\bar\partial(\xi_1dz_1+\xi_2dz_2)$$
and
$$\omega_2=\bar\partial\zeta=\bar\partial(\zeta_1dz_1+\zeta_2dz_2),$$
where $(z_1,z_2)$ is the local coordinate on $S$. Let
 $$A=\left(%
\begin{array}{c}
   \xi_1+\sqrt{-1}\zeta_1\\
  \xi_2+\sqrt{-1}\zeta_2  \\
\end{array}%
\right).$$ Using matrix $A$ we can calculate the  curvature $R_u$
of metric $\omega_u$ and $R_u\wedge R_u$. Let $g=(g_{i\bar j})$ if
$\omega_S=\frac{\sqrt{-1}}{2}g_{i\bar j}dz_i\wedge dz_{\bar j}$.
We can prove
\begin{theo}
$(V=\pi^*E,\pi^*F_{H_0},X,\omega_u)$ is the solution of
Strominger's system if and only if the function $u$ of $S$
satisfies the equation
\begin{equation}\label{107}
\bigtriangleup
e^u\cdot\frac{\omega^2_S}{2!}+\partial\dbar(e^{-u}\textup{tr}(\dbar
A\wedge \partial A^*\cdot g^{-1}))+\partial\dbar u\wedge
\partial\dbar u=0.
\end{equation}
In particular, when $\omega_2=n\omega_1$,$n\in \mathbb{Z}$,
$(V,\pi^*F_{H_0},X,\omega_u)$ is the solution to Strominger's
system if and only if smooth function $u$ on $S$ satisfies the
equation:
\begin{equation}\label{108}
\bigtriangleup\left(e^u+\frac{(1+n^2)}
{4}\parallel\omega_1\parallel_{\omega_S}^2e^{-u}\right)-8\frac{\det(u_{i\bar
j})}{\det(g_{i\bar j})}=0.
\end{equation}
\end{theo}
Actually we can prove  that
$$\text{tr}(\bar\partial A\wedge\partial A^*\cdot g_S^{-1})$$
is a globally well-defined (1,1)-form on $S$. In particular, when
$\omega_2=n\omega_1$, $n\in \mathbb{Z}$,
$$\text{tr}(\bar\partial A\wedge\partial A^*\cdot g_S^{-1})
=\sqrt{-1}\frac{1+n^2}{4}\parallel\omega_1\parallel_{\omega_S}^2\omega_S.$$
Let $f=\frac{(1+n^2)} {4}\parallel\omega_1\parallel^2$. If we let
$g'_{i\bar j}=(e^u-fe^{-u})g_{i\bar j}-4u_{i\bar j}$, then we can
rewrite the equation (\ref{108}) as
\begin{equation*}
\frac{\det g'_{i\bar j}}{\det g_{i\bar j}}=(e^u-fe^{-u})^2+
2\{(e^u+fe^{-u})\mid\bigtriangledown u\mid^2+e^{-u}\bigtriangleup
f- 2e^{-u}\bigtriangledown u\cdot\bigtriangledown f\}
\end{equation*}
We solve equation (\ref{107}) by the continuity method \cite{Yau}.
We will  prove
\begin{theo}
There is  an unique solution of equation (\ref{107}) under the
elliptic condition
$e^u\omega_S+\sqrt{-1}e^{-u}\textup{tr}(\bar\partial
A\wedge\partial A^*\cdot g_S^{-1})-2\sqrt{-1}\partial\bar\partial
u>0$ and the normalization $\int_S e^{-u}=A<<1$. If
$\omega_2=n\omega_1$, then
\begin{equation*}
A<\min \left\{1,C_1^{-1}\left(\max\{(7^{\frac{1}{3}},
(2C_1)^2,(1+\sup f),16(\max R_{i\bar jk\bar
l}+1)\}\right)^{-\frac{2}{B}}\right\}
\end{equation*}
where $C_1$ depends only on $S$ (it can be written by   P. Li's
notation in \cite{PL}) and constant $B$ is
\begin{equation*}
B=\prod_{\beta=1}^{\infty}\left(1-\frac{1}{2^{\beta}}\right)>0
\end{equation*}
\end{theo}
Actually we can get the estimate $\inf u\geq -\ln
C_1-\frac{B}{2}\ln A$. So if $A<C_1^{-\frac{2}{B}}$, then  $\inf
u>0$. This is important in our estimate.

Fix the solution $u$ of equation (\ref{107}). Then according to
theorem 2, we get the reducible solution
$(V,F_{\pi^*H_0},X,\omega_u)$.  It can be extended to a family of
irreducible solution by  perturbing around it. So we follow
Li-Yau's method \cite{LY} and get the following
\begin{theo}
 Let $(E,H_0,S,\omega_S)$ be as before. Fix
its holomorphic structure $D_0''$. Then there is a smooth
deformation $D_s''$ of $(E,D_0'')$ so that there are
Hermitian-Yang-Mills metric $H_s$ on $(E,D_s)$ and smooth function
$\phi_s$ on $S$ such that
$$\left(V=\pi^*E,\pi^*D_s'',\pi^*H_s,\pi^*(e^{u+\phi_s}\omega_S)+\frac{\sqrt{-1}}{2}\theta
\wedge\bar\theta\right)$$ are the irreducible solutions to
strominger's system on $X$ and so that $\lim_{s\rightarrow
0}\phi_s=0$ and $\lim_{s\rightarrow 0}H_s$ is a  regular reducible
hermitian Yang-Mills connection on $E=L_1\oplus L_2\oplus TS$.
\end{theo}

The organization of the paper is as follows: sect. 2 is a review
of some results of paper \cite{GP}. In sect. 3 we calculate
$\text{tr}R\wedge R$. Then we can construct the reducible solution
(Theorem 2) in sect. 4 and get irreducible solution (theorem 4) in
sect. 5. From sect. 6, we solve the equation (\ref{107}) (Theorem
3) by continuity method. At first, we prove the openness in sect.
6. We do the estimates up to third order from sect. 7 to sect. 10.
In order to write everything down clearly and easily, we do
estimates only to equation (\ref{108}). Then we summarize the
estimates in sect. 11 to get the closeness for equation
(\ref{108}). Finally, in sect. 12, we explain why  we can easily
generalize our estimates for equation (\ref{108}) to estimates for
equation (\ref{107}).

 {\bf Acknowledgement.}\ \ J.X. Fu would like
to thank Professor Jun Li for interesting discussions and helps.
The Proposition 15 is due to Jun Li.  J.X. Fu is supported by NSFC
grant 10471026. S.T. Yau is supported partially by NSF grants
DMS-0306600 and DMS-0074329.

\section{Geometric Modules} In this section, we take Goldestein and
Prokushkin's geometric model for complex non-Kahler manifolds with
$SU(3)$ structure \cite{GP}. We organize their some  results as
the following:
\begin{theo}\cite{GP}
Let $(S,\omega_S,\Omega_S)$ be a Calabi-Yau\ 2-fold with  a
non-vanishing holomorphic $(2,0)-$form $\Omega_S$. Let $\omega_1$
and $\omega_2$ be closed 2-forms on $S$ satisfying the
following conditions:\\
(1). $\omega_1$ and $\omega_2$ are anti-self dual $(1,1)$-forms,
 $\ast\omega_1=-\omega_1, \ast\omega_2=-\omega_2$, which are equivalent to
\begin{equation}\label{201}
 \omega_1\wedge \omega_S=0,\ \ \ \  \omega_2\wedge \omega_S=0.
 \end{equation}
(2). $\frac{\omega_1}{2\pi}$ and $\frac{\omega_2}{2\pi}$ represent
integral cohomology classes.\\
Then there is a hermitian 3-fold $X$ such that $\pi:X\rightarrow
S$ is a holomorphic $T^2$-fibration over $S$ such that following holds: \\
1. For any 1-forms $\alpha$ and $\beta$ defined on some open
subset of $S$ and satisfying $d\alpha=\omega_1$ and
$d\beta=\omega_2$ there are local coordinates $x$ and $y$ on $X$
such that $dx+idy$ is a holomorphic form on $T^2$-fibers and the
metric on $X$ has the following form:
\begin{equation}\lab{202}
g_0=\pi^*g+(dx+\pi^*\alpha)^2+(dy+\pi^*\beta)^2
\end{equation}
where $g$ is the Calabi-Yau metric on $S$ corresponding to Kahler
form
$\omega_S$.\\
2. $X$ admits a nowhere vanishing holomorphic $(3,0)$-form with
unit length:
$$\Omega=((dx+\pi^*\alpha)+i(dy+\pi^*\beta))\wedge\pi^*\Omega_S$$
3. If either $\omega_1$ or $\omega_2$ represent a non-trivial
cohomological class then $X$ admits no Kahler metric. \\
4. But $X$  is a balanced manifold \cite{MI}. Actually hermitian
form
\begin{equation}\label{2011}
\omega_0=\pi^*\omega_S+(dx+\pi^*\alpha)\wedge(dy+\pi^*\beta) ;
\end{equation}
corresponding to the metric (\ref{202}) is balanced, i.e., $d\omega^2_0=0$;\\
5. Furthermore,  for any smooth function $u$ on $S$, the hermitian
metric
\begin{equation*}
\omega_u=\pi^*(e^u\omega_S)+(dx+\pi^*\alpha)\wedge(dy+\pi^*\beta)
\end{equation*}
is also balanced.
\end{theo}

Goldestein and Prokushkin also
have studied the cohomology of this
non-K\"{a}hler manifold $X$:
$$h^{1,0}(X)=h^{1,0}(S),$$
$$h^{0,1}(X)=h^{0,1}(S)+1;$$
In particular
$$h^{0,1}(X)=h^{1,0}(X)+1.$$
Moreover,
\begin{eqnarray*}
b_1(X)&=&b_1(S)+1,\ \ \text{when}\ \ \omega_2=n\omega_1,\\
b_1(X)&=&b_1(S),\ \ \ \ \ \ \ \text{when}\ \ \omega_2\neq
n\omega_1;\\
 b_2(X)&=&b_2(S)-1,\ \ \text{when}\ \
\omega_2=n\omega_1,\\
b_2(X)&=&b_2(S)-2,\ \  \text{when}\ \ \omega_2\neq n\omega_1
\end{eqnarray*}
and $$ b_3(X)=0.$$

\section{The calculation of $\text{tr}R\wedge R$}
In order to calculate the curvature $R$ and $\text{tr}R\wedge R$,
we should express the Hermitian metric (\ref{202}) under some
basis of  holomorphic (1,0) vector fields. So at first we should
write down the complex structure on $X$. Let
$\{U,z_j=x_j+\sqrt{-1}y_j,j=1,2\}$ be a local coordinate in $S$.
The horizontal lifts of vector fields $\frac{\partial}{\partial
x_j}$ and $\frac{\partial}{\partial y_j}$ which are in the kernel
of $dx+\pi^*\alpha$ and $dy+\pi^*\beta$ are
\begin{equation*}
X_j=\frac{\partial}{\partial
x_j}-\alpha\left(\frac{\partial}{\partial
x_j}\right)\frac{\partial}{\partial
x}-\beta\left(\frac{\partial}{\partial
x_j}\right)\frac{\partial}{\partial y}\ \ \ \ \text{for} \ \
j=1,2,
\end{equation*}
\begin{equation*}
Y_j=\frac{\partial}{\partial
y_j}-\alpha\left(\frac{\partial}{\partial
y_j}\right)\frac{\partial}{\partial
x}-\beta\left(\frac{\partial}{\partial
y_j}\right)\frac{\partial}{\partial y}\ \ \ \ \text{for} \ \
j=1,2.
\end{equation*}
Then the complex structure $\tilde{I}$ on $X$ is defined as
\begin{eqnarray*}
\tilde{I}X_j&=&Y_j,\ \ \ \  \tilde{I}Y_j=-X_j, \ \ \ \ \text{for}
\ \ j=1,2,\\
\tilde{I}\frac{\partial}{\partial x}&=&\frac{\partial}{\partial
y}, \ \ \ \ \tilde{I}\frac{\partial}{\partial
y}=-\frac{\partial}{\partial x}.
\end{eqnarray*}
Let
\begin{eqnarray*}
U_j&=&X_j-\sqrt{-1}\tilde{I}X_j=X_j-\sqrt{-1}Y_j,\\
U_0&=&\frac{\partial}{\partial x}
-\sqrt{-1}\tilde{I}\frac{\partial}{\partial x}
=\frac{\partial}{\partial x}-\sqrt{-1}\frac{\partial}{\partial y}.
\end{eqnarray*}
Then $\{U_j,U_0\}$ is the basis of the (1,0) vector fields on $X$.
Under this basis, the metric (\ref{202}) takes the following
hermitian matrix:
\begin{equation}\label{203}
\left(\begin{array} {cc} (g_{i\bar j}) & 0\\
0 & 1
\end{array}\right)
\end{equation}
because $U_1$ and $U_2$ are in the kernel of $dx+\pi^*\alpha$ and
$dy+\pi^*\beta$. Let
\begin{equation}\label{2012}
 \theta=dx+\sqrt{-1}dy+\pi^*(\alpha+\sqrt{-1}\beta)
 \end{equation}
It's easily checked that
$\{\pi^*d\overline{z}_j,\overline\theta\}$ annihilates the
$\{U_j,U_0\}$ and  is the basis of  (0,1) forms on $X$. So
$\{\pi^*dz_j,\theta\}$ are (1,0) forms on $X$. Certainly
$\pi^*dz_j$ are holomorphic (1,0) forms and $\theta$ is not. So we
should construct another holomorphic (1,0) form on $X$. Because
$\omega_1$ and $\omega_2$ are harmonic forms on $S$,
$\overline\partial\omega_1=\overline\partial\omega_2=0$. Locally
we can find (1,0) forms $\xi=\xi_1dz_1+\xi_2dz_2$ and
$\zeta=\zeta_1dz_1+\zeta_2dz_2$ on $S$ , where $\xi_i$ and
$\zeta_j$ are smooth complex functions on some open set of $S$,
such that $\omega_1=\overline
\partial\xi$ and $\omega_2=\overline\partial\zeta$. Let
\begin{eqnarray*}
\theta_0&=&\theta-\pi^*(\xi+\sqrt{-1}\zeta)\\
&=&(dx+\sqrt{-1}dy)+\pi^*(\alpha+\sqrt{-1}\beta)-\pi^*(\xi+\sqrt{-1}\zeta)
\end{eqnarray*}
We claim that $\theta_0$ is the holomorphic (1,0) form. By our
construction, $\theta_0$ is the (1,0) form. So we should only
explain that $\theta_0$ is holomorphic.  Because $\theta$ is a
(1,0)-form on $X$, then $\partial \theta$ is a (2,0)-form. But
$d\theta=d(dx+\sqrt{-1}dy+\pi^*(\alpha+\sqrt{-1}\beta))=\pi^*(\omega_1+\sqrt{-1}\omega_2)$
is a  (1,1) form on $X$. So
\begin{equation}\label{204}
\partial\theta=0\ \
\text{and}\ \
\overline\partial\theta=d\theta=\pi^*(\omega_1+i\omega_2).
\end{equation}
Thus we have
\begin{eqnarray*}
\overline\partial\theta_0&=&\overline\partial\theta-\overline\partial\pi^*(\xi+\sqrt{-1}\zeta)\\
&=&\pi^*(\omega_1+\sqrt{-1}\omega_2)-\pi^*(\omega_1+\sqrt{-1}\omega_2)=0
\end{eqnarray*}
So $\theta_0$ is the holomorphic (1,0) form and
$\{\pi^*dz_j,\theta_0\}$ is the basis of holomorphic (1,0) forms
on $X$. Therefore we can construct the basis of  holomorphic
vector fields, which is
 dual to the  basis  of $\{\pi^*dz_j,\theta_0\}$. Let
$$\varphi_j=\xi_j+\sqrt{-1}\zeta_j\ \ \ \ \text{for}\ \ \ \ j=1,2$$
and
$$\tilde U_j=U_j+\varphi_jU_0 \ \ \ \ \text{for}\ \ \ \ j=1,2$$
Then it's easily checked that $\{\tilde U_j,U_0\}$ is dual to
$\{\pi^*dz_j,\theta_0\}$ because  $U_j$ is in the kernel of
$\theta$. So it's the basis of holomorphic (1,0) vector fields.
Under this basis, the metric $g_0$ takes  the following hermitian
matrix:
\begin{eqnarray}\label{205}
H_X=\left(\begin{array}{ccc} g_{1\bar 1}+\mid \varphi_1\mid^2 &
g_{1\bar2}+\varphi_1\overline \varphi_2& \varphi_1\\
g_{2\bar1}+\varphi_2\overline \varphi_1 & g_{2\bar2}+\mid \varphi_2\mid^2 & \varphi_2\\
\overline \varphi_1& \overline \varphi_2 &
1\end{array}\right)=\left(\begin{array}{cc} g+A\cdot A^* & A\\
A^* & 1 \end{array}\right)
\end{eqnarray}
where $ g$ is the Calabi-Yau metric on $S$ and
$A=(\varphi_1,\varphi_2)^t$.

According to Strominger's explain in ref \cite{Str}, when the
manifold is not Kahler, we should take the curvature of Hermitian
connection on the holomorphic tangent bundle $T'X$. Using the
metric (\ref{205}), we can easily calculate the connection and
curvature. By directly calculation, we get the curvature
$$
R=\overline\partial(\partial H_X\cdot
H_X^{-1})=\left(\begin{array} {cc} R_{1\bar 1}&
R_{1\bar2}\\R_{2\bar1} & R_{2\bar2}
\end{array}\right)$$
where
\begin{eqnarray*}
R_{1\bar1}&=&R_S+\overline\partial A\wedge(\partial A^*\cdot
g^{-1})+A\cdot \overline\partial(\partial A^*\cdot g^{-1})\\
R_{1\bar2}&=&-R_SA+(\partial g\cdot g^{-1})\wedge\overline
\partial A-\overline\partial A\wedge(\partial A^*\cdot
g^{-1})A\\
&&-A\overline\partial(\partial A^*\cdot g^{-1})A+A(\partial
A^*\cdot g^{-1})\wedge\overline\partial
A+\overline\partial\partial A\\
R_{2\bar1}&=& \overline\partial(\partial A^*\cdot g^{-1})\\
R_{2\bar2}&=& -\overline\partial(\partial A^*\cdot
g^{-1})A+(\partial A^*\cdot g^{-1})\wedge \overline
\partial A
\end{eqnarray*}
and $R_S$ is the curvature of Calabi-Yau metric $g$ on $S$. It is
easily checked that $\text{tr}(\overline\partial A\wedge(\partial
A^*\cdot g^{-1})+A\cdot \overline\partial(\partial A^*\cdot
g^{-1}))-\overline\partial(\partial A^*\cdot g^{-1})A+(\partial
A^*\cdot g^{-1})\wedge \overline
\partial A=0$. So $\text{tr}R=\pi^*\text{tr}R_S$.
\begin{prop}\cite{GP2}
The Ricci forms of the hermitian  connections on $X$ and $S$ have
the relation $\text{tr}R=\pi^*\text{tr}R_S$.
\end{prop}
\begin{rema}
In the above calculation, we don't use the condition that the
metric $g$ on $S$ is Calabi-Yau. \end{rema}
 Next we should
calculate the $\text{tr}R\wedge R$.
\begin{prop}
\begin{equation}\label{301}
\textup{tr} R\wedge R=\pi^*(\textup{tr}R_S\wedge
R_S+2\textup{tr}\partial\overline\partial(\overline\partial
A\wedge
\partial A^*\cdot g^{-1})).
\end{equation}
\end{prop}
\begin{proof}
We take the following trick. Fix the point $p\in S$ and pick $A$
such that $A(p)=0$, e.g., we can take the gauge transformation to
get this point. So when we calculate the $\text{tr} R\wedge R$ at
the point $p$, all terms containing the factor $A$ will vanish.
Thus
\begin{eqnarray*}
&&\text{tr} R\wedge R\\
&=&\text{tr} R_S\wedge R_S+2\text{tr}R_S\wedge \overline\partial
A\wedge(\partial A^*\cdot g^{-1})\\
&&+2\text{tr}\partial g\cdot g^{-1}\wedge\overline\partial
A\wedge\dbar(\partial A^*\cdot g^{-1})+2\text{tr}\overline
\partial \partial A\wedge\overline\partial(\partial A^*\cdot g^{-1})\\
&&+\text{tr}\overline\partial A\wedge((\partial A^*\cdot
g^{-1})\wedge\overline\partial A\wedge(\partial A^*\cdot
g^{-1}))\\
&&+((\partial A^*\cdot g^{-1})\wedge\overline \partial
A\wedge(\partial A^*\cdot g^{-1}))\wedge\overline\partial A\\
&=&\text{tr} R_S\wedge R_S+2\text{tr}R_S\wedge
\overline\partial A\wedge(\partial A^*\cdot g^{-1})\\
&&+2\text{tr}\partial g\cdot g^{-1}\wedge\overline\partial
A\wedge\dbar(\partial A^*\cdot g^{-1})+2\text{tr}\overline
\partial\partial A\wedge\overline\partial(\partial A^*\cdot g^{-1})
\end{eqnarray*}
We finish the proof of this proposition by proof of the following
two claims.

 {\bf Claim 1.}
\begin{eqnarray*}
\textup{tr}\partial\overline\partial(\overline\partial A\wedge
\partial A^*\cdot g^{-1})&=&
\textup{tr}R_S\wedge
\overline\partial A\wedge(\partial A^*\cdot g^{-1})\\
&&+\textup{tr}\partial g\cdot g^{-1}\wedge\overline\partial
A\wedge \dbar(\partial
A^*\cdot g^{-1})\\
&&+\textup{tr}\overline
\partial\partial A\wedge\overline\partial(\partial A^*\cdot g^{-1})
\end{eqnarray*}
\begin{proof}
\begin{eqnarray*}
&&\text{tr}\partial\overline\partial(\overline\partial A\wedge
\partial A^*\cdot g^{-1})\\
&=&-\text{tr}\partial(\overline\partial A\wedge
\overline\partial(\partial
A^*\cdot g^{-1}))\\
&=&\text{tr}\overline\partial\partial A\wedge
\overline\partial(\partial A^*\cdot
g^{-1})+\text{tr}\overline\partial A\wedge
\partial\overline\partial(\partial A^*\cdot g^{-1})\\
&=&\text{tr}\overline\partial\partial A\wedge
\overline\partial(\partial A^*\cdot
g^{-1})+\text{tr}\overline\partial A\wedge
\overline\partial (\partial A^*\wedge\partial g^{-1})\\
&=&\text{tr}\overline\partial\partial A\wedge
\overline\partial(\partial A^*\cdot
g^{-1})-\text{tr}\overline\partial A\wedge
\overline\partial(\partial A^*\cdot g^{-1}\wedge\partial g\cdot g^{-1})\\
&=&\text{tr}\overline\partial\partial A\wedge
\overline\partial(\partial A^*\cdot
g^{-1})-\text{tr}\overline\partial A\wedge
\overline\partial(\partial A^*\cdot g^{-1})\wedge\partial g\cdot
g^{-1}\\
&&+\text{tr}\overline\partial A\wedge
(\partial A^*\cdot g^{-1})\wedge\dbar(\partial g\cdot g^{-1})\\
&=&\text{tr}\overline\partial\partial A\wedge
\overline\partial(\partial A^*\cdot
g^{-1})-\text{tr}\overline\partial A\wedge
\overline\partial(\partial A^*\cdot g^{-1})\wedge\partial g\cdot
g^{-1}\\
&&+\text{tr}\overline\partial A\wedge
(\partial A^*\cdot g^{-1})\wedge R_S\\
&=&\text{tr}(\overline\partial\partial A\wedge
\overline\partial(\partial A^*\cdot
g^{-1}))+\text{tr}(R_S\wedge\overline\partial A\wedge
\partial A^*\cdot
g^{-1}) \\
&&+\text{tr}(\partial g\cdot g^{-1}\wedge\overline\partial A\wedge
\overline\partial(\partial A^*\cdot g^{-1})
\end{eqnarray*}
\end{proof}

\noindent{\bf Claim 2.} $\textup{tr}(\dbar A\wedge \partial
A^*\cdot g^{-1})$ is the well-defined (1,1)-form on $S$.
\begin{proof}
We take local coordinates $(U,z_i)$ and $(W,w_j)$ on $S$ such that
$U\cap W\neq\emptyset$. Let Jacobi matrix $J=(\frac{\partial
w_i}{\partial z_j})$. We can let
$$(\omega_1+\sqrt{-1}\omega_2)\mid_U=\overline\partial(\varphi_1dz_1+\varphi_2dz_2)=\overline\partial
\varphi_1\wedge dz_1+\dbar \varphi_2\wedge dz_2$$
$$(\omega_1+\sqrt{-1}\omega_2)\mid_W=\overline\partial(\gamma_1dw_1+\gamma_2dw_2)=\overline\partial
\gamma_1\wedge dw_1+\dbar \gamma_2\wedge dw_2$$
 Then on $U\cap W$,
\begin{eqnarray*}
\left(\begin{array}{cc}\dbar \gamma_1 &
\dbar \gamma_2\end{array}\right)\wedge\left(\begin{array}{c} dw_1\\
dw_2\end{array}\right)
&=& \left(\begin{array}{cc}\dbar \varphi_1 & \dbar \varphi_2\end{array}\right) \wedge\left(\begin{array}{c} dz_1\\
dz_2\end{array}\right)
\end{eqnarray*}
So
\begin{eqnarray}\label{302}
\left(\begin{array}{cc}\dbar \varphi_1 & \dbar
\varphi_2\end{array}\right)=\left(\begin{array}{cc}\dbar \gamma_1
& \dbar \gamma_2\end{array}\right)J
\end{eqnarray}
On the other hand, we have
\begin{equation}\label{303}
g(z)=J^t g(w)\overline{J}
\end{equation}
where $g(z)=(g_{i\bar j}(z))$ and $g(w)=(g_{i\bar j}(w))$ denote
the metrics on $U$ and $W$ respectively. Then on $U\cap W$, from
(\ref{302}),(\ref{303}), we calculate
\begin{eqnarray*}
&&\text{tr}\left(\begin{array}{c} \dbar \gamma_1\\ \dbar
\gamma_2\end{array}\right)\wedge\left(\begin{array}{cc}
\partial \bar \gamma_1& \partial\bar \gamma_2\end{array}\right)\cdot
g^{-1}(w)\\
&=&\text{tr}\left(\begin{array}{c} \dbar \gamma_1\\ \dbar
\gamma_2\end{array}\right)\wedge\left(\begin{array}{cc}
\overline{\overline\partial \gamma_1}& \overline{\overline\partial
\gamma_2}\end{array}\right)\cdot g^{-1}(w)\\
&=&\text{tr}(J^t)^{-1}\left(\begin{array}{c} \dbar \varphi_1\\
\dbar \varphi_2\end{array}\right)\wedge\left(\begin{array}{cc}
\overline{\overline\partial \varphi_1}&
\overline{\overline\partial
f_2}\end{array}\right)\bar J^{-1}\cdot \bar J\cdot g^{-1}(z)\cdot J^t\\
&=&\text{tr}J^t\cdot (J^t)^{-1}\left(\begin{array}{c} \dbar \varphi_1\\
\dbar \varphi_2\end{array}\right)\wedge\left(\begin{array}{cc}
\overline{\overline\partial \varphi_1}&
\overline{\overline\partial
\varphi_2}\end{array}\right)\cdot g^{-1}(z)\\
&=&\text{tr}\left(\begin{array}{c} \dbar \varphi_1\\ \dbar
\varphi_2\end{array}\right)\wedge\left(\begin{array}{cc}
\partial\bar \varphi_1& \partial
\bar \varphi_2\end{array}\right)\cdot g^{-1}(z)\\
\end{eqnarray*}
which proves that  $\text{tr}(\dbar A\wedge \partial A^*\cdot
g^{-1})$ is the well-defined (1,1) form on $S$.
\end{proof}
\end{proof}
Although $\text{tr}\dbar A\wedge \partial A^*\cdot g^{-1}$ is the
well-defined (1,1) form on $S$, we can not express it by
$\omega_1$ and $\omega_2$. But in the  particular case, we can do
it.
\begin{prop}
When $\omega_2=n\omega_1$, $n\in \mathbb{Z}$,
\begin{equation}\label{119}
\textup{tr}\dbar A\wedge \partial A^*\cdot
g^{-1}=\frac{\sqrt{-1}}{4}(1+n^2)\parallel
\omega_1\parallel^2_{\omega_S}\omega_S
\end{equation}
where $\omega_S$ is the given Calabi-Yau metric on $S$.
\end{prop}
\begin{proof}
We recall that locally, we have
\begin{eqnarray*}
\omega_1&=&\dbar \xi,\ \ \ \ \xi=\xi_1dz_1+\xi_2dz_2,\\
\omega_2&=&\dbar \zeta, \ \ \ \ \zeta=\zeta_1dz_1+\zeta_2dz_2,\\
\varphi_j&=&\xi_j+\sqrt{-1}\zeta_j,\ \ \ \ \text{for}\ \ j=1,2\\
A&=&\left(\begin{array}{c}\varphi_1\\ \varphi_2\end{array}\right),
\, \ A^*=\left(\begin{array}{cc}\bar \varphi_1&\bar
\varphi_2\end{array}\right).
\end{eqnarray*}
When $\omega_2=n\omega_1$,we take $\zeta=n\xi$. Then
$\dbar\zeta_j=n\dbar\xi_j$. So
\begin{eqnarray*}
\dbar A=\left(\begin{array}{c}\dbar \varphi_1\\ \dbar
\varphi_2\end{array}\right)=(1+n\sqrt{-1})\left(\begin{array}{c}\dbar
\xi_1\\ \dbar \xi_2\end{array}\right)
\end{eqnarray*}
and
\begin{equation*}
\partial A^*=\left(\begin{array}{cc}\partial \bar \varphi_1& \partial\bar \varphi_2\end{array}\right)=
(1-n\sqrt{-1})\left(\begin{array}{cc}\partial\bar\xi_1&
\partial\bar \xi_2\end{array}\right)
\end{equation*}
Using above equalities, locally  we calculate
\begin{equation}\label{0301}
\begin{aligned}
&\text{tr}\dbar A\wedge \partial A^*\cdot
g^{-1}\\
=&(1+n^2)\text{tr}\left(\begin{array}{c}\dbar \xi_1\\ \dbar
\xi_2\end{array}\right)\wedge\left(\begin{array}{cc}\partial\bar\xi_1&
\partial\bar \xi_2\end{array}\right)\cdot g^{-1}\\
=&\frac{1+n^2}{\det g}\text{tr}
\left(\begin{array}{c}\frac{\partial \xi_1}{\partial\bar z_i}d\bar z_i\\
\frac{\partial \xi_2}{\partial\bar z_i}d\bar
z_i\end{array}\right)\wedge\left(\begin{array}{cc}\overline{\frac{\partial\xi_1}{\partial\bar
z_j}}dz_j &\overline{\frac{\partial\xi_2}{\partial\bar z_j}}dz_j
\end{array}\right)\cdot
\left(\begin{array}{cc}g_{2\bar2}&  -g_{1\bar2}\\
-g_{2\bar1}& g_{1\bar1}\end{array}\right)\\
=&\frac{1+n^2}{\det g}\text{tr}
\left(\begin{array}{c}\frac{\partial \xi_1}{\partial\bar z_i}\\
\frac{\partial \xi_2}{\partial\bar
z_i}\end{array}\right)\wedge\left(\begin{array}{cc}\overline{\frac{\partial\xi_1}{\partial\bar
z_j}} &\overline{\frac{\partial\xi_2}{\partial\bar z_j}}
\end{array}\right)\cdot
\left(\begin{array}{cc}g_{2\bar2}&  -g_{1\bar2}\\
-g_{2\bar1}& g_{1\bar1}\end{array}\right)d\bar z_i\wedge dz_j
\end{aligned}
\end{equation}
In order to get the global formula, we need some formulas about
$\omega_1$. Because $\omega_1$ is real, we have
\begin{equation}\label{305}
\overline{\frac{\partial\xi_i}{\partial\bar z_j}}=-
\frac{\partial\xi_j}{\partial\bar z_i}\ \ \ \ \text{for}\ \
i,j=1,2
\end{equation}
From $\omega_1$ is anti-self-dual, i.e.,  $\omega_1\wedge
\omega_S=0$, locally we have
\begin{equation}\label{306}
g_{1\bar1}\frac{\partial\xi_2}{\partial\bar z_2}
+g_{2\bar2}\frac{\partial\xi_1}{\partial\bar
z_1}-g_{1\bar2}\frac{\partial\xi_2}{\partial\bar z_1}
-g_{2\bar1}\frac{\partial\xi_1}{\partial\bar z_2}=0
\end{equation}
Because
\begin{equation}\label{307}
\omega_1\wedge\omega_1=-\omega_1\wedge\ast\omega_1=
-\omega_1\ast\bar\omega_1=-\parallel\omega_1\parallel^2\frac{\omega^2_S}{2!},
\end{equation}
locally we also have
 \begin{equation}\label{308}
\frac{1}{\det(g)}\left(\frac{\partial\xi_1}{\partial\bar
z_1}\frac{\partial\xi_2}{\partial\bar
z_2}-\frac{\partial\xi_1}{\partial\bar
z_2}\frac{\partial\xi_2}{\partial\bar
z_1}\right)=\frac{1}{8}\parallel\omega_1\parallel^2
\end{equation}
Now using above (\ref{305}), (\ref{306}) and (\ref{308}), we can
calculate the component of $d\bar z_1\wedge d z_1$ in
(\ref{0301}):
\begin{equation}\label{309}
\begin{aligned}
&\frac{1+n^2}{\det(g)}\left(g_{2\bar2}\frac{\partial\xi_1}{\partial\bar
z_1}\overline{\frac{\partial\xi_1}{\partial\bar
z_1}}-g_{2\bar1}\frac{\partial\xi_1}{\partial\bar
z_1}\overline{\frac{\partial\xi_2}{\partial\bar
z_1}}-g_{1\bar2}\frac{\partial\xi_2}{\partial\bar
z_1}\overline{\frac{\partial\xi_1}{\partial\bar z_1}}-g_{1\bar1}
\frac{\partial\xi_2}{\partial\bar
z_1}\overline{\frac{\partial\xi_2}{\partial\bar z_1}}\right)\\
=&\frac{1+n^2}{\det(g)}\left(g_{2\bar1}\frac{\partial\xi_1}{\partial\bar
z_1}\frac{\partial\xi_1}{\partial\bar
z_2}+g_{1\bar2}\frac{\partial\xi_2}{\partial\bar
z_1}\frac{\partial\xi_1}{\partial\bar
z_1}-g_{2\bar2}\left(\frac{\partial\xi_1}{\partial\bar
z_1}\right)^2-g_{1\bar1} \frac{\partial\xi_2}{\partial\bar
z_1}\frac{\partial\xi_1}{\partial\bar z_2}\right)\\
=&\frac{1+n^2}{\det(g)}\left(\frac{\partial\xi_1}{\partial\bar
z_1}\left(g_{2\bar1} \frac{\partial\xi_1}{\partial\bar
z_2}+g_{1\bar2}\frac{\partial\xi_2}{\partial\bar
z_1}\right)-g_{2\bar2}\left(\frac{\partial\xi_1}{\partial\bar
z_1}\right)^2-g_{1\bar1} \frac{\partial\xi_2}{\partial\bar
z_1}\frac{\partial\xi_1}{\partial\bar z_2}\right)\\
=&\frac{1+n^2}{\det(g)}\left(\frac{\partial\xi_1}{\partial\bar
z_1}\left(g_{1\bar1} \frac{\partial\xi_2}{\partial\bar
z_2}+g_{2\bar2}\frac{\partial\xi_1}{\partial\bar
z_1}\right)-g_{2\bar2}\left(\frac{\partial\xi_1}{\partial\bar
z_1}\right)^2-g_{1\bar1} \frac{\partial\xi_2}{\partial\bar
z_1}\frac{\partial\xi_1}{\partial\bar z_2}\right)\\
=&\frac{1+n^2}{\det(g)}g_{1\bar1}\left(\frac{\partial\xi_1}{\partial\bar
z_1} \frac{\partial\xi_2}{\partial\bar z_2}-
\frac{\partial\xi_2}{\partial\bar
z_1}\frac{\partial\xi_1}{\partial\bar z_2}\right)\\
=&\frac{1+n^2}{8}\parallel\omega_1\parallel^2g_{1\bar1}
\end{aligned}
\end{equation}
As the same reason, the components of $d\bar z_2\wedge dz_1$,
$d\bar z_1\wedge dz_2$  and $d\bar z_2\wedge dz_2$ in (\ref{0301})
are $\frac{1+n^2}{8}\parallel\omega_1\parallel^2g_{1\bar2}$,
$\frac{1+n^2}{8}\parallel\omega_1\parallel^2g_{2\bar1}$  and
$\frac{1+n^2}{8}\parallel\omega_1\parallel^2g_{2\bar2}$
respectively. So  we obtain
\begin{eqnarray*}
&&\text{tr}\dbar A\wedge \partial A^*\cdot
g_S^{-1}\\
&=&\frac{1+n^2}{8}\parallel\omega_1\parallel^2(g_{1\bar1}d\bar
z_1\wedge dz_1+g_{1\bar2}d\bar z_2\wedge dz_1+g_{2\bar1}d\bar
z_1\wedge dz_2+g_{2\bar2}d\bar z_2\wedge dz_2)\\
&=&\frac{\sqrt{-1}}{4}(1+n^2)\parallel\omega_1\parallel^2\omega_S.
\end{eqnarray*}
\end{proof}

\section{Constructing the reducible solution to Strominger's system}
We take the 3-dimensional hermitian manifolds $(X,\omega_0,\tilde
I )$ as described in section 2. Let $\omega_S$ is the Calabi-Yau
metric on $S$. From (\ref{2011}) and (\ref{2012}), the Kahler form
$\omega_0$ is
$$
\omega_0=\pi^*\omega_S+\frac{\sqrt{-1}}{2}\theta\wedge\bar\theta.$$
Using (\ref{204}) and the facts that $\parallel\Omega\parallel=1$
and $\omega_1$ and $\omega_2$ are anti-self-dual, we have
\begin{equation}\label{0401}
\begin{aligned}
&\ d(\parallel\Omega\parallel_{\omega_0}\omega_0^2)\\
=&\ d\omega^2_0=d(\pi^*\omega^2_S+
\sqrt{-1}\pi^*\omega_S\wedge\theta\wedge\bar\theta)\\
=&\ \sqrt{-1}\pi^*\omega_S\wedge
d\theta\wedge\bar\theta-\sqrt{-1}\pi^*\omega_S\wedge
\theta\wedge d\bar\theta\\
=&\ \sqrt{-1}\pi^*\omega_S\wedge
(\omega_1+\sqrt{-1}\omega_2)\wedge\bar\theta-\sqrt{-1}\pi^*\omega_S\wedge (\omega_1-\sqrt{-1}\omega_2)
\wedge\theta\\
=&\ 0
\end{aligned}
\end{equation}
According to Lemma 1, $(\omega_0,\Omega)$ is the solution of
equation (\ref{104}).
 Let $u$ be the smooth function on $S$ and take
\begin{equation}\label{401}
\omega_u=\pi^*(e^u\omega_S)+\frac{\sqrt{-1}}{2}\theta\wedge\bar\theta
\end{equation}
Then
$$\parallel\Omega\parallel^2_{\omega_u}=\frac{\omega_0^3}{\omega^3_u}=\frac{1}{e^{2u}}$$
and
\begin{eqnarray*}
\parallel\Omega\parallel_{\omega_u}\omega^2_u&=&e^{-u}(e^{2u}\omega_S^2+
\sqrt{-1}e^u\omega_S\wedge\theta\wedge\bar\theta)\\
&=&\omega_0^2+(e^u-1)\omega^2_S
\end{eqnarray*}
From (\ref{0401}), we obtain
$$d(\parallel\Omega\parallel_{\omega_u}\omega^2_u)=d\omega^2_0+d(e^u-1)\wedge\omega_S^2=0$$
and we have proven the following
\begin{lemm}\cite{GP}
The metric (\ref{401}) defined over $X$ satisfies the equation
(\ref{105}) and so satisfies the equation (\ref{104}).
\end{lemm}
Now we construct the solutions to Strominger's system. Because
$\frac{\omega_1}{2\pi},\frac{\omega_2}{2\pi}\in
H^{1,1}(S,\mathbb{R})\cap H^2(S,\mathbb{Z})$, there are
holomorphic line bundles $L_1$ and $L_2$ over $S$ such that their
curvatures of hermitian connections are $\sqrt{-1}\omega_1$ and
$\sqrt{-1}\omega_2$ respectively. Corresponding to these
curvatures, there exist hermitian  metrics $h_1$ and $h_2$ on
$L_1$ and $L_2$ because $S$ is Kahler. Let
\begin{equation}
E=L_1\oplus L_2\oplus T'S
\end{equation}
and let
$$H_0=(h_1,h_2,\omega_S).$$
 The curvature $F_{H_0}$ of $E$ is
\begin{eqnarray}\label{403}
F_{H_0}=\left(\begin{array}{ccc}\sqrt{-1}\omega_1& & \\
& \sqrt{-1}\omega_2& \\
& & R_S\end{array}\right)
\end{eqnarray}
where $R_S$ is the curvature of $T'S$ corresponding to the
hermitian metric $\omega_S$. Let $V=\pi^*E$, and let
$F_{\tilde{H}_0}=\pi^*F_{H_0}$. We try to make
$(V,F_{\tilde{H}_0},X,\omega_u)$ to be the solution to
Strominger's system.
\begin{lemm}
For any smooth function $u$ on $S$,
$(V,F_{\tilde{H}_0},X,\omega_u)$  satisfies the equations
(\ref{101}), (\ref{102}) and (\ref{104}).
\end{lemm}
\begin{proof}By lemma 10, equation (\ref{104}) is
satisfied. Because $\omega_1$, $\omega_2$ and $R_S$ are (1,1)
forms on $S$, $F_{{\tilde H}_0}$ is the (1,1) form on $X$. So
$F_{\tilde H_0}^{2,0}=F_{\tilde H_0}^{0,2}=0$. We also have
\begin{eqnarray*}
F_{\tilde H_0}\wedge\omega^2_u=\pi^*(F_{H_0}\wedge
e^u\omega_S)\wedge(\pi^*
(e^{u}\omega_S)+\sqrt{-1}\theta\wedge\bar\theta)=0
\end{eqnarray*}
by   facts $\omega_1\wedge\omega_S=\omega_2\wedge\omega_S=0$ and
$R_S\wedge\omega_S=0$.
\end{proof}
So we only need to consider the equation (\ref{103}). We take the
factor $\alpha'=\frac{1}{2}$\footnote{We can take $\alpha'=1$,
only if we take the   metric  $2\omega_u$.}, then $(V,F_{\tilde
H_0},X,\omega_u)$ should satisfy the equation
\begin{equation}\lab{404}
\sqrt{-1}\partial\dbar \omega_u= \frac{1}{2}(\text{tr} F_{\tilde
H_0}\wedge F_{\tilde H_0}-\text{tr} R_u\wedge R_u)
\end{equation}
here $R_u$ denotes the curvature of Hermitian connection on $T'X$
corresponding to the hermitian metric $\omega_u$. Now we calculate
each term in equation (\ref{404}). The  Laplace operator
$\bigtriangleup$ on $S$ is define by
$\bigtriangleup=\dbar^*\circ\dbar$ associate the Calabi-Yau metric
$\omega_S$. So for any smooth function $\psi$ on $S$,
\begin{equation}\label{405}
\sqrt{-1}\partial\dbar \psi\wedge\omega_S=\bigtriangleup
\psi\cdot\frac{\omega^2_S}{2!}
\end{equation}
{\bf Claim 3.} $\sqrt{-1}\partial\dbar\omega_u=\bigtriangleup
e^u\cdot\frac{\omega_S^2}{2!}+\frac{1}{2}(\parallel\omega_1\parallel^2+
\parallel\omega_2\parallel^2)\frac{\omega^2_S}{2!}$
\begin{proof}
Using (\ref{204}), (\ref{405}) and (\ref{307}), we get
\begin{eqnarray*} \sqrt{-1}\partial\overline{\partial}\omega_u
&=&\sqrt{-1}\partial\overline{\partial}(e^u\omega_S+\frac{\sqrt{-1}}{2}\theta\wedge\overline\theta)\\
&=&\sqrt{-1}\partial\dbar
e^u\wedge \omega_S-\frac{1}{2}\dbar\theta\wedge\partial\bar\theta\\
&=&\bigtriangleup
e^u\cdot\frac{\omega^2_S}{2!}-\frac{1}{2}(\omega_1+\sqrt{-1}\omega_2)
\wedge(\omega_1-\sqrt{-1}\omega_2)\\
&=&\bigtriangleup
e^u\cdot\frac{\omega^2_S}{2!}-\frac{1}{2}(\omega_1\wedge\omega_1+\omega_2\wedge\omega_2)
\\
&=&\bigtriangleup
e^u\cdot\frac{\omega^2_S}{2!}+\frac{1}{2}(\parallel\omega_1\parallel^2+
\parallel\omega_2\parallel^2)\frac{\omega^2_S}{2!}
\end{eqnarray*}
\end{proof}
\noindent {\bf Clam 4.} $\text{tr}F_{H_0}\wedge
F_{H_0}=(\parallel\omega_1\parallel^2+
\parallel\omega_2\parallel^2)\frac{\omega^2_S}{2!}+\text{tr}R_S\wedge
R_S$
\begin{proof}
From (\ref{403}), we have
\begin{eqnarray*}
\text{tr}F_{H_0}\wedge
F_{H_0}&=&-\omega_1\wedge\omega_1-\omega_2\wedge\omega_2+\text{tr}R_S\wedge
R_S\\
&=&(\parallel\omega_1\parallel^2+
\parallel\omega_2\parallel^2)\frac{\omega^2_S}{2!}+\text{tr}R_S\wedge
R_S
\end{eqnarray*}
\end{proof}
\noindent {\bf Claim 5.} $\textup{tr}R_u\wedge
R_u=\pi^*\textup{tr}R_S\wedge R_S+2\pi^*(\partial\dbar u\wedge
\partial\dbar u)+2\pi^*(\partial\dbar(e^{-u}\textup{tr}(\dbar
A\wedge \partial A^*\cdot g_S))\wedge\omega).$
\begin{proof}
Actually in the proof of the Proposition 8 we don't use the
condition that $\omega_S$ is Kahler. So if we replace metric $g$
by $e^ug$,  we can still get:
\begin{equation}\label{0402}
\begin{aligned}
\text{tr}R_u\wedge R_u=&\pi^*(\text{tr}R_S^u\wedge
R_S^u+2\text{tr}\partial\dbar(\dbar A\wedge \partial A^*\cdot
(e^{u}g)^{-1}))\\
=&\pi^*(\text{tr}R_S^u\wedge
R_S^u+2\partial\dbar(e^{-u}\text{tr}(\dbar A\wedge
\partial A^*\cdot g^{-1})))
\end{aligned}
\end{equation}
here $R_S^u$ denotes the curvature of hermitian connection of
$T'S$ corresponding to the hermitian metric $e^ug$. So
\begin{eqnarray*}
R^u_S&=&\dbar(\partial(e^ug)\cdot(e^ug)^{-1})\\
&=&\dbar(\partial u\cdot I+\partial g\cdot g^{-1})\\
&=&\dbar\partial u\cdot I+R_S
\end{eqnarray*}
and
\begin{equation}\label{0403}
\begin{aligned}
\text{tr}R_S^u\wedge R_S^u=&\text{tr}R_S\wedge R_S+2\partial\dbar
u\wedge \partial\dbar u+2\partial\dbar u\wedge \text{tr}R_S\\
=&\text{tr}R_S\wedge R_S+2\partial\dbar u\wedge \partial\dbar u
\end{aligned}
\end{equation}
here we use the fact that $\text{tr}R_S=0$ because the hermitian
metric $g$ is the Calabi-Yau metric on $S$. Inserting (\ref{0403})
into (\ref{0402}), we have proven the claim.
\end{proof}

From Claim 3, Claim 4, and Claim 5, we can rewrite the equation
(\ref{404}) as
\begin{equation}\label{406}
\bigtriangleup
e^u\cdot\frac{\omega^2_S}{2!}+\partial\dbar(e^{-u}\text{tr}(\dbar
A\wedge \partial A^*\cdot g^{-1}))+\partial\dbar u\wedge
\partial\dbar u=0
\end{equation}
Now we can prove
\begin{theo}
$(V,F_{\pi^*H_0},X,\omega_u)$ is the solution of Strominger's
system if and only if the function $u$ of $S$ satisfies the
equation (\ref{406}). In particular, when $\omega_2=n\omega_1$,
$n\in \mathbb{Z}$, $(V,F_{\pi^*H_0},X,\omega_u)$ is the solution
to Strominger's system if and only if smooth function $u$ on $S$
satisfies the equation:
\begin{equation}\label{407}
\bigtriangleup\left(e^u+\frac{(1+n^2)}
{4}\parallel\omega_1\parallel_{\omega_S}^2e^{-u}\right)\frac{\omega_S^2}{2!}+\partial\dbar
u\wedge\partial\dbar u=0
\end{equation}
or
\begin{equation}\label{408}
\bigtriangleup\left(e^u+\frac{(1+n^2)}
{4}\parallel\omega_1\parallel_{\omega_S}^2e^{-u}\right)-8\frac{\det(u_{i\bar
j})}{\det(g_{i\bar j})}=0
\end{equation}
\end{theo}
\begin{proof}
We only need to prove the second part of theorem.  When
$\omega_2=n\omega_1$, from Proposition 9 and (\ref{405}), we have
\begin{eqnarray*}
\partial\dbar(e^{-u}\text{tr}(\dbar A\wedge \partial A^*\cdot
g^{-1}))&=&\frac{\sqrt{-1}}{4}(1+n^2)\partial\dbar(e^{-u}\parallel\omega_1\parallel^2)
\wedge\omega_S\\
&=&\frac{(1+n^2)}{4}\bigtriangleup(e^{-u}\parallel\omega_1\parallel^2)\frac{\omega^2_S}{2!}
\end{eqnarray*}
So the equation (\ref{407}) follows from equation (\ref{406}).
Equation (\ref{408}) is derived from
\begin{eqnarray*}
\partial\dbar u\wedge\partial\dbar u&=&2\det(u_{i\bar
j})dz_1\wedge d\bar z_1\wedge dz_2\wedge d\bar z_2\\
&=&-8\frac{\det(u_{i\bar j})}{\det(g_{i\bar
j})}\frac{\omega_S^2}{2!}.
\end{eqnarray*}
\end{proof}

\section{irreducible solution}
We have constructed the following reducible solution:
$(V,\pi^*H_0,X,\omega_u)$ if $u\in C^{\infty}(S,\mathbb{R})$ is
the solution of the equation
\begin{equation}\label{001}
\sqrt{-1}\partial\bar\partial e^u\wedge
\omega_S+\partial\bar\partial (e^{-u}\text{tr}(\bar\partial
A\wedge\partial A^*\cdot g^{-1}))+\partial\bar\partial u\wedge
\partial\bar\partial u=0
\end{equation}
satisfying elliptic condition
\begin{equation}\label{002}
e^u\omega_S+\sqrt{-1}e^{-u}\text{tr}(\bar\partial A\wedge
\partial A^*\cdot g^{-1})-2\sqrt{-1}\partial\bar\partial u>0
\end{equation}
and normalization
$$\int_S e^{-u}=A.$$
From the next section to the end of the paper, we will prove that
the equation (\ref{001}) has a unique solution if $A<<1$. So in
this section, we assume that $u$ is the solution of equation
(\ref{001}). We will obtain the irreducible solution by perturbing
around the reducible solution $(V,\pi^*H_0,X,\omega_u)$. We follow
the Li-Yau's method \cite{LY}.

 Let $D''_s$ be a family of holomorphic
structures on $E$ over $S$, $H$ be a hermitian metric on $E$ over
$S$ and $\omega$ be a hermitian metric on $S$. We want to look for
conditions on $D_s''$, $H$ and and function $\phi$ such that under
these conditions $(V,\pi^*D''_s,\pi^*H,X,\omega_{u+\phi})$ is the
solution to Strominger's solution, where
$$\omega_{u+\phi}=\pi^*(e^{u+\phi}\omega)+\frac{\sqrt{-1}}{2}\theta\wedge\bar\theta.$$

Fix the metric $H_0$ as the reference metric on $E$ over $S$. Then
for any hermitian metric $H$ on $E$, we can define a smooth
endomorphism $h$ on  $E$ by
$$<s,t>_{H}=<s\cdot h,t>_{H_0}.$$
Under this isomorphism, we define $\mathcal{H}(E)_1$ be the space
of all hermitian metric on $E$ whose corresponding endomorphism
has determinant one.
 Let
$\mathcal{C}(\omega_S)=\{e^{\phi}\omega_S\}$ be the space of all
hermitian metrics on $S$ which are conformal to $\omega_S$. Let
$\Endo^0 E$  be the vector bundle of traceless hermitian
anti-symmetric endomorphisms of $(E,H_0)$. Let
$\mathcal{H}_0(S)=\left\{\psi\frac{\omega^2}{2!}\mid\int_S
\psi\frac{\omega^2}{2!}=0\right\}$. We define the operator
$${\bf L}={\bf L_1}\oplus{\bf
L_2}:\mathcal{H}_1(E)\times\mathcal{C}(\omega_S)\rightarrow
\Omega^4_{\mathbb{R}}(\Endo^0 E)\oplus \mathcal{H}_0(S)$$ by
\begin{equation}\label{003}
{\bf L_1}(h,e^\phi\omega_S)=e^\phi
h^{-\frac{1}{2}}F_hh^{\frac{1}{2}}\wedge \omega_S \end{equation}
\begin{equation}\label{004}
\begin{aligned}
{\bf
L_2}(h,e^\phi\omega_S)=&\sqrt{-1}\partial\bar\partial(e^{u+\phi}\omega_S)
+\partial\bar\partial(e^{-u-\phi}\text{tr}(\bar\partial
A\wedge\partial A^*\cdot
g^{-1}_{\omega_S}))\\
&+\partial\bar\partial (u+\phi)\wedge\partial\bar\partial (u+\phi)
-\frac{1}{2}(\text{tr}F_h\wedge F_h -\text{tr}F_{H_0}\wedge
F_{H_0})
\end{aligned}
\end{equation}
Because $F_h=F_{H_0}+D_0''(D_0'h\cdot h^{-1})$ and
$F_{H_0}\wedge\omega_S=0$, then
$$e^\phi h^{-1/2}F_h
h^{1/2}\wedge\omega_S=e^\phi D_0''(D_0'h\cdot
h^{-1})\wedge\omega_S$$ and from the notation of paper \cite{UY},
we get
$$\text{tr}(e^\phi h^{-1/2}F_h
h^{1/2}\wedge\omega_S) =e^\phi D_0''(D_0'\text{tr}\log
h)\wedge\omega_S=0$$ because $\det h=1$. So the image of ${\bf
L_1}$ lies in $\Omega_{\mathbb{R}}^4(\Endo^0 E)$. As to ${\bf
L_2}$, according to $\partial\bar\partial$-lemma on K3 surface,
the image of ${\bf L_2}$ lies in $R(dd_c)$. Because for any
$\sqrt{-1}\partial\bar\partial\alpha\in
R(\sqrt{-1}\partial\bar\partial)$, we can write
$\sqrt{-1}\partial\bar\partial\alpha=\psi\frac{\omega_S^2}{2!}$
for some function $\psi$ such that $\int_S
\psi\frac{\omega^2}{2!}=0$. So the image of ${\bf L_2}$ lies in
$\mathcal{H}_0(S)$. Therefore the operator ${\bf L}$ is
well-defined.
\begin{prop}
If $(h,e^\phi\omega_S)\in\ker {\bf L}$, then
$(V,\pi^*D_0'',\pi^*h,\omega_{u+\phi})$ is the solution of
Strominger's system.
\end{prop}
\begin{proof}
According to paper \cite{LY},
$(V,\pi^*D''_0,\pi^*h,X,\omega_{u+\phi})$ is the solution to
Strominger's system if and only if $(\pi^*h,\omega_{u+\phi})$ lies
in the kernel of the operator:
$${ \bf{\tilde L}}={\bf{\tilde L_1}}\oplus{\bf {\tilde L_2}}\oplus{\bf{\tilde
L_3}} :\mathcal{H}(V)_1\times\mathcal{H}(X)\rightarrow
\Omega_{\mathbb{R}}^6(\Endo^0 V)\oplus R(dd_c)\oplus
R(d^*_{\omega_0})$$ defined by
\begin{eqnarray*}
{\bf {\tilde L_1}}(\tilde h,\tilde\omega)&=&\tilde
h^{-\frac{1}{2}}F_{\tilde h}{\tilde h}^{\frac{1}{2}}\wedge
\tilde\omega^2\\
{\bf {\tilde L_2}}(\tilde
h,\tilde\omega)&=&\sqrt{-1}\partial\bar\partial\tilde\omega-\frac{1}{2}(\text{tr}F_{\tilde
h}\wedge
F_{\tilde h}-\text{tr}R_{\tilde\omega}\wedge R_{\tilde\omega})\\
{\bf {\tilde L_3}}(\tilde
h,\tilde\omega)&=&\ast_{\omega_0}d(\parallel\Omega\parallel_{\tilde\omega}\tilde\omega^2)
\end{eqnarray*}
We want  to reduce above operator ${\bf {\tilde L}}$ to vector
bundle $E$ over $S$.  When $(\tilde h,\tilde
\omega)=(\pi^*h,\omega_{u+\phi})$,
\begin{eqnarray*}
{\bf {\tilde L_1}}(\pi^*h,\omega_{u+\phi})
&=&\pi^*(h^{-\frac{1}{2}}F_hh^{\frac{1}{2}})\wedge(\omega_{u+\phi})^2\\
&=&\pi^*(h^{-\frac{1}{2}}F_hh^{\frac{1}{2}}\wedge(e^{u+\phi}\omega_S))
\wedge(\pi^*(e^{u+\phi}\omega_S)+\sqrt{-1}\theta
\wedge\bar\theta)\\
&=&\pi^*e^u\cdot\pi^*{\bf
L_1}(h,e^\phi\omega_S)\wedge(\pi^*(e^{u+\phi}\omega_S)+\sqrt{-1}\theta
\wedge\bar\theta),
\end{eqnarray*}
then $(h,e^\phi\omega_S)\in\ker {\bf L_1}$ if and only if
$(\pi^*h,\omega_{u+\phi})\in \ker{\bf {\tilde L_1}}$.

Next we consider ${\bf{\tilde L_2}}$. When $(\tilde
h,\tilde\omega)=(\pi^*h,\omega_{u+\phi})$, by Proposition 8 and as
explained in section 4, we have
\begin{equation*}
\text{tr}R_{\omega_{u+\phi}}\wedge R_{\omega_{u+\phi}}
=\pi^*(\text{tr}R_{e^{(u+\phi)}\omega_S}\wedge
R_{e^{(u+\phi)}\omega_S}+2\partial\bar\partial(e^{-u-\phi}\text{tr}(\bar\partial
A\wedge\partial A^*\cdot g^{-1}_{\omega_S})))
\end{equation*}
Meanwhile  we have
$$R_{e^{(u+\phi)}\omega_S}=\bar\partial\partial( u+\phi)\cdot I+R_{\omega_S}$$
and thus
\begin{equation*}
\text{tr}R_{e^{(u+\phi)}\omega_S}\wedge
R_{e^{(u+\phi)}\omega_S}=2\partial\bar\partial
(u+\phi)\wedge\partial\bar\partial(
u+\phi)+\text{tr}R_{\omega_S}\wedge R_{\omega_S}
\end{equation*}
because $\text{tr}R_{\omega_S}=0$. Then we see that
\begin{eqnarray*}
\text{tr}R_{\omega_{u+\phi}}\wedge R_{\omega_{u+\phi}}
&=&\pi^*\{2\partial\bar\partial (u+\phi)\wedge\partial\bar\partial
(u+\phi)+\text{tr}(R_{\omega_S}\wedge
R_{\omega_S})\\
&&+2\partial\bar\partial(e^{-u-\phi}\text{tr}(\bar\partial
A\wedge\partial A^*\cdot g^{-1}_{\omega_S}))\}
\end{eqnarray*}
Using above equality, we can obtain
\begin{eqnarray*}
{\bf {\tilde
L_2}}(\pi^*h,\omega^u)&=&\pi^*\{\sqrt{-1}\partial\bar\partial(e^{(u+\phi)}\omega_S)
+\partial\bar\partial(e^{-u-\phi}\text{tr}(\bar\partial
A\wedge\partial A^*\cdot
g^{-1}_{\omega_S}))\\
&&+\partial\bar\partial (u+\phi)\wedge\partial\bar\partial
(u+\phi) -\frac{1}{2}(\text{tr}F_h\wedge F_h -\text{tr}
F_{H_0}\wedge F_{H_0})\}\\
&=&\pi^*({\bf L_2}(h,e^\phi\omega_S))
\end{eqnarray*}
So ${\bf {\tilde L_2}}(\pi^*h,e^u\omega)=0$ if and only if ${\bf
L_2}(h,\omega)=0$.

As to ${\bf{\tilde L_3}}$, by Lemma 10, we always have
 ${\bf \tilde
L_3}(\pi^*h,\omega_{u+\phi})=\ast_0d(\parallel\Omega\parallel_{\omega_{u+\phi}}\omega_{u+\phi}^2)=0$.
\end{proof}

Certainly, the statement that $(V,\pi^*D_0,\pi^*H_0,\omega_u)$ is
the solution is equivalent to say that  $(I,\omega_S)\in\ker{\bf
L}$. Now we follow Li and Yau's paper \cite{LY}. Let
$\mathbb{R}^{+2}=\{T=(T_1,T_2)\in\mathbb{R}^2\mid T_i>0\}$; let
$I_1$, $I_2$ and $I_3$ be the identity endomorphisms of $L_1$,
$L_2$ and $T'S$ respectively. Then the assignment
$$T=(T_1,T_2)\in \mathbb{R}^{+2}\longmapsto h_T=T_1I_1\oplus
T_2I_2\oplus T^{-1/2}_1T^{-1/2}_2I_3$$ associated each $T\in
\mathbb{R}^{+2}$ to a hermitian endomorphism of $E$. Obviously,
the hermitian curvature $F_{h_T}=F_{H_0}$. Therefore we still have
$(h_T,\omega_S)\in \ker {\bf L}$.

Pick an integer $k$ and a large $p$ and endow the domain and the
target of ${\bf L_1}\oplus{\bf L_2}$ the Banach space structures
as indicated:
$$\mathcal{H}_{1}(E)_{L^p_k}\times\mathcal{C}(\omega_S)_{L^p_k}\rightarrow
\Omega_\mathbb{R}^{4}(\text{End}^0
E)_{L_{k-2}^p}\oplus\mathcal{H}_0(S)_{L_{k-2}^p}
$$
${\bf L_1}\oplus{\bf L_2}$ becomes a smooth operator and its
linearized operator $\delta{\bf L_1}\oplus\delta{\bf L_2}$ at a
solution $(h_T,\omega_S)$ becomes a linear map
$$\Omega^0({\text{Her}^0} E)_{L_k^p}\oplus\{\phi\omega_S\}_{L_k^p}\rightarrow
\Omega_\mathbb{R}^{4}(\text{End}^0 E)_{L_{k-2}^p}
\oplus\mathcal{H}_0(S)_{L_{k-2}^p}
$$
Here we used $\text{Her}^0 E$ to denote the
$\mathbb{R}$-sub-vector bundle of $\text{End}E$ consisting of
traceless pointwise $<,>$-hermitian symmetric endormorphisms of
$E$ and the canonical isomorphisms $T_{h_T}\cH_1(E)_{L^p_k}\cong
\Omega^0({\text{ Her}^0} E)_{L_k^p}$. Clearly we also have
$T_{\omega_S}\cC(\omega_S)_{L_k^p}\cong \{\phi\omega_S\}_{L_k^p}$.

To study the kernel and the cokernel of $\delta{\bf
L_1}\oplus\delta{\bf L_2}$ at a trivial solution $(h_T,\omega_S)$
we will first look at the linear map
\begin{equation*}
F(\delta h)= D_0'' D'_{0,h_T}(\delta h)\wedge\omega_S:
\Omega^0({\text{Her}^0} E)_{L^p_k}\rightarrow
\Omega^{4}_{\mathbb{R}}(\text{End}^0 E)_{L_{k-2}^p}.
\end{equation*}
Here according to our convention, $D_{h_T}=D_{0,h_T}\pri\oplus
D_0''$ is the hermitian connection of $(E,D_0'',h_T)$ for a
$T=(T_1,T_2)\in \mathbb{R}^{+2}$. Since $(E,D_0'')=L_1\oplus
L_2\oplus TS$ and $\deg L_i=0$, the above is a linear elliptic
operator of index $0$ whose kernel is
$$V_0=\{M_1\cdot I_1\oplus M_2\cdot I_2\oplus -(M_1/2+M_2/2) I_3\}
$$
and whose cokernel is
\begin{equation}\label{006}
V_1=\omega_S^2\cdot V_0 \sub \Omega^{4}_\mathbb{R}(\text{End}^0
E)_{L_{k-2}^p}.
\end{equation}
We let $\bP$ be the obvious projection
$${\bf P}: \Omega_{\mathbb{R}}^{4}(\text{End}^0 E)_{L_{k-2}^p} \rightarrow \Omega_{\mathbb{R}}^{4}
(\text{End}^0 E)_{L_{k-2}^p}/V_1.
$$
\begin{prop}\label{p3.3}
Let $(S,\omega_S)$, $\Omega_S$, $H_0$ and $T=(T_1,T_2)\in
\mathbb{R} ^{+2}$ be as before. Then  the linear operator
\begin{eqnarray*}
&&{\bf P}\circ\delta{\bf L_1}(h_T,\omega_S) \oplus\delta{\bf
L_2}(h_T,\omega_S):\\
&&\Omega^0({\text{Her}^0}
E)_{L_k^p}\oplus\{\phi\omega_S\}_{L_k^p}\rightarrow
\Omega_{\mathbb{R}}^{4}(\Endo^0 E)_{L_{k-2}^p}
/V_1\oplus\mathcal{H}_0(S)_{L_{k-2}^p}
\end{eqnarray*}
 is
surjective.
\end{prop}
\begin{proof}
Because $F_{h_T}=F_{H_0}$ and $\delta \omega=\phi\omega_S$ for
some smooth function $\phi$,
$${h_T}^{-\frac{1}{2}}F_{h_T}h_T^{\frac{1}{2}}\wedge\delta\omega=0.$$
Then
\begin{eqnarray*}
\delta{\bf L_1}(h_T,\omega_S)(\delta h,\delta
\omega)=D''_0D'_{h_T}\delta h\wedge \omega_S
\end{eqnarray*}
So
$${\bf P}\circ\delta{\bf L_1}(h_T,\omega_S):\Omega^0({\text{Her}}
E)_{L_k^p}\oplus\{\phi\omega_S\}_{L_k^p} \rightarrow
\Omega_{\mathbb{R}}^{4}(\text{End}^0 E)_{L_{k-2}^p}/V_1$$ is
surjective and
\begin{equation} \label{007} \ker
{\bf P}\circ\delta{\bf L_1}(h_T,\omega_S)=V_0\oplus
\{\phi\omega_S\}_{L_k^p}
\end{equation}

On the other hand, when $\delta\omega=\phi\omega_S$,
\begin{eqnarray*}
\delta{\bf L_2}(h_T,\omega_S)(\delta h,\phi
\omega_S)&=&\sqrt{-1}\partial\bar\partial(e^u\phi\omega_S)
-\partial\bar\partial(e^{-u}\phi\textup{tr}(\bar\partial
A\wedge\partial
A^*\cdot g_S^{-1}))\\
&&+2\partial\bar\partial u\wedge \partial\bar\partial \phi
-\textup{tr}\delta F_{h_T}(\delta h)\wedge F_{h_T}
\end{eqnarray*}
Because $\textup{tr}\delta F_{h_T}(\delta h)\wedge F_{h_T}\in
\mathcal{H}_0(S)$, we only need to prove that $\delta{\bf
L_2}(h_T,\omega_S):0\oplus\{\phi\omega_S\}_{L_k^p} \rightarrow
\mathcal{H}_0(S)_{L_{k-2}^p}$ is surjective because we have
(\ref{007}). So we should solve the following equation:
\begin{equation}\label{008}
\sqrt{-1}\partial\bar\partial(e^u\phi\omega_S)
-\partial\bar\partial(e^{-u}\phi\textup{tr}(\bar\partial
A\wedge\partial A^*\cdot g_S^{-1})) +2\partial\bar\partial u\wedge
\partial\bar\partial \phi=\psi\frac{\omega_S^2}{2!}
\end{equation}
for any $\psi\in L_{k-2}^p$ such that $\int \psi=0$.  If we define
the linear operator $L$ from $L_k^p$ to $L_{k-2}^p$ by
$$L(\phi)=\sqrt{-1}\partial\bar\partial(e^u\phi\omega_S)
-\partial\bar\partial(e^{-u}\phi\textup{tr}(\bar\partial
A\wedge\partial A^*\cdot g_S^{-1})) +2\partial\bar\partial u\wedge
\partial\bar\partial \phi$$
then $L$ is the linearization of equation (\ref{002}) and we will
prove $\ker L^*=\mathbb{R}$ and $\dim \ker L=1$ in section 6. So $
\mathcal{H}_0(S)\perp \ker L^*$. From the  elliptic condition, the
operator $L$ is elliptic. So the equation (\ref{008}) has solution
$\phi\in L_k^p$ for any $\psi\in L^p_{k-2}$ such that
$\int\psi=0$. Thus we have proven that $P\circ\delta
L_1\oplus\delta L_2$ is surjective and it's kernel is  to
$V_0\oplus \ker L$.
\end{proof}

Now we deform the holomorphic structure $D_0''$. The following
proposition  is due to Jun Li.
\begin{prop}
There is a family $D_s''$ of deformations of holomorphic
structures of $E$ so that its $k$-th order for $k<m$
Kodaira-Spencer class $\kappa$ all vanish while its $m$-th order
Kodaira-Spencer class has non-vanishing summands in
$H^1(L_i^\vee\otimes TS)$ and $H^1(TS^\vee\otimes L_i)$.
\end{prop}
\begin{proof}(Given by Jun Li)
Because $\deg L_i=0$ and $TS$ is slope stable with respect to the
Hodge class $\omega_S$, $L_i^\vee\otimes TS$ has no global
sections. Using the Serre duality, $H^2(L_i\otimes TS)=0$ as well.
Thus to compute $H^1(L_i^\vee\otimes TS)$, we use Riemann-Roch for
K3 surfaces
$$\chi(L_i^\vee\otimes TS)=\frac{1}{2}c_1(L_i^\vee\otimes
TS)^2+2\chi(\mathcal{O}_S)-c_2(L_i^\vee\otimes TS)$$ Because
$c_1(L_i^\vee\otimes TS)=-2c_1(L_i)$ and $c_2(L^\vee_i\otimes
TS)=c_2(TS)+c_1(L_i)^2=20+c_1(L_i)^2$, we have
$$h^1(L_i^\vee\otimes TS)=16-c_1(L_i)^2\geq 16.$$
Here the last inequality follows from $c_1(L_i)\cdot \omega_S=0$
and the Hodge index theorem.

For the same reason, we have
$$h^1(TS^\vee\otimes L_i)\geq 16.$$

Now consider the vector bundle
$$E=L_1\oplus L_2\oplus TS.$$
The extension group $\textup{Ext}^1(E,E)$ has summands
\begin{equation}\label{010}
\textup{Ext}^1(L_i,TS) \and \textup{Ext}^1(TS,L_i),
\end{equation}
which are $H^1(L_i^\vee\otimes TS)$ and $H^1(TS^\vee\otimes L_i)$,
respectively; hence are positive dimension. Thus we can find a
direction $\eta\in \textup{Ext}^1(E,E)$ that has non-trivial
components in the desired factors (\ref{010}).

It remains to show that $\eta$ can be realized as the tangent of
an actual deformation. But this may not be true since the
obstruction to deformation of $E$ is the traceless part of
$\textup{Ext}^2(E,E)$, which is isomorphic two copies of
$H^2(\mathcal{O}_S)\cong\mathbb{C}$. What is true is that there is
a family of deformations of holomorphic structures of $E$, denoted
by $\bar\partial_s$ so that
$$\frac{d^k}{ds^k}\bar\partial_s\mid_{s=0}=0,\ \ \ k<m$$
and
$$\frac{d^m}{ds^m}\bar\partial_s\mid_{s=0}\neq 0$$
has non-trivial exponents in (\ref{010}).
\end{proof}

Actually, we can define the Kuranish map $K:\mathcal{U}\rightarrow
\text{Ext}^2(E,E)$, where $\mathcal{U}$ is some open neighborhood
of origin in $\text{Ext}^1(E,E)$. $K$ is the holomorphic map and
the complex analytic variety $\mathcal{X}=K^{-1}(0)$ is the
parametric space of all holomorphic structures on $E$ near
$D_0''$. Considering the dimensions of $\text{Ext}^1(E,E)$ and
$\text{Ext}^2(E,E)$, we can choose an element
$$\eta=\left(%
\begin{array}{ccc}
  0 &0 &C_{13} \\
  0 & 0 &C_{23} \\
  C_{31} &C_{32}& C_{33}
\end{array}%
\right) \in\text{Ext}^1(\mathcal{E},\mathcal{E})$$ such that
$C_{i3}\neq 0$ and $C_{3j}\neq 0$ for $1\leq i,j\leq 2 $ and
$\eta$ belongs to the tangent cone of $\mathcal{X}$ at the point
$D_0''$. So there is a curve $D_s''$ of degree $m$ of  smooth
deformation of the holomorphic structure $D_0''$. If we write
$$D_s''=D_0''+A_s, \qquad A_s\in \Omega^{0,1}(\text{End}^0 E),
$$then  $A^{(k)}_0=0$ for $k<m$ and $A^{(m)}_0=\eta$.
We assume that $C_{ij}$ are $D_0''$-harmonic. Because
$\text{Pic}S$ is discrete, we can assume further that
$\text{tr}A_s=0$ for all $s$.

With the connection forms $A_s$, the metric $H_0$ and Kahler form
$\omega_S$ so chosen, we can now define operators
$${\bf L_{s,1}}\oplus{\bf L_{s,2}}:\mathcal{H}_1(E)_{L_k^p}\times
\mathcal{C}(\omega_S)_{L_k^p}\rightarrow
\Omega^4_{\mathbb{R}}(\Endo^0 E)_{L_{k-2}^p}\oplus
\mathcal{H}_0(S)_{L_{k-2}^p}$$ with ${\bf L_{s,i}}$ defined as in
(\ref{003}) and (\ref{004}) of which the curvature form $F_h$ is
replaced by the hermitian curvature of $(E,D''_s, h)$:
$$F_{s,h}=D_{s,h}\circ D_{s,h}.$$
From paper \cite{LY}, we have
\begin{equation*}
F_{s,h}=F_s+D''_s(D_s'h\cdot h)
\end{equation*}
and
\begin{equation*}
F_s=F_0+(D''_0+D_0')(A_s-A_s^*)-(A_s-A^*_s)\wedge(A_s-A^*_S)
\end{equation*}
Because $\text{tr}A_s=0$, $\det h=1$ and $F_0\wedge \omega_S=0$,
$$\text{tr}{\bf
L_{s,1}}(h,e^\phi\omega_S)=e^{\phi}\text{tr}h^{-1/2}F_{s,h}h^{1/2}\wedge
\omega_S=0$$ So ${\bf L_{s,1}}$ still lies in
$\Omega^4_{\mathbb{R}}(\text{End}^0 E)_{L^p_{k-2}}$. Let $\bP$ be
the projection from $\Omega^4_{\mathbb{R}}(\text{End}^0
E)_{L_{k-2}^p}\oplus\mathcal{H}_0(S)_{L_{k-2}^p}$ to
$\Omega^4_{\mathbb{R}}(\text{End}^0 E)_{L_{k-2}^p}/V_1\oplus
\mathcal{H}_0(S)_{L_{k-2}^p}$. Because we have proven that the
linearized operator of ${\bf P}\circ{\bf L_{0,1}}\oplus{\bf
L_{0,2}}$ is surjective at $(h_{T_0},\omega_S)$. Hence by the
implicit theorem, for sufficiently small $s$ there are smooth
solutions $(h_{s,T},\omega_{s,T})$ to ${\bf P}\circ{\bf
L_{s,1}}\oplus{\bf L_{s,2}}=0$ near $(h_{T_0},\omega_S)$. We can
assume that the solutions $(h_{s,T},\omega_{s,T})$ can be
parameterized by
$$(s,T)\in [0,a)\times B_\epsilon (T_{0,1})\times
B_\epsilon(T_{0,2})$$ where $T_0=(T_{0,1},T_{0,2})$. For
simplicity, we denote by $F_{s,T}$ the curvature of the hermitian
vector bundle $(E,D_s'',h_{s,T})$. By our construction, it
satisfies
$${\bf L_{s,1}}(H_{s,T},\omega_{s,T})\equiv 0\!\!\mod V_1,\quad
{\bf L_{s,2}}(H_{s,T},\omega_{s,T})=0.
$$
Hence to find solutions to ${\bf L_{s,1}}\oplus{\bf L_{s,2}}=0$ it
suffices to investigate the vanishing loci of the functional $\bf
r_i(s,\cdot)$ from $B_\epsilon(T_{0,i})$ to the Lie algebra
$u(L_i)$ defined by
\begin{equation}\label{011}
\brr_i(s,T)=\int_X\bigl[\bL_{s,T}(H_{s,T},\omega_{s,T})\bigr]_{i}\wedge\omega_S
\end{equation}
where $[\cdot]_i$ is the projection from
$\Omega_{\mathbb{R}}^{\bullet}(\Endo^0 E)$ to
$\Omega_{\mathbb{R}}^{\bullet}(u( L_i))$ and  $u(L_i)$ is the
bundle of $<\,,>$-hermitian  antisymmetric endomorphisms of $L_i$.

We shall compute ${\bf r}_i^{(k)}(0,T)$ for all $T$ and for all
$k\leq 2m$. Because $\omega_{s,T}\in \mathcal{C}(\omega_S)$, we
can write $\omega_{s,T}={\phi_{s,T}}\omega_S$ for some positive
functions $\phi_{s,T}$ on $S$ such that $\phi_{0,T}=1$. Then we
have
\begin{equation}\label{012}
\frac{d^k}{ds^k}\mid_{s=0}\omega_{s,T}=\phi^{(k)}_{0,T}\omega_S.
\end{equation}
On the other hand, since $(h_{s,T},\omega_{s,T})$ are solutions to
${\bf L_{s,1}}\oplus{\bf L_{s,2}}=0 \mod V_1$, there is a function
${\bf c}(s,T)$ taking values in $V_1$ with ${\bf c}(0,T)=0$ so
that
\begin{equation}\label{013}
F_{s,T}\wedge \omega_{s,T}=h_{s,T}^{-\frac{1}{2}} {\bf c}(s,T)
h_{s,T}^{\frac{1}{2}}.
\end{equation}
We can write
$${\bf c}(s,T)=\text{diag}\left(M_1(s,T),M_2(s,T),-(M_1(s,T)/2+M_2(s,T)/2)I_3\right)\cdot\omega_S^2$$
where $M_i(s,T)$ is the function only depending on $s$ and $T$.

At first, we compute ${\bf r}_i^{(k)}(0,T)$ for any $T$ and $k\leq
m-1$. When $k\leq m-1$, $A^{(k)}_0=0$. Then
\begin{equation}\label{014}
F^{(k)}_{0,T}=\sum_{l=0}^{k}C_k^lD_0''[D_0'h_{0,T}^{(l)}\cdot
(h^{-1}_{0,T})^{(k-l)}].
\end{equation}
Because $D_0'h_T=0$, $\dot{F}_{0,T}=D_0''[D_0'\dot{h}_{0,T}\cdot
h^{-1}_T]$ and
$$\frac{d}{ds}\mid_{s=0}\left(h_{s,T}^{-1/2}F_{s,T}h^{1/2}_{s,T}\right)=h^{-1/2}_{0,T}
\dot{F}_{0,T}h^{1/2}_{0,T}.$$ We also have
$F_{0,T}\wedge\omega_S=0$. Combining these  equalities with
(\ref{012}),
\begin{equation*}
\dot \brr_i(0,T)=\int_S
T_i^{-1/2}[\dot{F}_{0,T}]_iT_i^{1/2}\wedge\omega_S+\int_ST^{-1/2}[F_{0,T}]_iT^{1/2}\wedge\dot{\omega}_S=0
\end{equation*}
On the other hand, taking derivative of $s$ at $s=0$ to
(\ref{013}) and couple ${\bf c}(0,T)=0$,  we have
\begin{equation*}
\dot{F}_{0,T}\wedge \omega_S=h_T^{1/2}\dot{\bf c}(0,T)h^{-1/2}_T
\end{equation*}
and then
\begin{equation*}
[\dot{F}_{0,T}]_i\wedge \omega_S=[\dot{\bf
c}(0,T)]_i=\dot{M}_i(0,T)\omega^2_S
\end{equation*}
From
\begin{equation*}
\int_S\dot{M}_i(0,T)\omega_S^2=\int_S[\dot{F}_{0,T}]_i\wedge\omega_S=0
\end{equation*}
we get $\dot{M}_i(0,T)=0$. So we get
\begin{equation*}
\dot{\bf c}(0,T)=0,\and
\dot{F}_{0,T}\wedge\omega_S=(D_0''D_0'\dot{h}_{0,T})\cdot
h^{-1}_{0,T}\wedge\omega_S=0
\end{equation*}
Thus
\begin{equation*}
\dot{h}_{0,T}\in V_0\and D_0'\dot{h}_{0,T}=0.
\end{equation*}
In this way, we can prove that
\begin{equation}\label{015}
{\bf r}^{(k)}_i(0,T)=0,\ \ {\bf c}^{(k)}(0,T)=0,\ \
F^{(k)}_{0,T}\wedge\omega_S=0\ \ D_0'h^{(k)}_{0,T}=0\ \  \text{for
any}\ \ k\leq m-1.
\end{equation}
When $k=m$, because of (\ref{015}) and (\ref{012}), we obtain
\begin{equation*}
\frac{d^m}{ds^m}\mid_{s=0}(h_{s,T}^{-1/2}F_{s,T}h^{1/2}_{s,T}\wedge\omega_{s,T})
=h^{-1/2}_{0,T}F_{0,T}^{(m)}h^{1/2}_{0,T}\wedge\omega_S
\end{equation*}
and
\begin{equation*}
F_{0,T}^{(m)}=(D_0''+D_0')(A_0^{(m)}-A_0^{*(m)})+D_0''[D_0'h^{(m)}_{0,T}]\cdot
h^{-1}_{0,T}
\end{equation*}
Then we get
\begin{equation*}
{\bf r}_i^{(m)}(0,T)=\int_S[F_{0,T}^{(m)}]_i\wedge\omega_S=0
\end{equation*}
Because $F_{0,T}^{(k)}\wedge\omega_S=0$, ${\bf c}^{(k)}(0,T)=0$
for $k\leq m-1$, from (\ref{013}), we get
\begin{equation*}
F_{0,T}^{(m)}\wedge\omega_S=h_T^{1/2}{\bf c}^{(m)}(0,T)h_T^{-1/2}
\end{equation*}
So using the same method of case $k=1$, we still have
\begin{equation*}
{\bf c}^{(m)}(0,T)=F^{(m)}_{0,T}\wedge \omega_S=0
\end{equation*}

When $k<m$,  $A^{(k)}_0=0$  and we have proven
$D_0'h_{0,T}^{(k)}=0$. By direct computation, we see
\begin{equation*}
F_{0,T}^{(m+k)}=(D_0''+D_0')(A_0^{(m+k)}-A_0^{*(m+k)})+D_0''\left(\frac{d^{m+k}}{ds^{m+k}}
\mid_{s=0}(D_s'h_{s,T}\cdot h_{s,T}^{-1})\right)
\end{equation*}
Then we still can get
\begin{equation*}
{\bf r}_i^{(m+k)}(0,T)=0,\ \ {\bf c}^{(m,k)}(0,T)=0\ \
F_{0,T}^{(m+k)}\wedge\omega_S=0\ \ \text{for}\ \ k<m.
\end{equation*}
At last we compute ${\bf r}^{(2m)}(0,T)$. Directly computing, we
get
\begin{eqnarray*}
F^{(2m)}_{0,T}&=&(D_0''+D_0')(A_0^{(2m)}-A^{*(2m)}_0)-C^m_{2m}(A_0^{(m)}-A_0^{*(m)})\wedge
(A_0^{(m)}-A_0^{*(m)})\\
&&-C^m_{2m}[A_0^{(m)},d^m/ds^{m}\mid_{s=0} (D_s'h_{s,T}\cdot
h^{-1}_{s,T})]\\
&&+D''_0(d^{2m}/ds^{2m}\mid_{s=0}(D_s'h_{s,T}\cdot h^{-1}_{s,T})
\end{eqnarray*}
and
\begin{equation*}
[A_0^{(m)},d^m/ds^{m}\mid_{s=0} (D_s'h_{s,T}\cdot h^{-1}_{s,T})]=
[A_0^{(m)},[A_0^{*(m)},h_{0,T}]h_{0,T}^{-1}]+
[A_0^{(m)},D_0'h_{0,T}^{(m)}\cdot h_{0,T}^{-1}]
\end{equation*}
Then from
$d^{2m}/ds^{2m}\mid_{s=0}(h^{-1/2}_{s,T}F_{s,T}h^{1/2}_{s,T}\wedge\omega_S)=h^{-1/2}_{0,T}
F^{(2m)}_{0,T}h^{1/2}_{0,T}\wedge\omega_S$, we see
\begin{eqnarray*}
{\bf r}_i^{(2m)}(0,T)&=&-C^m_{2m}\int_S
[(A_0^{(m)}-A_0^{*(m)})\wedge
(A_0^{(m)}-A_0^{*(m)})]_i\\
&&-C^m_{2m}\int_S[[A_0^{(m)},[A_0^{*(m)},h_{0,T}]h_{0,T}^{-1}]]_i\\
&&-C^m_{2m}\int_S [[A_0^{(m)},D_0'h_{0,T}^{(m)}\cdot
h_{0,T}^{-1}]]_i
\end{eqnarray*}
The last term is zero because $D_0''^*A_0^{(m)}=0$ and Lemma 2.3
in paper \cite{LY}. Using
$$ A_0^{(m)}=\left(%
\begin{array}{ccc}
  0&0 & C_{13} \\
  0&0&C_{23}  \\
  C_{31}&C_{32}&C_{33}
\end{array}%
\right) \and
H_{0,T} =\left(%
\begin{array}{ccc}
  T_1 & 0&0 \\
  0 &T_2 &0\\
  0&0&T^{-1/2}_1T_2^{-1/2}I_3
\end{array}%
\right)
$$
one computes
\begin{equation*}
[[A_0^{(m)},[A_0^{*(m)},h_{0,T}]h_{0,T}^{-1}]]_1=(1-T^{-3/2}_1T^{-1/2}_2)C_{13}\wedge
C_{13}^* +(1-T^{3/2}_1T_2^{1/2})C_{31}^*\wedge C_{31}
\end{equation*}
and \begin{equation*} -[(A_0^{(m)}-A_0^{*(m)})\wedge
(A_0^{(m)}-A_0^{*(m)})]_1=C_{13}\wedge C^*_{13}+C_{31}^*\wedge
C_{31}
\end{equation*}
Therefore
\begin{equation*}
{\bf r}^{(2m)}_1(0,T)=-C_{2m}^m\int_S
(T^{-3/2}_1T^{-1/2}_2C_{13}\wedge C_{13}^*
+T^{3/2}_1T_2^{1/2}C_{31}^*\wedge C_{31})\wedge\omega_S
\end{equation*}
As the same reason, we have
\begin{equation*}
{\bf r}^{(2m)}_2(0,T)=-C^m_{2m}\int_S
(T^{-1/2}_1T^{-3/2}_2C_{23}\wedge C_{23}^*
+T^{1/2}_1T_2^{3/2}C_{32}^*\wedge C_{32})\wedge\omega_S
\end{equation*}
Let $B_{i3}=\sqrt{-1}\int_SC_{i3}\wedge C_{i3}^*\wedge \omega_S$
and $B_{3i}=-\sqrt{-1}\int_S C_{3i}^*\wedge C_{3i}\wedge\omega_S$
for $i=1,2$. By our assumption, we have $B_{i3}>0$ and $B_{3i}>0$.
Then
\begin{eqnarray*}
{\bf r}^{(2m)}_1(0,T)&=&\sqrt{-1}C^m_{2m}\{
B_{13}T^{-3/2}_1T^{-1/2}_2-B_{31} T^{3/2}_1T^{1/2}_2\} \\
{\bf r}^{(2m)}_2(0,T)&=&\sqrt{-1}C^m_{2m}\{
B_{23}T^{-1/2}_1T^{-3/2}_2-B_{32}T^{1/2}_1T^{3/2}_2\} \\
\end{eqnarray*}
Clearly, ${\bf r}_i^{(2m)}(0,T_0)=0$ if
\begin{equation}
T_0=(T_{01},T_{02})=\left(\left(\frac{B_{13}}{B_{31}}\right)^{3/8}\left(\frac{B_{23}}{B_{32}}\right)^{-1/8}
,\left(\frac{B_{13}}{B_{31}}\right)^{-1/8}\left(\frac{B_{23}}{B_{32}}\right)^{3/8}\right)
\end{equation}
Then we define the map $G:B_\epsilon(T_0)\rightarrow S^1(1)$ by
\begin{equation*}
G(T)=\frac{({\bf r}^{2m}_1(0,T),{\bf
r}^{2m}_2(0,T))}{\parallel({\bf r}^{2m}_1(0,T),{\bf
r}^{2m}_2(0,T))\parallel}
\end{equation*}
or
\begin{equation}
G((T_1,T_2))=\frac{(B_{13}T^{-1}_1-B_{31}
T^{2}_1T^{1}_2,B_{23}T^{-1}_2-B_{32}T^{1}_1T^{2}_2)}{\parallel
(B_{13}T^{-1}_1-B_{31}
T^{2}_1T^{1}_2,B_{23}T^{-1}_2-B_{32}T^{1}_1T^{2}_2)
\parallel}
\end{equation}
Then it is easily proven that for $\epsilon$ small enough (here
$\epsilon<1$), $G$ is homotopic to
$$G_1(T)=\frac{(B_{13}T^{-1}_1,B_{23}T^{-1}_2)}{\parallel
(B_{13}T^{-1}_1,B_{23}T^{-1}_2)
\parallel}=\frac{(B_{13}T_2,B_{23}T_1)}{\parallel
(B_{13}T_2,B_{23}T_1)
\parallel}$$
 Thus $\deg G=\deg G_1=-1$.  Then as discussion of the proof of
 theorem 4.3 in paper \cite{LY}, we see
that the map for small $\epsilon$ enough, $$({\bf r}_1,{\bf
r}_2)(s,.):B_{\epsilon}(T_0)\rightarrow \mathbb{R}^2,\ \ \
s\in(0,a')
$$
attains value $0\in \mathbb{R}^2$ for all $s\in (0,a')$ in
$B_\epsilon(T_0)$. So for sufficiently small $s$, there are
solution $(h_s,e^\phi\omega_S)$ to ${\bf L_s}=0$ near
$(h_{T_0},\omega_S)$. From our definition of ${\bf L_{s,1}}$, we
know that $h_{s,T}$ is the hermitian-yang-mills solution on
$(E,D_s)$. From the proposition 13 for ${\bf L_s}$, we have gotten
the irreducible solution of stromingers system on non-Kahler
manifold $X$. Actually we have proven
\begin{theo}
Let $(E,H_0,S,\omega_S)$ be as before. Fix its  holomorphic
structure $D_0''$. Then there is a smooth deformation $D_s''$ of
$(E,D_0'')$ so that there are hermitia-yang-mills  metric $H_s$ on
$(E,D_s)$ and smooth function $\phi_s$ on $S$ such that
$$\left(V=\pi^*E,\pi^*D_s'',\pi^*H_s,\tilde \omega_s=\pi^*(e^{u+\phi_s}\omega_S)+\frac{\sqrt{-1}}{2}\theta
\wedge\bar\theta\right)$$ are the irreducible solutions to
strominger's system on $X$ and so that $\lim_{s\rightarrow
0}\phi_s=0$ and $\lim_{s\rightarrow 0}H_s$ is a  regular reducible
hermitian Yang-Mills connection on $E=L_1\oplus L_2\oplus TS$.
\end{theo}

\section{Openness}
We should  solve the following equation
\begin{equation}\lab{501}
\sqrt{-1}\partial\bar\partial
e^u\wedge\omega_S+\partial\bar\partial(e^{-u}\text{tr}(\bar\partial
A\wedge\partial A^*\cdot g^{-1}))+\partial\bar\partial
u\wedge\partial\bar\partial u=0
\end{equation}
 by
the continuity method. More precisely we introduce a parameter
$t\in [0,1]$ into the equation and consider the following equation
with parameter
\begin{equation}\label{502}
\sqrt{-1}\partial\bar\partial e^u\wedge\omega_S+{\bf
t}\partial\bar\partial(e^{-u}\text{tr}(\bar\partial
A\wedge\partial A^*\cdot g^{-1}))+\partial\bar\partial
u\wedge\partial\bar\partial u=0
\end{equation}
We need the following
\begin{equation}\label{503}
\text{Elliptic condition}:\ \ e^u\omega_S+\sqrt{-1}{\bf
t}e^{-u}\text{tr}(\bar\partial A\wedge\partial A^*\cdot
g^{-1})-2\sqrt{-1}\partial\dbar u>0,
\end{equation}
and
\begin{eqnarray}
\lab{504} \text{Normalization}:\ \ \ \ \int_S e^{-u}=A, \ \ \ \
\int_S 1=1
\end{eqnarray}
So we consider the solution in the following  space
\begin{equation}\label{506}
B=\{u\in C^{k+2,\alpha}\mid u \ \ \text{satisfies the conditions
(\ref {503}) and (\ref{504})}\}
\end{equation}
Let $C_0^{k,\alpha}=\{\psi\in C^{k,\alpha}\mid\int \psi=0\}$.
 Let \begin{equation*} {\bf T}=\{t\in [0,1]\mid \text{for}\ \
t \ \ \text{the equation (\ref{502}) admits a solution}\}
\end{equation*}
Now $0\in {\bf T}$ with the solution $u=-\ln A$. In this section
we prove
\begin{lemm}
${\bf T}$ is open.
\end{lemm}
\begin{proof}
Let $t_0\in {\bf T}$ and $u(t_0)$ is the solution of the equation
(\ref{502}). Define the linear operator $L$ from $C^{k+2,\alpha}$
to $C^{k,\alpha}$:
\begin{equation}\lab{507} L(\phi)=\ast_{\omega_S}
(\sqrt{-1}\partial\bar\partial
(e^{u_{t_0}}\phi)\wedge\omega_S-{\bf
t_0}\partial\bar\partial(e^{-u_{t_0}}\phi\text{tr}(\bar\partial
A\wedge\partial A^*\cdot g^{-1}))+2\partial\bar\partial u_{t_0}
\wedge\partial\bar\partial\phi).
\end{equation}
The principle part of operator $\ast_{\omega_S} L$ is
\begin{equation*}
\sqrt{-1}\partial\bar\partial
\phi\wedge(e^{u_{t_0}}\omega_S+{\sqrt{-1}\bf
t_0}e^{-u_{t_0}}\text{tr}(\bar\partial A\wedge\partial A^*\cdot
g^{-1})-2\sqrt{-1}\partial\bar\partial u_{t_0}).
\end{equation*}
From elliptic condition (\ref{503}), we get:
\begin{equation}\lab{0501}
\omega'_{t_0}=e^{u_{t_0}}\omega_S+{\sqrt{-1}\bf
t_0}e^{-u_{t_0}}\text{tr}(\bar\partial A\wedge\partial A^*\cdot
g^{-1})-2\sqrt{-1}\partial\bar\partial u_{t_0}>0
\end{equation}
So  $L$ is the linear elliptic operator.  Now Consider the
operator $G$ mapping $u$ in $B$ near $u_{t_0}$ to
$C^{k,\alpha}_0$:
$$\ast_{\omega_S} (\sqrt{-1}\partial\bar\partial e^u\wedge\omega_S+{\bf
t}\partial\bar\partial(e^{-u}\text{tr}(\bar\partial
A\wedge\partial A^*\cdot g^{-1}))+\partial\bar\partial
u\wedge\partial\bar\partial{ u}). $$
 Then the differential
$dG$ of $G$ at $u_{t_0}$ evaluated at the derivation $\phi$ is
$L(\phi)$\footnote{We have the formula: $-8\frac{\det(u_{i\bar
j})}{\det(g_{i\bar j})}=\ast\partial\dbar u\wedge\partial\dbar
u$.}. So $dG=L\mid_{T_{u_{t_0}}B}$, where $T_{u_{t_0}}B=\{\phi\in
C^{k+2,\alpha}\mid\int e^{-u_{t_0}}\phi=0\}$ is the  tangent space
of $B$ at $u_0$.  Before proving $dG$ is invertible, we introduce
the following elliptic operator $P$ (ref. P.224-227 in \cite{LT}).
 Because $\omega'_{t_0}$ is real and positive, $\omega_{t_0}'$ can be
 taken as the hermitian (not Kahler !) metric on
$S$. Let
\begin{equation}\label{509}
P=\sqrt{-1}\Lambda_{\omega_{t_0}'}\partial\dbar
\end{equation}
Then $P$ is the elliptic operator on $S$. As the operator
$\bigtriangleup$ of Kahler metric, it satisfies
\begin{equation}\label{510}
\sqrt{-1}\partial\dbar
\psi\wedge\omega_{t_0}'=P(\psi)\frac{\omega'^2_{t_0}}{2!}
\end{equation}
for any smooth function $\psi$ on $S$.  Furthermore, operator $P$
and its adjoint operator $P^*$ have the following properties:
\begin{lemm}(ref. \cite{LT})
1. $\ker(P)=\mathbb{C}$;\\
2. $\dim \ker(P^*)=1$, and every function
$\phi\in\ker(P^*\mid_{C^\infty(S,\mathbb{R})})$ has constant sign.
\\
3.
$C^\infty(S,\mathbb{R})=\textup{Im}(P\mid_{C^\infty(S,\mathbb{R})})\oplus\mathbb{R}$
\end{lemm}
Certainly, when  $k$ is big enough, the operator $P$ acting on
$C^{k,\alpha}$ and $P^*$ acting on $C^{k+2,\alpha}$ also have the
above properties. Now from the definitions of operators $L$, $L^*$
and $P$, for any $\psi\in C^{k,\alpha}(S,\mathbb{R})$,
\begin{eqnarray*}
&&\int L^*(\psi)\phi\frac{\omega^2_S}{2!}\\
&=&\int \psi\cdot L(\phi)\frac{\omega^2_S}{2!}\\
&=&\int \psi\cdot\{\sqrt{-1}\partial\bar\partial
(e^{u_{t_0}}\phi)\wedge\omega_S-{\bf
t_0}\partial\bar\partial(e^{-u_{t_0}}\phi\text{tr}(\bar\partial
A\wedge\partial A^*\cdot g^{-1}))+2\partial\bar\partial u_{t_0}
\wedge\partial\bar\partial\phi\}
\\
&=&\int \phi\sqrt{-1}\partial\partial
\psi\wedge(e^{u_{t_0}}\omega_S+{\sqrt{-1}\bf
t_0}e^{-u_{t_0}}\text{tr}(\bar\partial A\wedge\partial A^*\cdot
g^{-1})-2\sqrt{-1}\partial\bar\partial u_{t_0})\\
 &=&\sqrt{-1}\int \phi\partial\dbar \psi\wedge \omega'_{t_0}\\
&=&\int\phi\cdot P(\psi)\frac{\omega'^2_{t_0}}{2!}\\
&=&\int P^*(\phi)\psi\frac{\omega'^2_{t_0}}{2!}
\end{eqnarray*}
 Then from Lemma 18, $\ker L^*=\ker P=\mathbb{R}$ and $\ker L=\ker P^*=
 \{\mathbb{R}\phi_0\}$, where $\phi_0$ is some function has
constant sign. It is clearly that $\ker L\cap T_{u_{t_0}}B=0$. So
$dG$ is injective. Because $L$ is linear elliptic, it
 is well known that the condition for $L(\phi)=\psi$ to have a weak
 solution on $S$ is that $\psi\perp \ker L^*$. The Schauder theory makes sure that
 $\phi\in C^{k+2,\alpha}$ when $\psi\in C^{k,\alpha}$.
 Now for any $\psi\in C_0^{k,\alpha}$, we have
 $\psi\perp\ker L^*$. So there is a $\phi_1$ such that $L(\phi_1)=\psi$.
 Take $c_0=-\frac{\int e^{-u_{t_0}}\phi_1}{\int e^{-u_{t_0}}\phi_0}$,
then $\phi_1+c_0\phi_0\in T_{u_{t_0}}B$ and
$L(\phi_1+c_0\phi_0)=0$. So $dG$ is surjective. Hence $dG$  of $G$
at $u_{t_0}$ is invertible. Thus we can use the implicity function
theorem to get the openness of  the set ${\bf T}$.
\end{proof}

\section{Zero order estimate}
From this section to section 10, we do estimates up to third order
to equation (\ref{108}).  Let
$f=\frac{1+n^2}{4}\parallel\omega_1\parallel^2$, where $\omega_1$
is the anti-self dual (1,1)-form on $S$, then the equation
(\ref{108}) is
\begin{equation}\label{00501}
\bigtriangleup\left(e^u+fe^{-u}\right)-8\frac{\det(u_{i\bar
j})}{\det(g_{i\bar j})}=0
\end{equation}
The elliptic condition is
\begin{eqnarray}\label{00503} \omega'=(e^u-tfe^{-u})\omega_S-2\sqrt{-1}\partial\dbar u>0,
\end{eqnarray}
and the normalization is
\begin{eqnarray}
\lab{00504} \int_S e^{-u}=A, \ \ \ \ \int_S 1=1
\end{eqnarray}

Timing elliptic condition $e^u-fe^{-u}>\bigtriangleup u$ by
$pe^{-pu}$, we get
\begin{equation*}
p(e^{-pu})(e^u-fe^{-u}) \geq p(e^{-pu})\bigtriangleup u
=-\bigtriangleup
(e^{-pu})+4\mid\bigtriangledown(e^{-u})^{\frac{p}{2}}\mid^2
\end{equation*}
 Integrating, we see that
\begin{equation}\lab{1402}
\int_S\mid\bigtriangledown(e^{-u})^{\frac{p}{2}}\mid^2\leq\frac{p}{4}\int_S(e^{-u})^{p-1}
\end{equation}
Applying the Sobolev inequality, we can find a constant $C$
depending only on $S$ such that
\begin{equation}\lab{1403}
\begin{aligned}
\left(\int_S (e^{-u})^{2p}\right)^{\frac{1}{2}}&\leq C\int_S
(e^{-u})^p+C\int_S\mid\bigtriangledown (e^{-u})^{\frac{p}{2}}\mid^2\\
&\leq C\int_S (e^{-u})^p+\frac{p}{4}C\int_S (e^{-u})^{p-1}
\end{aligned}
\end{equation}
In the following we use the constant in the generic sense. So $C$
may mean different constants in different equations. By
(\ref{1403}) and Holder inequality, we get
\begin{equation}\lab{1404}
\begin{aligned}
 \int_S (e^{-u})^{2p}&\leq C^2\left(\int_S
(e^{-u})^p\right)^2+C^2
p^2\left(\int_S (e^{-u})^{p-1}\right)^2\\
&\leq C^2\left(\int_S (e^{-u})^p\right)^2+C^2 p^2\left(\int_S
(e^{-u})^p\right)^{\frac{2(p-1)}{p}}
\end{aligned}
\end{equation}
We assume that
\begin{equation}\lab{0517}
\int_Se^{-u}=A<1 \end{equation}
We discuss the following two cases:\\
Case (1): For any $p\in \mathbb{Z}$, $\int_Se^{-pu}<1$. Then
$\left(\int_S(e^{-u})^p\right)^2<\left(\int_S(e^{-u})^p\right)^{\frac{2(p-1)}{p}}$
and from (\ref{1404}),
\begin{equation*}
\int_S(e^{-u})^{2p}\leq
C^2p^2\left(\int_S(e^{-u})^p\right)^{\frac{2(p-1)}{p}}
\end{equation*}
Let $2p=2^{\beta}$, then
\begin{eqnarray*}
\int_S (e^{-u})^{2^{\beta}}&\leq& C^2(2^{\beta-1})^2\left(\int_S
(e^{-u})^{2^{\beta-1}}\right)^{2(1-2^{-(\beta-1)})}\\
&\leq&\left(\prod_{b=1}^{\beta-1}C^{2^b}\right)\left(\prod_{b=1}^{\beta-1}
\left(2^{(\beta-b)}\right)^{2^b}\right)
\left(\int_S(e^{-u})^2\right)^
{2^{\beta-1}\cdot\prod_{k=1}^{\beta-1}(1-\frac{1}{2^k})}\\
&\leq& C^{2^\beta-2}\cdot
2^{2^{\beta+1}}\left(\int_S(e^{-u})^2\right)^
{2^{\beta-1}\cdot\prod_{k=1}^{\beta-1}(1-\frac{1}{2^k})}
\end{eqnarray*}
where the last inequality follows by
\begin{equation}\lab{1405}
\prod_{b=1}^{\beta-1} \left(2^{\beta-b}\right)^{2^b}\leq
2^{2^{\beta+1}}
\end{equation}
which can be derived from following calculation:
\begin{eqnarray*}
\prod_{b=1}^{\beta-1}\left(2^{\beta-b}\right)^{2^b}&=&\prod_{b=1}^{\beta-1}2^{2^b}
\prod_{b=1}^{\beta-1}
\left(2^{\beta-(b+1)}\right)^{2^{b}}\\
&=&2^{2^{\beta}-2}\left(\prod_{b=1}^{{\beta}-1}
\left(2^{\beta-(b+1)}\right)^{2^{b+1}}\right)^{\frac{1}{2}}\\
&=&2^{2^\beta-1-\beta}\left( \prod_{b=1}^{\beta-1}
\left(2^{\beta-b}\right)^{2^b}\right)^{\frac{1}{2}}\\
&\leq&2^{2^\beta}\left( \prod_{b=1}^{\beta-1}
\left(2^{\beta-b}\right)^{2^b}\right)^{\frac{1}{2}}
\end{eqnarray*}
So we get
\begin{equation*}
\left(\int (e^{-u})^{2^{\beta}}\right)^{\frac{1}{2^\beta}}\leq
C^{1-2^{1-\beta}}\cdot 2^{2}\left(\int_S(e^{-u})^2\right)^
{\frac{1}{2}\cdot\prod_{k=1}^{\beta-1}(1-\frac{1}{2^k})}
\end{equation*}
Let $\beta\rightarrow \infty$, we see that
\begin{equation}\lab{1406}
\exp(-\inf u)=\parallel e^{-u}\parallel_\infty\leq
C\left(\int_S(e^{-u})^2\right)^{\frac{B}{2}}
\end{equation}
where
\begin{equation}\lab{1407}
B=\prod_{\beta=1}^{\infty}(1-\frac{1}{2^\beta})>0
\end{equation}
To finish our estimate of $\inf u$, it suffices to estimate
$\parallel e^{-u}\parallel_2$. When  $p=2$ the inequality
(\ref{1402}) yields
\begin{equation}\lab{1408}
\int_S\mid\bigtriangledown(e^{-u})\mid^2\leq\frac{1}{2}\int_Se^{-u}
\end{equation}
 Now from normalizing condition (\ref{504}), we have $\int_S(e^{-u}-A)=0$. So
 by the Poincare inequality and (\ref{1408}), we have
 \begin{equation*}
 \int_S\mid(e^{-u}-A)\mid^2\leq C\int_S\mid\nabla
 (e^{-u}-A)\mid^2\leq CA
 \end{equation*}
 and
 \begin{equation}\lab{1409}
 \int_S (e^{-u})^2\leq A^2+CA\leq CA
 \end{equation}
 Combining (\ref{1406}) and (\ref{1409}), we get
 \begin{equation}\lab{1410}
\exp(-\inf u)=\parallel e^{-u}\parallel_\infty\leq
C_1A^{\frac{B}{2}}
\end{equation}
and
\begin{equation}\lab{1411}
\inf u\geq -\ln C_1-\frac{B}{2}\ln A
\end{equation}

 Case(2): There is a integer $p$ such that $\int_Se^{-pu}>1$. Let
$p_0$ be the first such integer. Then for any $p>p_0$, by holder
inequality,
$$\int_Se^{-pu}\geq\left(\int_Se^{-p_0u}\right)^{\frac{p}{p_0}}>1$$
From (\ref{1403}), we know that
\begin{eqnarray*}
\int_S(e^{-u})^{2p}&\leq& C^2p^2\left(\int_S(e^{-u})^p\right)^2\ \
\ \ \  \ \ \ \ \text{for}\ \ p\geq p_0\\
\int_S(e^{-u})^{2p}&\leq&
C^2p^2\left(\int_S(e^{-u})^p\right)^{\frac{2(p-1)}{p}}\ \ \
\text{for}\ \ p< p_0
\end{eqnarray*}
Now using above inequality, discussing as the case (1), we can get
the estimate of $\inf u$. Furthermore, we still have bound
(\ref{1411}) with the same $B$ satisfy (\ref{1407}).

 Next
we estimate $\sup_S u$. At first, we compute
\begin{equation}\lab{1412}
\begin{aligned}
&\int_SP(e^{ku})\frac{\det g'}{\det
g}\frac{\omega^2}{2!}\\
&=\int_S2g'^{i\bar j}\frac{\partial^2(e^{ku})}{\partial
z_i\partial\bar{z}_j}\frac{\det g'}{\det g}\frac{\omega^2}{2!}\\
&=2k^2\int_Sg'^{i\bar j}e^{ku}\frac{\partial u}{\partial
z_i}\frac{\partial u}{\partial \bar
z_j}\frac{\omega^2}{2!}+2k\int_S g'^{i\bar
j}e^{ku}\frac{\partial^2u}{\partial z_i\partial \bar
z_j}\frac{\det g'}{\det g}\frac{\omega^2}{2!}\\
&\geq k\int_S e^{ku}\left(2g'^{i\bar j}\frac{\partial^2u}{\partial
z_i\partial \bar z_j}\frac{\det g'}{\det
g}\right)\frac{\omega^2}{2!}
\end{aligned}
\end{equation}
But applying the equation,
\begin{equation}\lab{1413}
\begin{aligned}
&2g'^{i\bar j}\frac{\partial^2u}{\partial z_i\partial \bar
z_j}\frac{\det g'}{\det g}\\
&=\left(2g'_{2\bar 2}\frac{\partial^2u}{\partial z_1\partial \bar
z_1}+ 2g'_{1\bar 1}\frac{\partial^2u}{\partial z_2\partial \bar
z_2}- 2g'_{1\bar2}\frac{\partial^2u}{\partial z_2\partial \bar
z_1}-2g'_{2\bar1} \frac{\partial^2u}{\partial z_1\partial \bar
z_2}\right)\frac{1}{\det g}\\
&=2(e^u-fe^{-u})g^{i\bar j}\frac{\partial^2u}{\partial z_i\partial
\bar
z_j}-16\frac{\det u_{i\bar j}}{\det g}\\
&=(e^u-fe^{-u})(\bigtriangleup u) -2\bigtriangleup(e^u+fe^{-u})
\end{aligned}
\end{equation}
Inserting (\ref{1413}) into (\ref{1412}), we get
\begin{equation}\lab{1414}
\int_SP(e^{ku})\frac{\det g'}{\det g}\frac{\omega^2}{2!}\geq
k\int_S e^{ku}(e^u-fe^{-u})(\bigtriangleup
u)\frac{\omega^2}{2!}-2k\int_Se^{ku}\bigtriangleup(e^u+fe^{-u})\frac{\omega^2}{2!}\\
\end{equation}
On the other hand, using the definition of operator $P$, we have
\begin{equation}\lab{1415}
\begin{aligned}
&\int_SP(e^{ku})\frac{\det g'}{\det g}\frac{\omega^2}{2!}
=\int_SP(e^{ku})\frac{\omega'^2}{2!}=\int_S
\sqrt{-1}\partial\bar\partial(e^{ku})\wedge \omega'\\
&=\int_S\sqrt{-1}\partial\bar\partial(e^{ku})\wedge ((e^u-fe^{-u})
\omega-2\sqrt{-1}\partial\bar\partial u)\\
&= \int_S (e^u-fe^{-u})\bigtriangleup (e^{ku})
\frac{\omega^2}{2!}+2\int_S\partial\bar\partial(e^{ku})\wedge\partial\bar\partial u\\
&=k\int_Se^{ku}(e^u-fe^{-u})(\bigtriangleup
u)\frac{\omega^2}{2!}+k^2\int_S(e^u-fe^{-u})e^{ku}\mid\bigtriangledown
u\mid^2\frac{\omega^2}{2!}
\end{aligned}
\end{equation}
Combining (\ref{1414}) and (\ref{1415}), we see that
\begin{equation}\lab{1416}
\begin{aligned}
&k^2\int_S(e^u-fe^{-u})e^{ku}\mid\bigtriangledown u\mid^2\frac{\omega^2}{2!}\\
&\geq-2k\int_Se^{ku}\bigtriangleup(e^u+fe^{-u})\frac{\omega^2}{2!}\\
&=-2k\int_Se^{ku}(e^u-fe^{-u})(\bigtriangleup
u)\frac{\omega^2}{2!}-2k\int_S e^{ku}(e^u+fe^{-u}) \mid
\bigtriangledown u\mid^2\\
&-2k\int_Se^{(k-1)u} (\bigtriangleup f)+4k\int_Se^{(k-1)u}
\bigtriangledown u\cdot\bigtriangledown f\\
\end{aligned}
\end{equation}
 Meanwhile, by integrate by part,
\begin{equation}\lab{1417}
\begin{aligned}
&-2k\int_Se^{ku}(e^u-fe^{-u})(\bigtriangleup
u)\frac{\omega^2}{2!}\\
&=2k(k+1)\int_Se^{(k+1)u}\mid\bigtriangledown
u\mid^2-2k(k-1)\int_Sfe^{(k-1)u}\mid\bigtriangledown u\mid^2\\
&-\frac{2k}{k-1}\int_Se^{(k-1)u}\bigtriangleup
f-2k\int_Se^{(k-1)u}\bigtriangledown u\cdot\bigtriangledown f\\
\end{aligned}
\end{equation}
Inserting (\ref{1417}) into (\ref{1416}) and applying Schwarz'
inequality, we  get
 \begin{eqnarray*}
&& k^2\int_S(e^u-fe^{-u})e^{ku}\mid\bigtriangledown u\mid^2\frac{\omega^2}{2!}\\
&\geq&2k^2\int_Se^{(k+1)u}\mid\bigtriangledown
u\mid^2-2k^2\int_Se^{(k-1)u}f\mid\bigtriangledown
u\mid^2-k\int_Se^{(k-1)u}\mid\bigtriangledown u\mid^2\\
&&-2k(1+\frac{1}{k-1})\int_Se^{(k-1)u}\bigtriangleup f
-k\int_Se^{(k-1)u}\mid
\bigtriangledown f\mid^2\\
\end{eqnarray*}
and we find
\begin{equation}\lab{1418}
\begin{aligned}
&k\int_Se^{(k-1)u}\mid \bigtriangledown
f\mid^2+2k(1+\frac{1}{k-1})\int_Se^{(k-1)u}\bigtriangleup f\\
&\geq k^2\int_Se^{(k+1)u}\mid\bigtriangledown
u\mid^2-k^2\int_Se^{(k-1)u}f\mid\bigtriangledown
u\mid^2-k\int_Se^{(k-1)u}\mid\bigtriangledown u\mid^2\\
\end{aligned}
\end{equation}
If we take $A>0$ small enough such that
\begin{equation}\lab{1419}
C_1^{-1}A^{-\frac{B}{2}}>1+\sup f
\end{equation}
then from (\ref{1410}) we see that $\inf e^u>1+\sup f$ and we can
estimate
\begin{equation}\lab{1420}
\begin{aligned}
& k^2\int_Se^{(k+1)u}\mid\bigtriangledown
u\mid^2-k^2\int_Se^{(k-1)u}f\mid\bigtriangledown
u\mid^2-k\int_Se^{(k-1)u}\mid\bigtriangledown u\mid^2\\
\geq& Ck^2\int_S e^{(k+1)u}\mid\bigtriangledown
u\mid^2=C\frac{4k^2}{(k+1)^2}\int_S\mid\bigtriangledown(e^u)^{\frac{k+1}{2}}\mid^2
\end{aligned}
\end{equation}
(\ref{1418}) and (\ref{1420}) imply that for  all $k\geq 1$,
\begin{equation}\lab{1421}
\int_S\mid\bigtriangledown(e^u)^{\frac{k+1}{2}}\mid^2\leq
C(k+1)\int_Se^{(k-1)u}
\end{equation}
Now applying the Sobolev inequality and since  $\inf u>0$, we get
\begin{equation}
\begin{aligned}\lab{1422}
\left(\int (e^{u})^{2k}\right)^{\frac{1}{2}}&\leq C\int
(e^{u})^k+C\int\mid\bigtriangledown (e^{u})^{\frac{k}{2}}\mid^2\\
&\leq C\int (e^{u})^k+Ck\int (e^{u})^{k-2}\leq Ck\int_S(e^{u})^k
\end{aligned}
\end{equation}
Let $2k=2^\beta$. (\ref{1422}) implies
\begin{eqnarray*}
\int_S(e^u)^{2^\beta}&\leq&
C(2^{\beta-1})^2\left(\int_S(e^u)^{2^{\beta-1}}\right)^2\\
&\leq&C^{2^{\beta}-2}\prod_{b=1}^{\beta-1}
\left(2^{\beta-b}\right)^{2^b}\left(\int_S(e^u)^2\right)^{2^{\beta-1}}\\
&\leq&C^{2^{\beta}-2}2^{2^{\beta+1}}\left(\int_S(e^u)^2\right)^{2^{\beta-1}}
\end{eqnarray*}
where the last inequality follows by (\ref{1405}). Let
$\beta\rightarrow\infty$, then we get
\begin{equation}\lab{1423}
\sup u=\parallel u\parallel_\infty=C\left(\int_S
(e^u)^2\right)^{\frac{1}{2}}
\end{equation}
So we should estimate $\parallel e^u\parallel_2$. When  $k=1$,
inequality (\ref{1421}) yields
\begin{equation}\lab{1424}
\int_S\mid\bigtriangledown(e^u)\mid^2\leq C
\end{equation}
Let $M_u=\int_S e^u$, then $\int_S(e^u-M_u)=0$. Applying Poincare
inequality and (\ref{1424}), we get
\begin{eqnarray*}
\int_S (e^u)^2-\left(\int_S(e^u)\right)^2&=&\int_S(e^u-M_u)^2
\\
&\leq& C\int_S\mid\bigtriangledown
(e^u-M_u)\mid^2\\
&=&C\int_S\mid\bigtriangledown e^u\mid^2\leq C
\end{eqnarray*}
So there is a constant $C_2$  depending on $S$, $f$ and $A$
(recall in (\ref{1419})) such that
\begin{equation}\lab{1425}
\int_S (e^u)^2\leq\left(\int_S(e^u)\right)^2+C_2
\end{equation}
Let $U_1=\{x\in S\mid \exp(-u(x))\geq\frac{A}{2}\}$ and
$U_2=\{x\in S\mid\exp(-u(x))<\frac{A}{2}\}$. Then from
(\ref{1410}), we have
\begin{eqnarray*}
A=\int_Se^{-u}&=&\int_{U_1}e^{-u}+\int_{U_2}e^{-u}<\int_{U_1}e^{-\inf
u}+\int_{U_2}\frac{A}{2}\\
&=&e^{-\inf u} \ \text{Vol}(U_1)+\frac{A}{2}(1- \
\text{Vol}(U_2))\\ &\leq&
\left(C_1A^{\frac{B}{2}}-\frac{A}{2}\right) \
\text{Vol}(U_1)+\frac{A}{2}
\end{eqnarray*}
Because $0<B<1$ and $0<A<1$, we can choose $A$ small enough such
that \begin{equation}\lab{05171} A<(2C_1)^{\frac{1}{1-B/2}}
\end{equation} then we have
\begin{equation*}
\text{Vol}(U_1)\geq
\frac{\frac{A}{2}}{C_1A^{\frac{B}{2}}-\frac{A}{2}}=\frac{1}{2C_1A^{\frac{B}{2}-1}-1}>0
\end{equation*}
and so
\begin{equation}\lab{1426}
\text{Vol}(U_2)=(1-\text{Vol}(U_1))
<1-\frac{1}{2C_1A^{\frac{B}{2}-1}-1}<1
\end{equation}
Now we want to use (\ref{1425}) and (\ref{1426}) to estimate
$\parallel e^u\parallel_2$. Applying Young inequality and Holder
inequality, we compute
\begin{equation}\lab{1427}
\begin{aligned}
\left(\int_Se^u\right)^2&=\left(\int_{U_1}e^u+\int_{U_2}e^u\right)^2\\
&\leq\left(1+\frac{1}{\epsilon}\right)\left(\int_{U_1}e^u\right)^2+(1+\epsilon)
\left(\int_{U_2}e^u\right)^2\\
&\leq\left(1+\frac{1}{\epsilon}\right)\left(\int_{U_1}e^{2u}\right)\text{Vol}(U_1)
+(1+\epsilon)\text{Vol}(U_2)
\left(\int_{U_2}e^{2u}\right)\\
&\leq\left(1+\frac{1}{\epsilon}\right)\left(\frac{2}{A}\right)^2+(1+\epsilon)\text{Vol}(U_2)
\left(\int_{U_2}e^{2u}\right)\\
&\leq\left(1+\frac{1}{\epsilon}\right)\left(\frac{2}{A}\right)^2+(1+\epsilon)\text{Vol}(U_2)
\left(\int_Se^{2u}\right)\\
\end{aligned}
\end{equation}
Inserting (\ref{1425}) and (\ref{1426}) into (\ref{1427}), we have
\begin{eqnarray*}
\left(\int_Se^u\right)^2
&\leq&\left(1+\frac{1}{\epsilon}\right)\left(\frac{2}{A}\right)^2+
C_2(1+\epsilon)
\left(1-\frac{1}{2C_1A^{\frac{B}{2}-1}-1}\right)\\
&&+(1+\epsilon) \left(1-\frac{1}{2C_1A^{\frac{B}{2}-1}-1}\right)
\left(\int_Se^{u}\right)^2\\
\end{eqnarray*}
Taking $\epsilon$ small enough such that
\begin{equation*}
(1+\epsilon) \left(1-\frac{1}{2C_1A^{\frac{B}{2}-1}-1}\right)<1
\end{equation*}
then we get
\begin{equation*}
\left(\int_Se^u\right)^2<\frac{\left(1+\frac{1}{\epsilon}\right)\left(\frac{2}{A}\right)^2+
C_2(1+\epsilon) \left(1-\frac{1}{2C_1A^{\frac{B}{2}-1}-1}\right)}
{1-(1+\epsilon) \left(1-\frac{1}{2C_1A^{\frac{B}{2}-1}-1}\right) }
\end{equation*}
 Now estimate of
$\int_S(e^u)^2$ follows from (\ref{1425}) and  and then estimate
of $\sup u$ follows from (\ref{1423})

\section{gradient estimate}
Let $\ln \mid\bigtriangledown u\mid^2+\ln v(u)$ achieves the
maximum at the point $q_1\in S$, where $v$ is some positive
function of $u$. Then at the point $q_1$ we have
\begin{equation}\lab{1501}
\bigtriangledown(\mid\bigtriangledown
u\mid^2)=-\left(\frac{v'(u)}{v(u)}\mid\bigtriangledown
u\mid^2\right)\bigtriangledown u
\end{equation}
We may choose the normal coordinate at the point $q_1$ such that
$\frac{\partial u}{\partial z_1}\neq 0$ and $\frac{\partial
u}{\partial z_2}=0$. Actually because $u$ is real, we can assume
that $\frac{\partial u}{\partial x_1}>0$ and $\frac{\partial
u}{\partial y_1}=0$. Thus we can assume that at the point $q_1$,
\begin{equation}\lab{1502}
\partial_1u\partial_{\bar 1} u=\partial_1 u\partial_1 u=\partial_{\bar 1}
u\partial_{\bar 1} u=\frac{1}{2}\mid \bigtriangledown u\mid^2
\end{equation}
At the point $q_1$,  from (\ref{1501}) we can also get
\begin{equation}\lab{1503}
\partial_1\partial _1u+\partial_1\partial_{\bar 1}u=\partial_{\bar 1}\partial_{\bar 1}u+
\partial_1\partial_{\bar 1}u=-\frac{1}{2}\frac{v'(u)}{v(u)}\mid
\bigtriangledown u\mid^2
\end{equation}
and
\begin{equation}\lab{1504}
\partial_1\partial_2u+\partial_2\partial_{\bar 1}u=\partial_{\bar 1}\partial_{\bar
2}u+\partial_1\partial_{\bar 2}u=0
\end{equation}
By the direct calculation, using (\ref{1502}), we see that at the
point $q_1$,
\begin{equation}\lab{1505}
\begin{aligned}
& P(\ln \mid \bigtriangledown u\mid ^2)\mid\bigtriangledown
u\mid^2\frac{\det g'}{\det g}
\\
&=2g'^{i\bar j}\left\{\frac{\partial^2\mid \bigtriangledown
u\mid^2}{\partial z_i\partial \bar{z}_j}-\frac{\partial \mid
\bigtriangledown u\mid^2}{\partial
z_i}\frac{\partial\mid\bigtriangledown u\mid^2}{\partial
\bar{z}_j}\cdot\frac{1}{\mid\bigtriangledown
u\mid^2}\right\}\frac{\det g'}{\det g}\\
&= 4g'^{i\bar j}(\partial_i\partial_{\bar j}\partial_p
u\partial_{\bar p}u+\partial_i\partial_{\bar j}\partial_{\bar
p}u\partial_pu)\frac{\det g'}{\det g}\\
&-4g'^{i\bar j}(\partial_i\partial_{\bar 1}u\partial_{\bar
1}\partial_{\bar j}u+\partial_i\partial_1u\partial_1\partial_{\bar
j}u)\frac{\det g'}{\det g}\\
& +4g'^{i\bar j}(\partial_i\partial_{\bar
2}u\partial_{2}\partial_{\bar j}u)\frac{\det g'}{\det
g}+2g'^{i\bar j}R^{1\bar 1}\ _{i\bar j}\mid\bigtriangledown
u\mid^2\frac{\det g'}{\det g}\\
&+4g'^{i\bar j}(\partial_i\partial_{ 2}u\partial_{\bar
2}\partial_{\bar j}u)\frac{\det g'}{\det g}.
\end{aligned}
\end{equation}
The last term $4g'^{i\bar j}(\partial_i\partial_{
2}u\partial_{\bar 2}\partial_{\bar j}u)\frac{\det g'}{\det
g}\geq0$. So we should estimate the first four terms. By equation
and (\ref{1501}), the first term in (\ref{1505}) is
\begin{equation}\lab{1506}
\begin{aligned}
&4g'^{i\bar j}(\partial_i\partial_{\bar j}\partial_p
u\partial_{\bar p}u+\partial_i\partial_{\bar j}\partial_{\bar
p}u\partial_pu)\frac{\det g'}{\det g}\\
&=2(e^u-fe^{-u})\bigtriangledown\bigtriangleup u\cdot
\bigtriangledown u-16\left(\bigtriangledown\left(\frac{\det
u_{i\bar j}}{\det g_{i\bar j}}\right)\cdot \bigtriangledown
u\right)\\
&=2(e^u-fe^{-u})\{\bigtriangledown\bigtriangleup u\cdot
\bigtriangledown
u\}-2\{\bigtriangledown\bigtriangleup(e^u+fe^{-u})\cdot
\bigtriangledown u\}\\
&=-2(e^u+fe^{-u})\bigtriangleup u\mid\bigtriangledown
u\mid^2-2(e^u-fe^{-u})\mid \bigtriangledown
u\mid^4\\
& -4e^{-u}(\bigtriangledown u\cdot \bigtriangledown
f)\mid\bigtriangledown u\mid^2+2e^{-u}\bigtriangleup
f\mid\bigtriangledown u\mid^2\\
 &-2(e^u+fe^{-u})(\bigtriangledown\mid\bigtriangledown
u\mid^2\cdot \bigtriangledown u)+2e^{-u}\bigtriangleup
u(\bigtriangledown
u\cdot \bigtriangledown f)\\
 &-2e^{-u}(\bigtriangledown
u\cdot\bigtriangledown\bigtriangleup
f)+4e^{-u}(\bigtriangledown(\bigtriangledown
u\cdot\bigtriangledown f)\cdot \bigtriangledown u)\\
&=-2(e^u+fe^{-u})\bigtriangleup u\mid\bigtriangledown
u\mid^2-2(e^u-fe^{-u})\mid \bigtriangledown u\mid^4\\
&-4e^{-u}(\bigtriangledown u\cdot \bigtriangledown
f)\mid\bigtriangledown u\mid^2+2e^{-u}\bigtriangleup
f\mid\bigtriangledown u\mid^2\\
&+2\frac{v'(u)}{v(u)}(e^u+fe^{-u})\mid\bigtriangledown u\mid^4
+2e^{-u}\bigtriangleup u(\bigtriangledown u\cdot
\bigtriangledown f)\\
&-2e^{-u}(\bigtriangledown u\cdot\bigtriangledown\bigtriangleup
f)+4e^{-u}(\bigtriangledown(\bigtriangledown
u\cdot\bigtriangledown f)\cdot \bigtriangledown u)
\end{aligned}
\end{equation}
But from (\ref{1502}), (\ref{1503}) and (\ref{1504}),
\begin{equation}\lab{1507}
\begin{aligned}
&\bigtriangledown(\bigtriangledown u\cdot\bigtriangledown
f)\cdot\bigtriangledown u\\
&=\left\{-\frac{\sqrt{2}}{4}\frac{v'(u)}{v(u)}\mid\bigtriangledown
u\mid(\partial_1f+\partial_{\bar
1}f)+\frac{1}{2}(\partial_1\partial_1f+2\partial_1\partial_{\bar
1}f+\partial_{\bar 1}\partial_{\bar1}f)\right\}
\mid\bigtriangledown
u\mid^2\\
&\geq-(C_3\mid\bigtriangledown u\mid+C_3)\mid\bigtriangledown
u\mid^2
\end{aligned}
\end{equation}
where in the last inequality $C_3$  depends on $\sup u$, $\inf u$
and $f$. In the following we use the constant $C_3$ depending on
$\sup u$, $\inf u$, $f$ and $S$ in the generic sense. So $C_3$ may
mean different constants in the different equations. Inserting
(\ref{1507}) into (\ref{1506}) and applying Schwarz inequality,
then the first term in (\ref{1505}) is
\begin{equation}\lab{1508}
\begin{aligned}
&\ \ \ 4g'^{i\bar j}(\partial_i\partial_{\bar j}\partial_p
u\partial_{\bar p}u+\partial_i\partial_{\bar j}\partial_{\bar
p}u\partial_pu)\frac{\det g'}{\det g}\\
&\geq-2(e^u-fe^{-u})\mid \bigtriangledown
u\mid^4+2\frac{v'(u)}{v(u)}(e^u+fe^{-u})\mid\bigtriangledown
u\mid^4\\
&\ \ \ +2(e^u+fe^{-u})(e^u-fe^{-u}-\bigtriangleup
u)\mid\bigtriangledown
u\mid^2\\
&\ \ \ -(e^u-fe^{-u}-\bigtriangleup u)(C_3\mid\bigtriangledown
u\mid^{}+C_3)\\
&\ \ \ -C_3\mid\bigtriangledown u\mid^3-C_3\mid\bigtriangledown
u\mid^2-C_3\mid\bigtriangledown u\mid-C_3
\end{aligned}
\end{equation}
Next we compute the second term in (\ref{1505}):
\begin{equation}\lab{1509}
\begin{aligned}
&\ \ \ -4g'^{i\bar j}(\partial_i\partial_{\bar
1}u\partial_{\bar1}\partial_{\bar
j}u+\partial_i\partial_1u\partial_1\partial_{\bar j}u)
\frac{\det g'}{\det g}\\
&=-4(e^u-fe^{-u})(\partial_i\partial_{\bar
1}u\partial_{\bar1}\partial_{\bar
i}u+\partial_i\partial_1u\partial_1\partial_{\bar i}u)\\
&\ \ \ +2\times 8\frac{\det u_{i\bar
j}}{\det g_{i\bar j}}(\partial_{\bar 1}\partial_{\bar1}u+\partial_1\partial_1u)\\
&=-4(e^u-fe^{-u})(\partial_i\partial_{\bar
1}u\partial_{\bar1}\partial_{\bar
i}u+\partial_i\partial_1u\partial_1\partial_{\bar i}u)\\
&\ \ \ +2(e^u-fe^{-u})(\bigtriangleup u)(\partial_{\bar 1}\partial_{\bar1}u+\partial_1\partial_1u)\\
&\ \ \ +2\{(e^u+fe^{-u})\mid\bigtriangledown
u\mid^2+e^{-u}\bigtriangleup f-2e^{-u}\bigtriangledown
u\cdot\bigtriangledown f\}(\partial_{\bar
1}\partial_{\bar1}u+\partial_1\partial_1u).
\end{aligned}
\end{equation}
But the first two terms in (\ref{1509}) are equal to
\begin{equation}\lab{1510}
\begin{aligned}
&\ \ \ -4(e^u-fe^{-u})(\partial_i\partial_{\bar
1}u\partial_{\bar1}\partial_{\bar
i}u+\partial_i\partial_1u\partial_1\partial_{\bar i}u)\\
&\ \ \ +2(e^u-fe^{-u})(\bigtriangleup u)(\partial_{\bar 1}\partial_{\bar1}u+\partial_1\partial_1u)\\
&=4(e^u-fe^{-u})(\partial_{\bar 1}\partial_{\bar
1}u+\partial_1\partial_1u)\partial_2\partial_{\bar2}u\\
&\ \ \
-4(e^u-fe^{-u})(\partial_2\partial_{\bar1}u\partial_{\bar1}\partial_{\bar2}u+
\partial_2\partial_1u\partial_1\partial_{\bar2}
u)\\
\end{aligned}
\end{equation}
From (\ref{1503}) we have
\begin{equation}\lab{1511}
\partial_1\partial_1u+\partial_{\bar1}\partial_{\bar1}u=-\frac{v'(u)}{v(u)}\mid\bigtriangledown
u\mid^2-2\partial_1\partial_{\bar1}u
\end{equation}
Inserting (\ref{1511}) and (\ref{1504}) into (\ref{1510}), and
using equation and Schwarz inequality, we get
\begin{equation}\lab{1512}
\begin{aligned}
&\ \ \ -4(e^u-fe^{-u})(\partial_i\partial_{\bar
1}u\partial_{\bar1}\partial_{\bar
i}u+\partial_i\partial_1u\partial_1\partial_{\bar i}u)\\
&\ \ \ +2(e^u-fe^{-u})(\bigtriangleup u)(\partial_{\bar 1}\partial_{\bar1}u+\partial_1\partial_1u)\\
&=-4(e^u-fe^{-u})\frac{v'(u)}{v(u)}u_{2\bar 2}\mid\bigtriangledown
u\mid^2-(e^u-fe^{-u})^2(\bigtriangleup
u)\\
&\ \ \ -(e^u-fe^{-u})\{(e^u+fe^{-u})\mid\bigtriangledown
u\mid^2+e^{-u}\bigtriangleup f-2e^{-u}\bigtriangledown
u\cdot\bigtriangledown f\}
\\
&\geq-4(e^u-fe^{-u})\frac{v'(u)}{v(u)}u_{2\bar
2}\mid\bigtriangledown
u\mid^2+(e^u-fe^{-u})^2(e^u-fe^{-u}-\bigtriangleup
u)\\
&\ \ \ -C_3\mid\bigtriangledown u\mid^2
-C_3\mid\bigtriangledown u\mid-C_3\\
&\geq-4(e^u-fe^{-u})\frac{v'(u)}{v(u)}u_{2\bar
2}\mid\bigtriangledown u\mid^2-C_3\mid\bigtriangledown u\mid^2
-C_3\mid\bigtriangledown u\mid-C_3\\
\end{aligned}
\end{equation}
Applying (\ref{1511}) and Schwarz inequality, the third term in
(\ref{1509}) is
\begin{equation}\lab{1513}
\begin{aligned}
&\ \ \ 2\{(e^u+fe^{-u})\mid\bigtriangledown
u\mid^2+e^{-u}\bigtriangleup f-2e^{-u}\bigtriangledown
u\cdot\bigtriangledown f\}(\partial_{\bar
1}\partial_{\bar1}u+\partial_1\partial_1u)\\
&\geq-2(e^u+fe^{-u})\frac{v'(u)}{v(u)}\mid\bigtriangledown u\mid^4
-C_3\mid\bigtriangledown u\mid^3-C_3\mid\bigtriangledown u\mid^2\\
&\ \ \ -4\{(e^u+fe^{-u})\mid\bigtriangledown
u\mid^2+e^{-u}\bigtriangleup f-2e^{-u}\bigtriangledown
u\cdot\bigtriangledown f\}u_{ 1\bar
1}\\
\end{aligned}
\end{equation}
Inserting (\ref{1512}) and (\ref{1513}) into (\ref{1509}), we get
the estimate of second term in (\ref{1505})
\begin{equation}\lab{1514}
\begin{aligned}
&\ \ \ -4g'^{i\bar j}(\partial_i\partial_{\bar
1}u\partial_{\bar1}\partial_{\bar
j}u+\partial_i\partial_1u\partial_1\partial_{\bar j}u) \frac{\det
g'}{\det g}\\
&\geq-2(e^u+fe^{-u})\frac{v'(u)}{v(u)}\mid\bigtriangledown
u\mid^4-4(e^u-fe^{-u})\frac{v'(u)}{v(u)}u_{2\bar
2}\mid\bigtriangledown u\mid^2
\\
&\ \ \ -4\{(e^u+fe^{-u})\mid\bigtriangledown
u\mid^2+e^{-u}\bigtriangleup f-2e^{-u}\bigtriangledown
u\cdot\bigtriangledown f\}u_{ 1\bar 1}\\
&\ \ \ -C_3\mid\bigtriangledown u\mid^3-C_3\mid\bigtriangledown
u\mid^2 -C_3\mid\bigtriangledown u\mid-C_3
\end{aligned}
\end{equation}
The third term in (\ref{1505}) is
\begin{equation}\lab{1515}
\begin{aligned}
&\ \ \ \ 4g'^{i\bar
j}(\partial_i\partial_{\bar2}u\partial_2\partial_{\bar j}u)\cdot
\frac{\det g'}{\det g}\\
&=-4(e^u-fe^{-u})\det(u_{i\bar j})+2(e^u-fe^{-u})(\bigtriangleup
u)u_{2\bar 2}
\\
&\ \ \ -2\bigtriangleup(e^u+fe^{-u})\cdot u_{2\bar 2}\\
 &=-4(e^u-fe^{-u})\det(u_{i\bar
j})\\
&\ \ \ -2\{(e^u+fe^{-u})\mid\bigtriangledown
u\mid^2+e^{-u}\bigtriangleup f-2e^{-u}\bigtriangledown
u\cdot\bigtriangledown f\}u_{2\bar 2}\\
&\geq\frac{1}{2}(e^u-fe^{-u})^2(e^u-fe^{-u}-\bigtriangleup
u)-C_3\mid\bigtriangledown
u\mid-C_3\mid\bigtriangledown u\mid^2\\
&\ \ \ -2\{(e^u+fe^{-u})\mid\bigtriangledown
u\mid^2+e^{-u}\bigtriangleup f-2e^{-u}\bigtriangledown
u\cdot\bigtriangledown f\}u_{2\bar 2}\\
&\geq -2\{(e^u+fe^{-u})\mid\bigtriangledown
u\mid^2+e^{-u}\bigtriangleup f-2e^{-u}\bigtriangledown
u\cdot\bigtriangledown f\}u_{2\bar 2}\\
&\ \ \ -C_3\mid\bigtriangledown
u\mid-C_3\mid\bigtriangledown u\mid^2\\
\end{aligned}
\end{equation}
Combine following  two terms in (\ref{1514}) and (\ref{1515}):
\begin{equation*}
\begin{aligned}
&\ \ \ \ -4\{(e^u+fe^{-u})\mid\bigtriangledown
u\mid^2+e^{-u}\bigtriangleup f-2e^{-u}\bigtriangledown
u\cdot\bigtriangledown f\}u_{ 1\bar 1}\\
 &\ \ \ \ -2\{(e^u+fe^{-u})\mid\bigtriangledown
u\mid^2+e^{-u}\bigtriangleup f-2e^{-u}\bigtriangledown
u\cdot\bigtriangledown f\}u_{2\bar 2}\\
&\geq\frac{1}{2}\{(e^u+fe^{-u})\mid\bigtriangledown
u\mid^2+e^{-u}\bigtriangleup f-2e^{-u}\bigtriangledown
u\cdot\bigtriangledown f\}(e^u-fe^{-u}-4u_{ 1\bar
1})\\
&\ \ \ +(e^u+fe^{-u})(e^u-fe^{-u}-\bigtriangleup
u)-(e^u-fe^{-u}-\bigtriangleup
 u)(C_3\mid\bigtriangledown u\mid+C_3)\\
&\ \ \ -C_3\mid\bigtriangledown u\mid^3-C_3\mid\bigtriangledown
u\mid^2 -C_3\mid\bigtriangledown
u\mid-C_3\\
&\geq(e^u+fe^{-u})(e^u-fe^{-u}-\bigtriangleup
u)-(e^u-fe^{-u}-\bigtriangleup
 u)(C_3\mid\bigtriangledown u\mid+C_3)\\
&\ \ \ -C_3\mid\bigtriangledown u\mid^3-C_3 \mid\bigtriangledown
u\mid^2-C_3\mid\bigtriangledown
u\mid-C_3\\
\end{aligned}
\end{equation*}
where in the last inequality we have assumed that
\begin{equation*}
(e^u+fe^{-u})\mid\bigtriangledown u\mid^2+e^{-u}\bigtriangleup
f-2e^{-u}\bigtriangledown u\cdot\bigtriangledown f\geq 0
\end{equation*}
Otherwise we could have gotten  the estimate of
$\mid\bigtriangledown u\mid^2$ at the point $q_1$. Then combing
(\ref{1514}) and (\ref{1515}), we get
\begin{equation}\lab{1516}
\begin{aligned}
&\ \ \ \ -4g'^{i\bar j}(\partial_i\partial_{\bar 1}u\partial_{\bar
1}\partial_{\bar j}u+\partial_i\partial_1u\partial_1\partial_{\bar
j}u-\partial_i\partial_{\bar 2}u\partial_{2}\partial_{\bar
j}u)\frac{\det g'}{\det
g}\\
&\geq-2(e^u+fe^{-u})\frac{v'(u)}{v(u)}\mid\bigtriangledown
u\mid^4-4(e^u-fe^{-u})\frac{v'(u)}{v(u)}u_{2\bar
2}\mid\bigtriangledown u\mid^2\\
&\ \ \ -C_3\mid\bigtriangledown u\mid^3-C_3 \mid\bigtriangledown
u\mid^2-C_3\mid\bigtriangledown u\mid-C_3
\end{aligned}
\end{equation}
 Let $R=\supp\mid
R^{1\bar 1}\ _{i\bar j}\mid$.
 The forth
term is
\begin{equation}\lab{1517}
\begin{aligned}
&2g'^{i\bar j}R^{1\bar 1}\ _{i\bar j}\mid\bigtriangledown
u\mid^2\frac{\det g'}{\det g}=-8u^{i\bar j}R^{1\bar 1}\ _{i\bar
j}\mid\bigtriangledown u\mid^2\\
&\geq-8R\sum_{i,j=1}\mid u_{i\bar j}\mid\mid\bigtriangledown
u\mid^2\geq-8R\left((\bigtriangleup u)^2-8\det u_{i\bar
j}\right)^{\frac{1}{2}}\mid\bigtriangledown u\mid^2\\
&\geq-16R(e^u-fe^{-u}-\bigtriangleup u)\mid\bigtriangledown
u\mid^2-C_3\mid\bigtriangledown u\mid^2
\end{aligned}
\end{equation}
Inserting (\ref{1508}), (\ref{1516}) and (\ref{1517}) into
(\ref{1505}), we can see taht
\begin{equation}\lab{1518}
\begin{aligned}
&P(\ln \mid\bigtriangledown u\mid^2)\mid\bigtriangledown
u\mid^2\frac{\det g'}{\det g}
\\
&\geq\{(3(e^u+fe^{-u})-16R)\mid\bigtriangledown
u\mid^2-C_3\mid\bigtriangledown
u\mid-C_3\}(e^u-fe^{-u}-\bigtriangleup
u)\\
&+\frac{v'(u)}{v(u)}(e^u-fe^{-u})(e^u-fe^{-u}-4u_{2\bar 2})
\mid\bigtriangledown u\mid^2\\
&-2(e^u-fe^{-u})\mid\bigtriangledown
u\mid^4-C_3\mid\bigtriangledown u\mid^3-C_3\mid\bigtriangledown
u\mid^2 -C_3\mid\bigtriangledown
u\mid-C_3\\
\end{aligned}
\end{equation}
Next we compute
\begin{equation}\lab{1519}
\begin{aligned}
&P(\ln v)\frac{\det g'}{\det g}\\
&=2g'^{i\bar j}\left(\frac{v'}{v}\frac{\partial^2 u}{\partial
z_i\partial\bar z_j}+\frac{vv''-v'^2}{v^2}\frac{\partial
u}{\partial z_i}\frac{\partial u}{\partial\bar
z_j}\right)\frac{\det g'}{\det g}\\
&=\frac{v'}{v}\left\{(e^u-fe^{-u})\bigtriangleup
u-16\frac{\det(u_{i\bar j})}{\det
g}\right\}+\frac{vv''-v'^2}{v^2}g'^{1\bar1}\mid\bigtriangledown
u\mid^2\frac{\det g'}{\det g}\\
&=-\frac{v'(u)}{v(u)}(e^u-fe^{-u})\bigtriangleup
u+\frac{vv''-v'^2}{v^2}\mid\bigtriangledown
u\mid^2(e^u-fe^{-u}-4u_{2\bar
2})\\
&-2\frac{v'(u)}{v(u)}\{(e^u+fe^{-u})\mid\bigtriangledown
u\mid^2+e^{-u}\bigtriangleup f-2e^{-u}\bigtriangledown u\cdot
\bigtriangledown f\}\\
&\geq-\frac{v'(u)}{v(u)}(e^u-fe^{-u})(e^u-fe^{-u}-\bigtriangleup
u)-2\frac{v'(u)}{v(u)}(e^u+fe^{-u})\mid\bigtriangledown
u\mid^2\\
&+\frac{vv''-v'^2}{v^2}\mid\bigtriangledown
u\mid^2(e^u-fe^{-u}-4u_{2\bar 2})-C_3\mid\bigtriangledown
u\mid-C_3
\end{aligned}
\end{equation}
From (\ref{1518}) and (\ref{1519}) we get
\begin{equation}\lab{1520}
\begin{aligned}
&P(\ln \mid \bigtriangledown u\mid^2+\ln v)\mid\bigtriangledown
u\mid^2\frac{\det g'}{\det
g}\\
&=\left\{(2(e^u+fe^{-u})+\frac{v'(u)}{v(u)}(e^u-fe^{-u})\right\}(e^u-fe^{-u}-\bigtriangleup
u)\mid\bigtriangledown u\mid^2\\
&+\{(e^u+fe^{-u}-16R)\mid\bigtriangledown
u\mid^2-C_3\mid\bigtriangledown
u\mid-C_3\}(e^u-fe^{-u}-\bigtriangleup
u)\\
&+\left\{\frac{vv''-v'^2}{v^2}\mid\bigtriangledown u\mid^2+
\frac{v'(u)}{v(u)}(e^u-fe^{-u})\right\}(e^u-fe^{-u}-4u_{2\bar
2})\mid\bigtriangledown u\mid^2\\
&-2\left\{(e^u-fe^{-u})+\frac{v'(u)}{v(u)}(e^u+fe^{-u})\right\}\mid\bigtriangledown u\mid^4\\
& -C_3\mid\bigtriangledown u\mid^3-C_3\mid\bigtriangledown u\mid^2
-C_3\mid\bigtriangledown
u\mid-C_3\\
\end{aligned}
\end{equation}

Take
\begin{equation*}
v(u)=e^{4\sup u-2u}+e^{2u-4\sup u}>0
\end{equation*}
Then
\begin{eqnarray*}
v'(u)&=&-2e^{4\sup u-2u}+2e^{2u-4\sup u}<0\\
v''(u)&=&4e^{4\sup u-2u}+4e^{2u-4\sup u}=4v(u)>0
\end{eqnarray*}
So the factor  first term in  (\ref {1520}) is
\begin{equation}\lab{1521}
2(e^u+fe^{-u})+(e^u-fe^{-u})\frac{v'(u)}{v(u)}>0
\end{equation}
The factor of third term in (\ref{1520}) is:
\begin{equation}\lab{1522}
\begin{aligned}
&\frac{vv''-v'2}{v^2}\mid\bigtriangledown
u\mid^2+\frac{v'}{v}(e^u-fe^{-u})
\\
&=\frac{16}{v^2}\mid\bigtriangledown u\mid^2-\frac{e^ue^{4\sup
u-2u}}{e^{4\sup u-2u}}\\
&>\frac{16}{e^{4\sup u-2\inf u}+3}\mid\bigtriangledown
u\mid^2-e^{\sup u}
\end{aligned}
\end{equation}
Choose $A$ such that
\begin{equation}\lab{1523}
C_1^{-1}A^{-\frac{B}{2}}>7^{\frac{1}{3}}
\end{equation}
then $e^{\inf u}>\frac{1}{3}\ln 7$ and the coefficient of forth
term in (\ref{1520}) is
\begin{equation}\lab{1524}
\begin{aligned}
&-2\left\{(e^u-fe^{-u})+\frac{v'}{v}(e^u+fe^{-u})\right\}\\
&\geq2\frac{e^{4\sup u-u}-3e^{3u-u\sup u}}{e^{4\sup
u-2u}+e^{2u-4\sup
u}}\\
&\geq\frac{2e^{4\sup u-u}-6}{e^{4\sup u-u}+1}>1
\end{aligned}
\end{equation}
Choose $A$ such that
\begin{equation}\lab{1525}
C^{-1}_1A^{-\frac{B}{2}}>16 R+1
\end{equation}
then $e^{\inf u}>16R+1$. Applying all above inequalities, at last
we can see that at the point $q_1$
\begin{equation}\lab{1526}
\begin{aligned}
0&\geq P(\ln\mid\bigtriangledown u\mid^2+\ln v)
\mid\bigtriangledown u\mid^2\frac{det g'}{\det g}\\
&\geq(\mid\bigtriangledown u\mid^2-C_3\mid\bigtriangledown
u\mid-C_3)(e^u-fe^{-u}-\bigtriangleup u)\\
&+\left\{\frac{16}{e^{4\sup u-2\inf u}+1}\mid\bigtriangledown
u\mid^2-e^{\sup u}\right\}(e^u-fe^{-u}-4u_{2\bar2})
\mid\bigtriangledown u\mid^2\\
&+\mid\bigtriangledown u\mid^4-C_3\mid\bigtriangledown
u\mid^3-C_3\mid\bigtriangledown u\mid^2- C_3\mid\bigtriangledown
u\mid- C_3
\end{aligned}
\end{equation}
From above inequality, we can easily see that there is a constant
$C_4$ depending on $f$, $M$, $\sup u$ and $\inf u$ such that
$\mid\bigtriangledown u\mid^2(q_1)\leq C_4$.

Since $\ln\mid\bigtriangledown u\mid^2+\ln(e^{4\sup
u-2u}+e^{2u-\sup u})$ achieves its maximum at $q_1$, we get the
bound of $\mid\bigtriangledown u\mid^2$:
\begin{equation}
\begin{aligned}
\mid\bigtriangledown u\mid^2&\leq C_4\frac{\left(e^{4\sup
u-2u(q)}+e^{2u(q)-4\sup u}\right)}{\left(e^{4\sup
u-2u}+e^{2u-4\sup u}\right)}\\
&\leq C_4\frac{\left(e^{4\sup u-2\inf u}+e^{2\inf u-4\sup
u}\right)}{\left(e^{2\sup
u}+e^{-2\sup u}\right)}\\
\end{aligned}
\end{equation}

\section{Second order Estimate}
We now do the second order a priori estimate  of $u$. Since
$(e^u-fe^{-u})g_{i\bar j}-4\partial_i\partial_{\bar j}u$ is
positive definite, to get a second order estimate of $u$ it
sufficient to have an upper bound estimate of
$e^u-fe^{-u}-\bigtriangleup u$. We fix a point $q_2$ and choose
normal coordinate at that point for $g_{i\bar j}$. Let $g'_{i\bar
j}=(e^u-fe^{-u})g_{i\bar j}-4\partial_i\partial_{\bar j}u$. We
rewrite the equation as
\begin{equation}\lab{2501}
\frac{\det g'_{i\bar j}}{\det g_{i\bar j}}=F
\end{equation}
where
\begin{equation}\lab{2502}
F=(e^u-fe^{-u})^2+2\{(e^u+fe^{-u})^2\mid\bigtriangledown
u\mid^2+e^{-u}\bigtriangleup f-2e^{-u}\bigtriangledown
u\cdot\bigtriangledown f\}
\end{equation}
Differential (\ref{2501}), we have
\begin{equation}\lab{2503}
g^{i\bar j}\frac{\partial g'_{i\bar j}}{\partial z_k}=g^{i\bar
j}\frac{\partial g_{i\bar j}}{\partial
z_k}+\frac{1}{F}\frac{\partial F}{\partial z_k}
\end{equation}
 We differentiate (\ref{2503}) again and obtain
\begin{eqnarray*}
&&-g'^{i\bar q}g'^{p\bar j}\frac{\partial g'_{p\bar q}}{\partial
\bar z_l}\frac{\partial g'_{i\bar j}}{\partial z_k}+g'^{i\bar
j}\frac{\partial^2g'_{i\bar j}}{\partial z_k\partial\bar
z_l}\\
&=&-g^{i\bar q}g^{p\bar j}\frac{\partial g_{p\bar q}}{\partial
\bar z_l}\frac{\partial g_{i\bar j}}{\partial z_k}+g^{i\bar
j}\frac{\partial^2g_{i\bar j}}{\partial z_k\partial\bar
z_l}+\frac{1}{F}\frac{\partial^2F}{\partial z_k\partial\bar
z_l}-\frac{1}{F^2}\frac{\partial F}{\partial z_k}\frac{\partial
F}{\partial\bar z_l}
\end{eqnarray*}
or
\begin{equation}\lab{2504}
\begin{aligned}
-4g'^{i\bar j}\frac{\partial^{4}u}{\partial z_i\partial\bar
z_j\partial z_k\partial \bar z_l}&=g'^{i\bar q}g'^{p\bar
j}\frac{\partial g'_{p\bar q}}{\partial \bar z_l}\frac{\partial
g'_{i\bar j}}{\partial z_k}-g'^{i\bar
j}\frac{\partial^2((e^u-fe^{-u})g_{i\bar j})}{\partial
z_k\partial\bar z_l}\\
&\ \ \ \ -g^{i\bar q}g^{p\bar j}\frac{\partial g_{p\bar
q}}{\partial \bar z_l}\frac{\partial g_{i\bar j}}{\partial
z_k}+g^{i\bar j}\frac{\partial^2g_{i\bar j}}{\partial
z_k\partial\bar z_l}\\
&\ \ \ \ +\frac{1}{F}\frac{\partial^2F}{\partial z_k\partial\bar
z_l}-\frac{1}{F^2}\frac{\partial F}{\partial z_k}\frac{\partial
F}{\partial\bar z_l}
\end{aligned}
\end{equation}
Contracting (\ref{2504}) with $g^{kl}$ and using the fact that the
metric $g_{i\bar j}$ is Ricci-flat,
 we have
 \begin{equation}\lab{2505}
\begin{aligned}
P(-\bigtriangleup u)&=g^{k\bar l}g'^{i\bar j}g'^{p\bar
q}\frac{\partial g'_{i\bar q}}{\partial z_k}\frac{\partial
g'_{p\bar j}}{\partial \bar z_l}-\frac{1}{2}\bigtriangleup
(e^u-fe^{-u})\sum_{i=1}^{2}g'^{i\bar i}\\
 &\ \ \ \ +\frac{1}{2F}\bigtriangleup
F-\frac{1}{F^2}g^{k\bar l}\frac{\partial F}{\partial
z_k}\frac{\partial F}{\partial\bar z_l}+4g'^{i\bar
j}\frac{\partial^2g^{k\bar l}}{\partial z_i\partial\bar
z_j}\frac{\partial^2 u}{\partial z_k\partial\bar z_l}
\end{aligned}
\end{equation}
Timing $\frac{\det g'_{i\bar j}}{\det g_{i\bar j}}$ to above
equation and using (\ref{2503}) and  (\ref{2501}), we see
 \begin{equation}\lab{2506}
\begin{aligned}
P(-\bigtriangleup u)\frac{\det g'_{i\bar j}}{\det g_{i\bar
j}}&=g^{k\bar l}g'^{i\bar j}g'^{p\bar q}\left(\frac{\partial
g'_{i\bar q}}{\partial z_k}\frac{\partial g'_{p\bar j}}{\partial
\bar z_l}-\frac{\partial g'_{i\bar j}}{\partial z_k}\frac{\partial
g'_{p\bar q}}{\partial\bar z_l}\right)\frac{\det
g'_{i\bar j}}{\det g_{i\bar j}}\\
 &\ \ \ \ +\frac{1}{2}\bigtriangleup
F-\frac{1}{2}\bigtriangleup (e^u-fe^{-u})\sum_{i=1}^{2}g'^{i\bar
i}\frac{\det g'_{i\bar j}}{\det g_{i\bar j}}\\
&\ \ \ \ +4g'^{i\bar j}\frac{\partial^2g^{k\bar l}}{\partial
z_i\partial\bar z_j}\frac{\partial^2 u}{\partial z_k\partial\bar
z_l}
\end{aligned}
\end{equation}
Now we compute at the point $q_2$
\begin{equation}\lab{2507}
\begin{aligned}
&\ \ \ \ g^{k\bar l}g'^{i\bar j}g'^{p\bar q}\left(\frac{\partial
g'_{i\bar q}}{\partial z_k}\frac{\partial g'_{p\bar j}}{\partial
\bar z_l}-\frac{\partial g'_{i\bar j}}{\partial z_k}\frac{\partial
g'_{p\bar q}}{\partial\bar z_l}\right)\frac{\det g'_{i\bar
j}}{\det g_{i\bar j}}\\
&=g^{k\bar l}(g'^{i\bar j}g'^{p\bar q}-g'^{i\bar q}g'^{p\bar
j})\frac{\partial g'_{i\bar q}}{\partial z_k}\frac{\partial
g'_{p\bar j}}{\partial \bar z_l}\frac{\det g'_{i\bar j}}{\det
g_{i\bar j}}\\
&=g^{k\bar l}\left(\frac{\partial g'_{2\bar 1}}{\partial
z_k}\frac{\partial g'_{1\bar 2}}{\partial \bar z_l}+\frac{\partial
g'_{1\bar 2}}{\partial z_k}\frac{\partial g'_{2\bar 1}}{\partial
\bar z_l}-\frac{\partial g'_{1\bar 1}}{\partial z_k}\frac{\partial
g'_{2\bar 2}}{\partial \bar z_l}-\frac{\partial g'_{2\bar
2}}{\partial z_k}\frac{\partial g'_{1\bar 1}}{\partial \bar
z_l}\right)\\
&=16\sum_{i\neq j}\left(\frac{\partial^3 u}{\partial z_i\partial
\bar z_j\partial z_k}\frac{\partial^3 u}{\partial z_j\partial \bar
z_i\partial\bar z_k}-\frac{\partial^3 u}{\partial z_i\partial \bar
z_i\partial z_k}\frac{\partial^3 u}{\partial z_j\partial \bar
z_j\partial\bar z_k}\right)\\
&\ \ \ \ +4\sum_{i}\frac{\partial^3 u}{\partial z_i\partial \bar
z_i\partial z_k}\frac{\partial (e^u-fe^{-u})}{\partial \bar
z_k}+4\sum_{i}\frac{\partial^3 u}{\partial z_i\partial \bar
z_i\partial \bar z_k}\frac{\partial (e^u-fe^{-u})}{\partial
z_k}\\
&\ \ \ \ -2\frac{\partial (e^u-fe^{-u})}{\partial \bar
z_k}\frac{\partial (e^u-fe^{-u})}{\partial  z_k}\\
&= 16\sum_{i,j}\frac{\partial^3 u}{\partial z_i\partial \bar
z_j\partial z_k}\frac{\partial^3 u}{\partial z_j\partial \bar
z_i\partial\bar z_k}-16\sum_{i,j}\frac{\partial^3 u}{\partial
z_i\partial \bar z_i\partial z_k}\frac{\partial^3 u}{\partial
z_j\partial \bar z_j\partial\bar z_k}
\\& \ \ \ \
+2\bigtriangledown (\bigtriangleup
u)\cdot\bigtriangledown(e^u-fe^{-u})-\mid\bigtriangledown
(e^u-fe^{-u})\mid^2\\
&\geq-2\mid\bigtriangledown\bigtriangleup
u\mid^2+2\bigtriangledown (\bigtriangleup
u)\cdot\bigtriangledown(e^u-fe^{-u})-C_5
\end{aligned}
\end{equation}
where $C_5$ depends on $f$, $M$ and $u$ up to one order
derivation. In the following we will use $C_5$ in the generic
sense. Because we want to estimate the upper bound of
$e^u-fe^{-u}-\bigtriangleup u$, we assume that
$e^u-fe^{-u}-\bigtriangleup u$ achieves the maximum at point $q_2$
and we take the normal coordinate at this point for $g_{i\bar j}$.
So at the point $q_2$, we have
\begin{equation}\lab{2508}
\bigtriangledown \bigtriangleup u=\bigtriangledown(e^u-fe^{-u})
\end{equation}
Inserting (\ref{2508}) into (\ref{2507}) and then inserting
(\ref{2507}) into (\ref{2506}), we obtain
\begin{equation}\lab{2509}
\begin{aligned}
P(-\bigtriangleup u)\frac{\det g'_{i\bar j}}{\det g_{i\bar
j}}&\geq \frac{1}{2}\bigtriangleup F+4g'^{i\bar
j}\frac{\partial^2g^{k\bar l}}{\partial z_i\partial\bar
z_j}\frac{\partial^2 u}{\partial z_k\partial\bar z_l}\\
&-\frac{1}{2}\bigtriangleup (e^u-fe^{-u})\sum_{i=1}^{2}g'^{i\bar
i}\frac{\det g'_{i\bar j}}{\det g_{i\bar j}}-C_5
\end{aligned}
\end{equation}
At first, we deal with the second term in above inequality. Using
equation, Schwarze inequality, and the fact metric $g_{i\bar j}$
is Ricci-flat, we have
\begin{equation}\lab{2510}
\begin{aligned}
 4g'^{i\bar j}\frac{\partial^2g^{k\bar l}}{\partial
z_i\partial\bar z_j}\frac{\partial^2 u}{\partial z_k\partial\bar
z_l} &=-16\left\{\frac{\partial^2g^{k\bar l}}{\partial
z_1\partial\bar z_1}\frac{\partial^2 u}{\partial u_2\partial \bar
z_ 2}+\frac{\partial^2g^{k\bar l}}{\partial z_2\partial\bar
z_2}\frac{\partial^2 u}{\partial u_1\partial \bar
z_1}\right\}\frac{\partial^2 u}{\partial z_k\partial\bar z_l}\\
&\ \ \  -16\left\{\frac{\partial^2g^{k\bar l}}{\partial
z_1\partial\bar z_2}\frac{\partial^2 u}{\partial u_2\partial \bar
z_ 1}+\frac{\partial^2g^{k\bar l}}{\partial z_2\partial\bar
z_1}\frac{\partial^2 u}{\partial u_1\partial \bar z_
2}\right\}\frac{\partial^2 u}{\partial z_k\partial\bar z_l}\\
&\geq-64(\max R_{i\bar jk\bar l})\sum_{ij}\mid u_{i\bar j}\mid^2\\
&=-16(\max R_{i\bar jk\bar l})(\bigtriangleup u)^2
-16(\max R_{i\bar jk\bar l})\times 8\frac{\det u_{i\bar j}}{\det g_{i\bar j}}\\
&=-16(\max R_{i\bar jk\bar l})\{(\bigtriangleup
u)^2-\bigtriangleup(e^u+fe^{-u})\}\\
&\geq-16(\max R_{i\bar jk\bar l})(\bigtriangleup
u)^2-C_5\bigtriangleup u-C_5\
\end{aligned}
\end{equation}
We also have
\begin{equation}\lab{2511}
\sum_{i=1}^2g'^{i\bar i}\frac{\det g_{i\bar j}}{\det g_{i\bar
j}}=2(e^u-fe^{-u}-\bigtriangleup u)
\end{equation}
Inserting (\ref{2510}) and (\ref{2511}) into (\ref{2509}), we
obtain
\begin{equation}\lab{2512}
\begin{aligned}
&\ \ \ \ P(-\bigtriangleup u)\frac{\det g'_{i\bar j}}{\det
g_{i\bar j}}\\
&\geq \frac{1}{2}\bigtriangleup F-\bigtriangleup
(e^u-fe^{-u})(e^u-fe^{-u}-\bigtriangleup u)
\\
&\ \ \ \ -16(\max R_{i\bar jk\bar l})(\bigtriangleup
u)^2-C_5\bigtriangleup u-C_5\\
&\geq \frac{1}{2}\bigtriangleup F+\left\{(e^u+fe^{-u})-16(\max
R_{i\bar jk\bar l})\right\}(\bigtriangleup u)^2-C_5\bigtriangleup
u-C_5
\end{aligned}
\end{equation}
So we should compute
\begin{equation}\lab{2513}
\begin{aligned}
\bigtriangleup F&=\bigtriangleup
(e^u-fe^{-u})^2+2\bigtriangleup(e^u+fe^{-u})\mid\bigtriangledown u
\mid^2\\
&+2(e^u+fe^{-u})\bigtriangleup(\mid\bigtriangledown u \mid^2)
+2\bigtriangledown(e^u+fe^{-u})\cdot\bigtriangledown(\mid\bigtriangledown
u\mid^2)\\
&+2\bigtriangleup e^{-u}\bigtriangleup
f+2e^{-u}\bigtriangleup^2f+2\bigtriangledown
e^{-u}\cdot\bigtriangledown\bigtriangleup f -4\bigtriangleup
e^{-u}\bigtriangledown u\cdot\bigtriangledown
f\\
&-4e^{-u}\bigtriangleup(\bigtriangledown u\cdot\bigtriangledown
f)-4\bigtriangledown e^{-u}\cdot\bigtriangledown(\bigtriangledown
u\cdot\bigtriangledown f)\\
&\geq+2(e^u+fe^{-u})\bigtriangleup(\mid\bigtriangledown u \mid^2)
+2\bigtriangledown(e^u+fe^{-u})\cdot\bigtriangledown(\mid\bigtriangledown
u\mid^2)\\
&-4e^{-u}\bigtriangleup(\bigtriangledown u\cdot\bigtriangledown
f)-4\bigtriangledown e^{-u}\cdot\bigtriangledown(\bigtriangledown
u\cdot\bigtriangledown f)-C_5\bigtriangleup u-C_5
\end{aligned}
\end{equation}
Using (\ref{2508}) and the equation, we have
\begin{equation}\lab{2514}
\begin{aligned}
&\bigtriangleup \mid\bigtriangledown u\mid^2=4g^{k\bar
l}\frac{\partial^2}{\partial z_k\partial\bar z_l}\{g_{i\bar
j}u_{i}u_{\bar j}\}\\
&=4g^{k\bar k}g^{i\bar i}\left\{\frac{\partial ^3u}{\partial
z_i\partial z_k\partial \bar z_{ k}}\frac{\partial u}{\partial
\bar z_i}+\frac{\partial ^3u}{\partial \bar z_i\partial
z_k\partial \bar z_{ k}}\frac{\partial u}{\partial z_i}\right\}
\\
&+4g^{i\bar i}g^{k\bar k}\left\{\frac{\partial ^2u}{\partial
z_i\partial z_k}\frac{\partial ^2u}{\partial \bar z_i\partial \bar
z_k}+\frac{\partial ^2u}{\partial z_i\partial \bar z_{
k}}\frac{\partial^2 u}{\partial
\bar z_i\partial z_k}\right\}\\
&=\bigtriangledown\bigtriangleup u\cdot\bigtriangledown
u+(\bigtriangleup u)^2-8\frac{\det u_{i\bar j}}{\det g_{i\bar
j}}+4g^{i\bar j}g^{k\bar l}u_{ik}u_{\bar j\bar l}\\
&=\bigtriangledown(e^u-fe^{-u})\cdot\bigtriangledown
u+(\bigtriangleup u)^2-\bigtriangleup(e^u+fe^{-u})
+4g^{i\bar j}g^{k\bar l}u_{ik}u_{\bar j\bar l}\\
&\geq(\bigtriangleup u)^2+4\Gamma-C_5\bigtriangleup u-C_5\\
\end{aligned}
\end{equation}
where we let $\Gamma=g^{i\bar j}g^{k\bar l}u_{ik}u_{\bar j\bar l}$
(see next section). Using (\ref{2508}) and schwarz inequality, we
also have
\begin{equation}\lab{2515}
4e^{-u}\bigtriangleup(\bigtriangledown u\cdot\bigtriangledown
f)\geq -C_5\sum_{ij}\mid u_{i\bar
j}\mid-C_5\Gamma^{\frac{1}{2}}-C_5
\end{equation}
and
\begin{equation}\lab{2516}
\begin{aligned}
&\ \ \ \
2\bigtriangledown(e^u+fe^{-u})\cdot\bigtriangledown(\mid\bigtriangledown
u\mid^2)-4\bigtriangledown
e^{-u}\cdot\bigtriangledown(\bigtriangledown
u\cdot\bigtriangledown f)\\
&\geq -C_5\sum_{ij}\mid u_{i\bar
j}\mid-C_5\Gamma^{\frac{1}{2}}-C_5
\end{aligned}
\end{equation}
Then inserting (\ref{2514}), (\ref{2515}) and (\ref{2516}) into
(\ref{2513}), we see
\begin{equation}\lab{2517}
\begin{aligned}
\bigtriangleup F&\geq 2(e^u+fe^{-u})(\bigtriangleup
u)^2+8(e^u+fe^{-u})\Gamma-C_5\bigtriangleup u\\
&\ \ \ \ -C_5\sum_{ij}\mid u_{i\bar
j}\mid-C_5\Gamma^{\frac{1}{2}}-C_5\\
&\geq 2(e^u+fe^{-u})(\bigtriangleup u)^2-C_5\bigtriangleup
u-C_5\sum_{ij}\mid u_{i\bar j}\mid-C_5
\end{aligned}
\end{equation}
Inserting (\ref{2517}) into (\ref{2512}), we obtain
\begin{equation}\lab{2518}
\begin{aligned}
P(-\bigtriangleup u)\frac{\det g'_{i\bar j}}{\det g_{i\bar
j}}&\geq\left\{2(e^u+fe^{-u})-16(\max R_{i\bar jk\bar
l})\right\}(\bigtriangleup u)^2\\
& -C_5\bigtriangleup u-C_5\sum_{ij}\mid u_{i\bar j}\mid-C_5
\end{aligned}
\end{equation}
Next we compute
\begin{equation}\lab{2519}
\begin{aligned}
&\ \ \ \ P(e^u-fe^{-u})\frac{\det g'_{i\bar j}}{\det g_{i\bar
j}}\\
&=2g'^{k\bar l}\frac{\partial ^2(e^u-fe^{-u})}{\partial
z_k\partial\bar z_l}\frac{\det g'_{i\bar j}}{\det g_{i\bar j}}\\
&=\bigtriangleup (e^u-fe^{-u})-8\sum_{i\neq
j}\{\partial_i\partial_{\bar i}u\partial_j\partial_{\bar
j}(e^u-fe^{-u})-\partial_i\partial_{\bar
j}u\partial_j\partial_{\bar i}(e^u-fe^{-u})\}\\
&=\bigtriangleup
(e^u-fe^{-u})-2(e^u+fe^{-u})\bigtriangleup(e^u-fe^{-u})\\
&\ \ \ \ -8(e^{-u}-fe^{-u})\sum_{i\neq
j}\{\partial_i\partial_{\bar i}u\partial_ju\partial_{\bar j}u
-\partial_i\partial_{\bar j}u\partial_ju\partial_{\bar i}u\}\\
&\ \ \ \ -8e^{-u}\sum_{i\neq j}\{\partial_i\partial_{\bar
i}u(\partial_ju\partial_{\bar j}f+\partial_{\bar j}u\partial_jf)
-\partial_i\partial_{\bar j}u(\partial_ju\partial_{\bar i}f
+\partial_{\bar i}u\partial_jf)\}\\
&\ \ \ \ +8e^{-u}\sum_{i\neq j}\{\partial_i\partial_{\bar
i}u\partial_j\partial_{\bar j}
f-\partial_i\partial_{\bar j}u\partial_j\partial_{\bar i}f\}\\
&\geq-C_5\bigtriangleup u-C_5\sum_{ij}\mid u_{i\bar j}\mid-C_5
\end{aligned}
\end{equation}
Combining (\ref{2517}) and (\ref{2519}), we obtain
\begin{equation}\lab{2520}
\begin{aligned}
&\ \ \ \ P((e^u-fe^{-u}-\bigtriangleup u)\frac{\det g'_{i\bar
j}}{\det
g_{i\bar j}}\\
&\geq\left\{2(e^u+fe^{-u})-16(\max R_{i\bar jk\bar
l})\right\}(\bigtriangleup u)^2\\
&\ \ \ -C_5\bigtriangleup u-C_5\sum_{ij}\mid u_{i\bar j}\mid-C_5\\
&\geq(\bigtriangleup u)^2-C_5\bigtriangleup u-C_5\
\end{aligned}
\end{equation}
because we have chosen $A$ such that $A$ satisfies (\ref{1525}),
from which we can get $e^{\inf u}>16 R+1$. Because we assume that
$e^u-fe^{-u}-\bigtriangleup u$ achieves the maximum at point
$q_2$, then (\ref{2520}) reads as
\begin{equation}
(\bigtriangleup u)^2-C_5\bigtriangleup u-C\leq 0
\end{equation}
Then we can easily get the upper bound estimate of
$e^u-fe^{-u}-\bigtriangleup u$.

\section{third order estimate}
Let \begin{eqnarray*} \Gamma&=&g^{i\bar j}g^{k\bar l}u_{ik}u_{\bar
j\bar l}\\
\Theta&=&g'^{i\bar r}g'^{s\bar j}g'^{k\bar t}u_{i\bar jk}u_{\bar r
s\bar
t}\\
\Xi&=&g'^{i\bar j}g'^{k\bar l}g'^{p\bar q}u_{ikp}u_{\bar j\bar l
\bar  q}\\
\Phi&=&g'^{i\bar j}g'^{k\bar l}g'^{p\bar q}g'^{r\bar s}u_{i\bar l
pr}u_{\bar jk\bar q
\bar  s}\\
\Psi&=&g'^{i\bar j}g'^{k\bar l}g'^{p\bar q}g'^{r\bar s}u_{i\bar l
p\bar s}u_{\bar jk\bar qr}
\end{eqnarray*}
We want to compute
\begin{equation}\lab{1901}
P((\kappa_1-\bigtriangleup u)\Theta+\kappa_2(m-\bigtriangleup
u)\Gamma+\kappa_3\mid\bigtriangledown
u\mid^2\Gamma+\kappa_4\Gamma)
\end{equation}where all $\kappa_i$ are positive constants and $m$ is a fixed constant
such that $\kappa_1-\bigtriangleup u>1$ and $m-\bigtriangleup
u>0$.  In the following computation, when we use some basic
inequalities such as Young inequality, Schwarz inequality, we will
not mention it. We will use $m_i$ to denote some positive constant
which depend on $f$, $M$ and $u$ up to second order derivations
and use $C_6$ as the constant in generic sense. We take our
calculation at the point $q_3$ and pick the normal coordinate at
this point such that $g_{i\bar j}=\delta_{ij}$, $\partial g_{i\bar
j}/\partial z_k=\partial g_{i\bar j}/\partial \bar z_k=0$. Now we
compute every terms in (\ref{1901}). At first we compute
\begin{equation}\lab{1902}
\begin{aligned}
P(\mid\bigtriangledown
u\mid^2)&=4g'^{\alpha\bar\beta}\partial_\alpha\partial_{\bar\beta}(g^{i\bar
j}u_iu_{\bar
j})\\
&=4g'^{\alpha\bar\beta}g^{i\bar j}(u_{i\bar\beta\alpha}u_{\bar
j}+u_{i}u_{\bar j\alpha\bar\beta}+u_{i\bar\beta}u_{\bar
j\alpha}+u_{i\alpha}u_{\bar
j\bar \beta})-C_6\\
&=4g'^{\alpha\bar\beta}g^{i\bar j}\{u_{i\alpha}u_{\bar
j\bar\beta}+u_{i\bar\beta\alpha}u_{\bar j}+u_iu_{\bar
j\alpha\bar\beta}+u_{i\bar\beta}u_{\bar j\alpha}\}-C_6\\
&\geq m_1\Gamma-C_6\Theta^{\frac{1}{2}}-C_6
\end{aligned}
\end{equation}
Next we use equation (\ref{2504}) to compute:
\begin{equation}\lab{1903}
\begin{aligned}
P(-\bigtriangleup u)&=-4g'^{\alpha\bar\beta}g^{i\bar
j}\frac{\partial^{4}u}{\partial z_i\partial\bar z_j\partial
z_\alpha\partial \bar z_\beta}-4g'^{\alpha\bar\beta}
\frac{\partial^2g^{i\bar j}}{\partial z_\alpha\partial\bar z_j}
\frac{\partial^2u}{\partial z_i\partial\bar z_j}\\
&=16g'^{\alpha\bar p}g'^{q\bar\beta}g^{i\bar
j}u_{i\bar\beta\alpha}u_{\bar j q\bar
p}\\
&-4g'^{\alpha\bar p}g'^{q\bar\beta}g^{i\bar
j}(u_{i\bar\beta\alpha}\partial_{\bar j}(e^u-fe^{-u})g_{q\bar
p}+u_{\bar jq\bar
p}\partial_i(e^u-fe^{-u})g_{\alpha\bar\beta})\\
&+F^{-1}g^{i\bar j}\partial_i\partial_{\bar j}F-F^{-2}g^{i\bar
j}\partial_i F\partial_{\bar j}F-C_6\\
&\geq m_2\Theta-C_6g^{i\bar j}\partial_i\partial_{\bar
j}(\mid\bigtriangledown u\mid^2)-C_6g^{i\bar j}\partial_i
(\mid\bigtriangledown u\mid^2)
\partial_{\bar j}(\mid\bigtriangledown u\mid^2)\\
&\geq m_2\Theta-C_6\Gamma-C_6
\end{aligned}
\end{equation}
Certainly we should also calculate:
\begin{equation}\lab{1904}
\begin{aligned}
P(\Gamma)&=2g'^{\alpha\bar\beta}\partial_\alpha\partial_{\bar\beta}\Gamma \\
&=2g'^{\alpha\bar \beta}g^{i\bar j}g^{k\bar l}(u_{ik\bar\beta
\alpha}u_{\bar j\bar l}+u_{ik} u_{\bar j\bar
l\alpha\bar\beta}+u_{ik\alpha} u_{\bar j\bar
l\bar\beta}+u_{ik\bar\beta}u_{\bar j\bar l\alpha})
\\
&+2g'^{\alpha\bar\beta}\partial_{\alpha}\partial_{\bar\beta}
(g^{i\bar j}g^{k\bar l})(u_{ik}u_{\bar j\bar l})-C_6\Gamma\\
 &\geq
m_3\Xi+m_3\Theta-\epsilon_1\kappa_4^{-1}\Phi-C_6\kappa_4\epsilon_1^{-1}\Gamma
\end{aligned}\end{equation}
Combining (\ref{1902}) and (\ref{1904}), we can  estimate
\begin{equation}\lab{1905}
\begin{aligned}
&P(\mid\bigtriangledown u\mid^2\Gamma)\\
&=P(\mid\bigtriangledown u\mid^2)\Gamma+\mid\bigtriangledown
u\mid^2P(\Gamma)\\
 &+2g'^{\alpha\bar\beta}(\partial_\alpha(\mid\bigtriangledown u\mid^2)\partial_{\bar\beta}\Gamma
 +\partial_{\bar\beta}(\mid\bigtriangledown u\mid^2)\partial_\alpha \Gamma)
\\
&\geq
m_1\Gamma^2-C_6\Theta^{\frac{1}{2}}\Gamma-C_6\Gamma+\mid\bigtriangledown
u\mid^2(m_3\Xi+m_3\Theta-\epsilon_1\Phi-C_6\epsilon_1^{-1}\Gamma)\\
&-C_6\Gamma^{\frac{1}{2}}(\Theta^{\frac{1}{2}}+\Xi^{\frac{1}{2}}+\Gamma^{\frac{1}{2}})\Gamma^{\frac{1}{2}}
\\
&\geq
m_1\Gamma^2-\epsilon_1\kappa_3^{-1}\Phi-C_6\kappa_3\epsilon_1^{-1}\Gamma-C_6\Xi-C_6\Theta
 \end{aligned}
 \end{equation}
 Combining (\ref{1903}) and (\ref{1904}), we get
\begin{equation}\lab{1906}
\begin{aligned}
&\ \ \ \ P((m-\bigtriangleup u)\Gamma)\\
&=P(-\bigtriangleup u)\Gamma+(m-\bigtriangleup
u)P(\Gamma)-2g'^{\alpha\bar\beta}\{\partial_\alpha(\bigtriangleup
u)\partial_{\bar\beta}\Gamma+\partial_{\bar\beta}(\bigtriangleup u)\partial_{\alpha}\Gamma\}\\
&\geq (m_2\Theta-C_6\Gamma-C_6)\Gamma+(m-\bigtriangleup
u)(m_3\Xi+m_3\Theta-\epsilon_1\Phi-C_6\epsilon_1^{-1}\Gamma)\\
&\ \ \ -C_6\Theta^{\frac{1}{2}}(\Theta^{\frac{1}{2}}+
\Xi^{\frac{1}{2}}+\Gamma^{\frac{1}{2}})\Gamma^{\frac{1}{2}}\\
&\geq
m_2\Theta\Gamma-C_6\Gamma^2-\epsilon_3\kappa_2^{-1}\Phi-C_6\kappa_2\epsilon^{-1}\Gamma-C_6\Xi-C_6\Theta
 \end{aligned}
\end{equation}
Now we deal with
 \begin{equation}\lab{2003}
 \begin{aligned}
P((\kappa_1-\bigtriangleup u)\Theta)&=P(\kappa_1-\bigtriangleup
u)\Theta+(\kappa_1-\bigtriangleup
u)P(\Theta)\\
&-2g'^{\alpha\bar\beta}\{\partial_{\alpha}(\bigtriangleup
u)\partial_{\bar\beta}\Theta+\partial_{\bar\beta}(\bigtriangleup
u)\partial_{\alpha}\Theta\}
\end{aligned}
\end{equation}
Applying (\ref{1903}), we get
\begin{equation}\lab{2002}
P(\kappa_1-\bigtriangleup
u)\Theta=m_2S^2-C_6\Gamma\Theta-C_6\Theta
\end{equation}
Let $(\kappa_1-\bigtriangleup u)\Theta+\kappa_2(m-\bigtriangleup
u)\Gamma+\kappa_3\mid\bigtriangledown
u\mid^2\Gamma+\kappa_4\Gamma$ achieve the maximum at the point
$q_3$. Then at the point $q_3$, we have,
\begin{eqnarray*}
\partial_{\bar\beta}\Theta=-\frac{1}{\kappa_1-\bigtriangleup
u}\{\Theta\partial_{\bar\beta}(m-\bigtriangleup
u)+\kappa_2\partial_{\bar\beta}((m-\bigtriangleup
u)\Gamma)+\kappa_3\partial_{\bar\beta}(\mid\bigtriangledown
u\mid^2\Gamma)+\kappa_4\partial_{\bar\beta}\Gamma\}
\end{eqnarray*}
and
\begin{equation}\lab{1907}
\begin{aligned}
&\ \ \ \
g'^{\alpha\bar\beta}\{\partial_\alpha(\kappa_1-\bigtriangleup
u)\partial_{\bar\beta}\Theta+\partial_{\bar\beta}(\kappa_1-\bigtriangleup
u)\partial_{\alpha}\Theta\}\\
&=2\text{Re}\ g'^{\alpha\bar\beta}\frac{(\bigtriangleup
u)_{\alpha}}{\kappa_1-\bigtriangleup u}\{-(\bigtriangleup
u)_{\bar\beta}\Theta-\kappa_2(\bigtriangleup
u)_{\bar\beta}\Gamma+\kappa_3(\mid\bigtriangledown
u\mid^2)_{\bar\beta}\Gamma\\
&\ \ \ \ \ \ \ \ \ \ \ \ \ \ \ \ \ \ \ \ \ \ \ \ \ \ \
\ +[\kappa_2(m-\bigtriangleup u)+\kappa_3\mid\bigtriangledown u\mid^2+\kappa_4]\Gamma_{\bar\beta}\}\\
&\geq \frac{-C_6}{\kappa_1-\bigtriangleup
u}\Theta^{\frac{1}{2}}\times\{\Theta^{\frac{3}{2}}+\kappa_2\Theta^{\frac{1}{2}}\Gamma+\kappa_3\Gamma^{\frac{3}{2}}\\
&\ \ \ \ \ \ \ \ \ \ \ \ \ \ \ \ \ \ \ \ \ \ \ \
 \ \
+(\kappa_2+\kappa_3+\kappa_4)(\Theta^{\frac{1}{2}}+\Xi^{\frac{1}{2}}+\Gamma^{\frac{1}{2}})\Gamma^{\frac{1}{2}}\}\\
&\geq \frac{-C_6}{\kappa_1-\bigtriangleup
u}\{\Theta^2+(\kappa_2+\kappa_3+\kappa_4)(\Theta\Gamma+\Theta+\Gamma+\Xi)+\kappa_3\Gamma^2\}
\end{aligned}
\end{equation}
At last we should estimate of $P(\Theta)$. We follow  paper
\cite{Yau}. We can get:
\begin{equation}\lab{1908}
\begin{aligned}
P(\Theta)=&2g'^{\alpha\bar\beta}[\ \ \ 2g'^{i\bar a}g'^{b\bar
p}g'^{q\bar r}g'^{s\bar j}g'^{k\bar t}+2g'^{i\bar p}g'^{q\bar
a}g'^{b\bar
r}g'^{s\bar j}g'^{k\bar t}\\
&\ \ \ \ \ \ \ +2g'^{i\bar p}g'^{q\bar r}g'^{s\bar a}g'^{b\bar
j}g'^{k\bar t}+2g'^{i\bar p}g'^{q\bar r}g'^{s\bar
j}g'^{k\bar a}g'^{b\bar t}\\
&\ \ \ \ \ \ \ +\ g'^{i\bar a}g'^{b\bar r}g'^{s\bar p}g'^{q\bar
j}g'^{k\bar t}+\ g'^{i\bar r}g'^{s\bar
a}g'^{b\bar p}g'^{q\bar j}g'^{k\bar t}\\
&\ \ \ \ \ \ \  +\ g'^{i\bar r}g'^{s\bar p}g'^{q\bar a}g'^{b\bar
j}g'^{k\bar t}+\ g'^{i\bar r}g'^{s\bar
p}g'^{q\bar j}g'^{k\bar a}g'^{b\bar t}]\\
 &\ \ \ \ \ \ \ \ \ \ \ \ \ \ \ \ \ \times \partial_\alpha g'_{b\bar a}\partial_{\bar\beta}g'_{\bar
pq}u_{i\bar jk}u_{\bar rs\bar t}\ \ (\text{first class})
\\
&-2g'^{\alpha\bar\beta}[2g'^{i\bar p}g'^{q\bar r}g'^{s\bar
j}g'^{k\bar t}+g'^{i\bar
r}g'^{s\bar p}g'^{q\bar j}g'^{k\bar t}]\\
&\ \ \ \ \times [\partial_{\bar\beta}g'_{\bar pq}u_{i\bar
jk\alpha}u_{\bar rs\bar t}+ \partial_{\alpha}g'_{q\bar
p}u_{\bar r s\bar t\bar\beta}u_{i\bar j k}] \ \ (\text{second class})\\
&-2g'^{\alpha\bar\beta}[2g'^{i\bar p}g'^{q\bar r}g'^{s\bar
j}g'^{k\bar t}+g'^{i\bar
r}g'^{s\bar p}g'^{q\bar j}g'^{k\bar t}]\\
&\ \ \ \times [\partial_{\bar\beta}g'_{\bar pq}u_{i\bar jk}u_{\bar
rs\bar t\alpha}+\partial_\alpha g'_{q\bar
p}u_{ i \bar j k\bar\beta}u_{\bar r s\bar t}]\ \  (\text{third class})\\
&-2g'^{\alpha\bar\beta}[2g'^{i\bar p}g'^{q\bar r}g'^{s\bar
j}g'^{k\bar t}+g'^{i\bar r}g'^{s\bar p}g'^{q\bar j}g'^{k\bar
t}]\times
\partial_\alpha\partial_{\bar\beta}g'_{\bar pq}u_{i\bar jk}u_{\bar rs\bar
t}\ \ (\text{forth
class})\\
&+2g'^{\alpha\bar\beta}g'^{i\bar r}g'^{s\bar j}g'^{k\bar t}\times
[u_{i\bar jk\bar \beta\alpha}u_{\bar rs\bar t}+u_{i\bar jk}u_{\bar
rs\bar t\bar
\beta\alpha}]\ \ (\text{fifth class})\\
&+2g'^{\alpha\bar\beta}g'^{i\bar r}g'^{s\bar j}g'^{k\bar t}\times
[u_{i\bar jk\bar \beta}u_{\bar rs\bar t\alpha}+u_{i\bar
jk\alpha}u_{\bar rs\bar t\bar
\beta}]\ \ (\text{sixth class})\\
&-C_6\Theta
\end{aligned}
\end{equation}
Comparing with (A.8) in \cite{Yau}, we should deal with some
classes in (\ref{1908}). The first class is:
\begin{equation}\lab{1909}
\begin{aligned}
&\ \ \ \ 2g'^{\alpha\bar\beta}g'^{i\bar a}g'^{b\bar p}g'^{q\bar
r}g'^{s\bar j}g'^{k\bar t} \partial_\alpha g'_{b\bar
a}\partial_{\bar\beta}g'_{\bar
pq}u_{i\bar jk}u_{\bar rs\bar t}\\
&=2g'^{\alpha\bar\beta}g'^{i\bar a}g'^{b\bar p}g'^{q\bar
r}g'^{s\bar j}g'^{k\bar t} (-4u_{b\bar a\alpha})(-4u_{\bar
pq\bar\beta})u_{i\bar jk}u_{\bar rs\bar
t}\\
&\ \ \ -\epsilon_2(\kappa_1-\bigtriangleup u) ^{-1}\Theta^2-C_6
\end{aligned}
\end{equation}
The second class is
\begin{equation}\lab{1910}
\begin{aligned}
&\ \ \ \ -2g'^{\alpha\bar\beta}g'^{i\bar p}g'^{q\bar r}g'^{s\bar
j}g'^{k\bar t} \partial_{\bar\beta}g'_{\bar
pq}u_{i\bar jk\alpha}u_{\bar rs\bar t}\\
&=-4g'^{\alpha\bar\beta}\g'^{i\bar p}g'^{q\bar r}g'^{s\bar
j}g'^{k\bar t} \partial_{\bar\beta}((e^u-fe^{-u})g_{\bar
pq}-4u_{\bar pq})u_{i\bar
jk\alpha}u_{\bar rs\bar t}\\
&=-4g'^{\alpha\bar\beta}g'^{i\bar p}g'^{q\bar r}g'^{s\bar
j}g'^{k\bar t} (-4u_{\bar pq\bar\beta})u_{i\bar jk\alpha}u_{\bar
rs\bar t}
\\
&\ \ \ -\epsilon_1(\kappa_1-\bigtriangleup
u)^{-1}\Phi-(\kappa_1-\bigtriangleup u)\epsilon_1^{-1}\Theta
 \end{aligned}
\end{equation}
As the same reason, the third class is:
\begin{equation}\lab{1911}
\begin{aligned}
&\ \ \ \ -2g'^{\alpha\bar\beta}g'^{i\bar p}g'^{q\bar r}g'^{s\bar
j}g'^{k\bar t}\partial_\alpha g'_{q\bar p}u_{ i \bar j
k\bar\beta}u_{\bar r s\bar
t}\\
&\geq-2g'^{\alpha\bar\beta}g'^{i\bar p}g'^{q\bar r}g'^{s\bar
j}g'^{k\bar t}(-4u_{q\bar p\alpha})u_{ i \bar j k\bar\beta}u_{\bar
r s\bar
t}\\
&\ \ \ -\epsilon_1(\kappa_1-\bigtriangleup
u)^{-1}\Psi-(\kappa_1-\bigtriangleup u)\epsilon_1^{-1}\Theta
\end{aligned}
\end{equation}
Next we deal with the forth class. We take the normal coordinate
at the point $q_3$. Then according to section 1 in \cite{Yau}, by
direct calculation, we can get
\begin{equation*}
u_{\bar
pq\bar\beta\alpha}=\partial_{\bar\beta}\partial_\alpha\partial_{\bar
p}\partial_{q}u-u_{q\bar \gamma}R^{\bar\gamma}_{\bar
p\alpha\bar\beta}=\partial_{\bar\beta}\partial_\alpha\partial_{\bar
p}\partial_{q}u-C_6
\end{equation*}
So by (\ref{2504}),
\begin{equation}\lab{1912}
\begin{aligned}
&\ \ \ \ -2g'^{\alpha\bar\beta}g'^{i\bar p}g'^{q\bar r}g'^{s\bar
j}g'^{k\bar t} \partial_\alpha\partial_{\bar\beta}g'_{\bar
pq}u_{i\bar jk}u_{\bar rs\bar t}\\
&\geq-2g'^{\alpha\bar\beta}g'^{i\bar p}g'^{q\bar r}g'^{s\bar
j}g'^{k\bar t} (-4u_{\bar\beta\alpha\bar p q}
)u_{i\bar jk}u_{\bar rs\bar t}-C_6\Theta\\
&\geq-2g'^{i\bar p}g'^{q\bar r}g'^{s\bar j}g'^{k\bar
t}g'^{\alpha\bar a}g'^{b\bar\beta} (-4u_{\bar ab\bar p })
(-4u_{\alpha\bar\beta q})u_{i\bar jk}u_{\bar rs\bar t}-\epsilon \Theta^2-C_6\Theta\\
&\ \ \ \ -2g'^{i\bar p}g'^{q\bar r}g'^{s\bar j}g'^{k\bar t}\left\{
F^{-1}F_{q\bar p}-F^{-2}F_qF_{\bar p}\right\}u_{i\bar
jk}u_{\bar rs\bar t}\\
 &\geq-2g'^{i\bar p}g'^{q\bar r}g'^{s\bar
j}g'^{k\bar t}g'^{\alpha\bar a}g'^{b\bar\beta} (-4u_{\bar ab\bar p
})
(-4u_{\alpha\bar\beta q})u_{i\bar jk}u_{\bar rs\bar t}\\
&\ \ \ \ -C_6\Theta\Gamma-m_2/8(\kappa_1-\bigtriangleup
u)^{-1}\Theta^2-C_6(\kappa_1-\bigtriangleup u)\Theta
\end{aligned}
\end{equation}
Now we deal with the fifth term. By direct calculation, we have
\begin{eqnarray*}
u_{i\bar
jk\bar\beta\alpha}&=&\partial_{\alpha}\partial_{\bar\beta}
\partial_k\partial_{\bar j}\partial_{\bar
i}u+u_{p\bar j\bar \beta}R^{p}_{ik\alpha} +u_{p\bar
j\alpha}R^{p}_{ik\bar\beta}+u_{i\bar p
k}R^{p}_{j\alpha\beta}\\
&&+u_{p\bar j}\partial_\alpha\partial_{\bar\beta}(g^{p\bar
q}\partial_k(g_{i\bar q}))\\
&=&\partial_{\alpha}\partial_{\bar\beta}
\partial_k\partial_{\bar j}\partial_{\bar
i}u+u_{p\bar j\bar \beta}R^{p}_{ik\alpha} +u_{p\bar
j\alpha}R^{p}_{ik\bar\beta}+u_{i\bar p
k}R^{p}_{j\alpha\beta}+C_6\\
\end{eqnarray*}
Differentiating (\ref{2504}) and using above equality, we can deal
with the fifth class:
\begin{equation}\lab{1913}
\begin{aligned}
&\ \ \ \ g'^{\alpha\bar\beta}g'^{i\bar r}g'^{s\bar j}g'^{k\bar
t}u_{i\bar
jk\bar \beta\alpha} u_{\bar r s\bar t}\\
&=g'^{i\bar r}g'^{s\bar j}g'^{k\bar t}\left\{g'^{\alpha\bar
p}g'^{q\bar\beta}g'_{q\bar pk}u_{\alpha\bar\beta i\bar
j}-1/4(g'^{\alpha\bar p}g'^{q\bar \beta}g'_{\bar pq\bar
j}g'_{\alpha\bar\beta,i})_k\right\}u_{\bar rs\bar t}\\
&\ \ \ \ +\frac{1}{4}g'^{i\bar r}g'^{s\bar j}g'^{k\bar
t}\{-F^{-1}F_{i\bar jk}+F^{-2}(F_kF_{i\bar j}+F_iF_{\bar
jk}+F_{\bar j}F_{ik})\\
& \ \ \ \ \ \ \ \ \ \ \ \ \ \ \ \ \ \ \ \ \ \ \ \ \
-F^{-3}F_iF_{\bar
j}F_k\}u_{\bar rs\bar t}+C_6\Theta\\
&=g'^{i\bar r}g'^{s\bar j}g'^{k\bar t}g'^{\alpha\bar
\beta}(-4u_{q\bar pk})u_{\alpha\bar\beta i\bar j}u_{\bar rs\bar
t}+C_6\Theta\\
&\ \ \ \ +g'^{i\bar r}g'^{s\bar j}g'^{k\bar t}g'^{\alpha\bar
p}g'^{q\bar\beta}\{(-4u_{\alpha\bar\beta k})u_{\bar p q\bar
jk}+(-4u_{\bar pq\bar j})u_{\alpha\bar\beta ik})\}u_{\bar rs\bar t}\\
&\ \ \ \ -g'^{i\bar r}g'^{s\bar j}g'^{k\bar t}(g'^{\alpha\bar
a}g'^{b\bar p}g'^{q\bar\beta}+g'^{\alpha\bar p}g'^{q\bar
a}g'^{b\bar\beta})(-4u_{b\bar ak})(-4u_{\bar pq\bar
j})u_{\alpha\bar\beta i}u_{\bar rs\bar t}\\
&\ \ \ \ -\epsilon_1(\kappa_1-\bigtriangleup
u)^{-1}(\Phi+\Psi)-C_6\epsilon^{-1}_{1}(\kappa_1-\bigtriangleup
u)\Theta-C_6\Theta\Gamma-C_6\Gamma^2
\end{aligned}
\end{equation}
Inserting (\ref{1909})-(\ref{1913}) into (\ref{1908}),
diagonalizing and simplifying, then comparing to (A.8) and (A.9)
in \cite{Yau}, we obtain
\begin{equation}\lab{1914}
\begin{aligned}
P(\Theta)&\geq\sum g'^{i\bar i}g'^{j\bar j}g'^{k\bar
k}g'^{\alpha\bar\alpha}\times\mid u_{i\bar j k\bar\alpha}-4\sum_p
u_{i\bar pk}u_{\bar jp\bar\alpha}g'^{p\bar p}\mid^2\\
&\ \ \ \ +\sum g'^{i\bar i}g'^{j\bar j}g'^{k\bar
k}g'^{\alpha\bar\alpha}\times\mid u_{i\bar
jk\alpha}-4\sum_p(u_{i\bar p\alpha}u_{p\bar jk}+u_{i\bar
pk}u_{p\bar j\alpha})g'^{p\bar p}\mid^2\\
&\ \ \ \ -\frac{1}{\kappa_1-\bigtriangleup
u}\left\{2\epsilon_1\Phi+2\epsilon_1\Psi+(\epsilon^2+\frac{m_{2}}{4})\Theta^2\right\}\\
&\ \ \ \
-C_6\Theta\Gamma-C_6\Gamma^2-C_6\kappa_1\epsilon^{-1}_1\Theta-C_6
\\
&=\sum g'^{i\bar i}g'^{j\bar j}g'^{k\bar
k}g'^{\alpha\bar\alpha}\times\mid\sqrt{1-2\epsilon_1(\kappa_1-\bigtriangleup
u)^{-1}} u_{i\bar j
k\bar\alpha}\\
&\ \ \ \ \ \ \ \ \ \ \ \ \ \ \ \ \ \
 -4\left(\sqrt{1-2\epsilon_1(\kappa_1-\bigtriangleup
 u)^{-1}}\right)^{-1}
\sum_p
u_{i\bar pk}u_{\bar jp\bar\alpha}g'^{p\bar p}\mid^2\\
&+\sum g'^{i\bar i}g'^{j\bar j}g'^{k\bar
k}g'^{\alpha\bar\alpha}\times\mid
\sqrt{1-5\epsilon_1(\kappa_1-\bigtriangleup u)^{-1}}u_{i\bar
jk\alpha}\\
 &\ \ \ \ \ \ \ -4\left(\sqrt{1-5\epsilon_1(\kappa_1-\bigtriangleup
u)^{-1}}\right)^{-1}\sum_p(u_{i\bar p\alpha}u_{p\bar jk}+u_{i\bar
pk}u_{p\bar j\alpha})g'^{p\bar p}\mid^2\\
&\ \ +\frac{3\epsilon_1}{\kappa_1-\bigtriangleup
u}\Phi-\frac{m_2/4+\epsilon_2+C_6\epsilon_{1}}{\kappa_1-\bigtriangleup
u}\Theta^2-C_6\Theta\Gamma-C_6\Gamma-C_6\frac{\kappa_1-\bigtriangleup u}{\epsilon_1}\Theta \\
&\geq \frac{3\epsilon_1}{\kappa_1-\bigtriangleup
u}\Phi-\frac{m_2/4+\epsilon_2+C_6\epsilon_{1}}{\kappa_1-\bigtriangleup
u}\Theta^2-C_6\Theta\Gamma-C_6\Gamma-C_6\frac{\kappa_1-\bigtriangleup
u}{\epsilon_1}\Theta
\end{aligned}
\end{equation}
Inserting (\ref{2002}), (\ref{1907}) and (\ref{1914}) into
(\ref{2003}), and then inserting (\ref{2003}), (\ref{1904}),
(\ref{1905}) and (\ref{1906}) into (\ref{1901}), at last we obtain
\begin{equation}\lab{1915}
\begin{aligned}
&\ \ \ \ P((\kappa_1-\bigtriangleup
u)\Theta+\kappa_2(m-\bigtriangleup
u)\Gamma+\kappa_3\mid\bigtriangledown u\mid^2T+\kappa_4\Gamma)\\
&\geq\left(m_2-\frac{C_6}{\kappa_1-\bigtriangleup
u}-\frac{m^2}{4}-\epsilon_2-C_6\epsilon_1\right)\Theta^2\\
&+\left(m_2\kappa_2-C_6(\kappa_1-\bigtriangleup
u)-\frac{C_6}{\kappa_1-\bigtriangleup
u}(\kappa_2+\kappa_3+\kappa_4)-C_6\right)\Theta\Gamma\\
&+\left(m_1\kappa_3-\frac{C_6}{\kappa_1-\bigtriangleup u}\kappa_3-C_6\kappa_2\right)\Gamma^2\\
&+\left(m_3\kappa_4-\frac{C_6}{\kappa_1-\bigtriangleup u}(\kappa_2+\kappa_3+\kappa_4)-C_6(\kappa_2+\kappa_3)\right)\Xi\\
&-C_6(\kappa_i,\epsilon_1)(\Theta+\Gamma)
\end{aligned}
\end{equation}
Now we can think the generic constant $C$ is fixed, because we can
take the biggest one. Fix $\epsilon_1$ and $\epsilon_2$ such that
$\epsilon_2+C_6\epsilon_1=\frac{m_2}{4}$. Take $\kappa_1$ big
enough such that $\frac{C_6}{\kappa_1-\bigtriangleup
u}<\frac{m_2}{4}$, then
\begin{equation}\lab{1916}
\left(m_2-\frac{C_6}{\kappa_1-\bigtriangleup
u}-\frac{m^2}{4}-\epsilon_2-C_6\epsilon_1\right)\Theta^2>\frac{m_2}{4}\Theta^2
\end{equation}
Let
\begin{equation*}
k_i=\frac{\kappa_i}{\kappa_1-\bigtriangleup u}  \ \ \ \
\text{for}\ \ \ i=2,3,4.
\end{equation*}
We  choose $k_2$, $k_3$ and $k_4$ such that
$$k_2>\frac{C_6}{m_2}+1$$
$${k_3}>\left(\frac{C_6}{m_1}+1\right)k_2$$
and
$$k_4>C_6\frac{k_2+k_3}{m_3}+1.$$
Then if we choose $\kappa_1$ big enough, we have
\begin{equation}\lab{1917}
\begin{aligned}
 &\left(m_2\kappa_2-C_6(\kappa_1-\bigtriangleup
u)-\frac{C_6}{\kappa_1-\bigtriangleup
u}(\kappa_2+\kappa_3+\kappa_4)-C_6\right)\Theta\Gamma\\
&>(m_2(\kappa_1-\bigtriangleup
u)-C_6(k_2+k_3+k_4)-C_6)\Theta\Gamma>\frac{m_2}{2}\kappa_1\Theta\Gamma,
\end{aligned}
\end{equation}
\begin{equation}\lab{1918}
\begin{aligned}
&\left(m_1\kappa_3-\frac{C_6}{\kappa_1-\bigtriangleup
u}\kappa_3-C_6\kappa_2\right)\Gamma^2\\
&>(m_1k_2(\kappa_1-\bigtriangleup
u)-C_6k_3)\Gamma^2>\frac{m_1}{2}k_2\kappa_1\Gamma^2
\end{aligned}
\end{equation}
and
\begin{equation}\lab{1919}
\begin{aligned}
&\left(m_3\kappa_4-\frac{C_6}{\kappa_1-\bigtriangleup
u}(\kappa_2+\kappa_3+\kappa_4)-C_6(\kappa_2+\kappa_3)\right)\Xi\\
&>(m_3(\kappa_1-\bigtriangleup
u)-C_6(k_1+k_2+k_3))\Xi>\frac{m_3}{2}\kappa_1\Xi
\end{aligned}
\end{equation}
Inserting (\ref{1916}), (\ref{1917}), (\ref{1918}) and
(\ref{1919}), we see that
\begin{equation}\begin{aligned}
0&\geq P((\kappa_1-\bigtriangleup
u)\Theta+\kappa_2(m-\bigtriangleup
u)\Gamma+\kappa_3\mid\bigtriangledown u\mid^2\Gamma+\kappa_4\Gamma)\\
&\geq\frac{m_2}{4}\Theta^2+\frac{m_2}{2}\kappa_1\Theta\Gamma+
\frac{m_1}{2}k_2\kappa_1\Gamma^2+\frac{m_3}{2}\kappa_1\Xi
\end{aligned}
\end{equation}
Above inequality gives an estimate of the the quantity $\sup_S
\Theta$ and $\sup_S \Gamma$. This in turn gives the estimates of
$u_{i\bar jk}$ and $u_{ij}$ for all $i,j,k$.

\section{Solving the equation}
In conclusion, we have proved the following
\begin{prop}
Let $S$ be a K3 surface with Calabi-Yau metric $\omega_S$. Let $u$
be a real-valued function in $C^4(S)$ such that $\int_S
e^{-u}\frac{\omega^2}{2!}=A$ and
$(e^u-fe^{-u})\omega_S-2\sqrt{-1}\partial\bar\partial u$ defines
another hermitian metric on $S$. Suppose
$$\bigtriangleup(e^u-fe^{-u})=\frac{\det u_{i\bar j}}{\det
g_{i\bar j}}.$$ If
 \begin{equation}
 A<\min \left\{1, C_1^{-1}\left(\max\{7^{\frac{1}{3}},(2C_1)^2,(1+\sup f),
 16(\max R_{i\bar jk\bar l}+1)\}\right)^{-\frac{2}{B}}\right\}
 \end{equation}
where $C_1$ is a constant only depending on $S$, then there is a
constant $C_0$ depending only on $S$, $\sup f$,
$\sup\mid\bigtriangledown^l f\mid$, and $A$ such that $\sup_S\mid
u\mid<C_0$, $\sup_S\mid \bigtriangledown u\mid\leq C_0$,
$\sup_S\mid u_{i\bar j }\mid<C_0$, $\sup_S\mid u_{i\bar
jk}\mid<C_0$.
\end{prop}
By above Proposition, we see that $T$ is closed. Combining Lemma
17, we get the proof of Theorem 2.

\section{The general case}
Timing the elliptic condition
$e^u\omega_S+\sqrt{-1}e^{-u}\textup{tr}(\bar\partial
A\wedge\partial A^*\cdot g_S^{-1})-2\sqrt{-1}\partial\bar\partial
u>0$ by $pe^{-pu}$ and integrating, then we can get (\ref{1402}):
\begin{equation*}
\begin{aligned}
\int\mid\bigtriangledown
e^{-\frac{p}{2}u}\mid^2\frac{\omega_S^2}{2!}&<\frac{p}{4}\int
e^{-(p-1)u}\frac{\omega_S^2}{2!}+\frac{p}{4}\int
e^{-(p+1)u}\frac{\sqrt{-1}}{2} \text{tr}(\bar\partial
A\wedge\partial A^*\cdot
g^{-1})\wedge\omega_S\\
&\leq \frac{p}{4}\int e^{-(p-1)u}\frac{\omega_S^2}{2!}
\end{aligned}
\end{equation*}
because $\sqrt{-1}\text{tr}(\bar\partial A\wedge\partial A^*\cdot
g^{-1})\wedge \omega_S\leq 0$. Then we can follow the discussion
in section 7 to get the estimate $\inf u\geq -\ln
C_1-\frac{B}{2}\ln A$. If $A$ is small enough, we can get $\inf
u>0$ big enough. So the term $e^u$ is always  control the  term
such as $e^{-u}\mid\text{tr}(\bar\partial A\wedge\partial A^*\cdot
g^{-1})\mid$ and the all estimates can be derived as the
particular case.

\end{document}